# Reservoir Computing and Photoelectrochemical Sensors: A Marriage of Convenience

Gisya Abdi,[†⁋] Lulu Alluhaibi,[†⁋] Ewelina Kowalewska,[†⁋] Tomasz Mazur,[†⁋] Krzysztof Mech,[†⁋] Agnieszka Podborska,[†⁋] Andrzej Sławek,[†⁋] Hirofumi Tanaka[‡,§] and Konrad Szaciłowski[†]*

†AGH University of Science and Technology, Academic Centre for Materials and Technology, al. Mickiewicza 30, 30-059 Kraków, Poland

‡ Kyushu Institute of Technology, Research Center for Neuromorphic AI Hardware, 2-4 Hibikino, Wakamatsu, Kitakyushu 808-0196, Japan

§ Kyushu Institute of Technology, Graduate School of Life Science and Systems Engineering, 2-4 Hibikino, Wakamatsu, Kitakyushu 808-0196, Japan

* Correspondence should be addressed to Konrad Szaciłowski, e-mail: szacilow@agh.edu.pl

**KEYWORDS**: information processing, reservoir computing, photoelectrochemistry, semiconductors, electron transfer, energy transfer, intelligent systems

**ABSTRACT:** Sensing technology is an important aspect of information processing. Current development in artificial intelligence systems (especially those aimed at medical and environmental applications) requires a lot of data on the chemical composition of biological fluids or environmental samples. These complex matrices require advanced sensing devices, and photoelectrochemical ones seem to have potential to overcome at least some of the obstacles. Furthermore, the development of artificial intelligence (AI) technology for autonomous robotics requires technology mimicking human senses, also those operating at the molecular level, such as gustation and olfaction. Again, photoelectrochemical sensing can provide some suitable solutions. In this review, we introduce the idea of integration of photoelectrochemical sensors with some unconventional computing paradigm – reservoir computing. This approach should not only boost the performance of the sensors itself, but also open new pathways through science. Integration of sensing devices with computing systems will also contribute to a better understanding (or at least mimicking) of the human senses and neuromorphic sensory information processing. Although reservoir systems can be considered magic "black boxes" and their operation is at the same time simple and hard to comprehend, this combination is expected to open a new era of effective information harvesting and processing systems.

## 1. Introduction

Chemosensing can be regarded as an information processing process at numerous levels of abstraction. In the most simplistic approach, sensing can be regarded as the translation of molecular syntactic information (presence or absence of chemical species, or their concentration) into pragmatic information of industrial, medical, or environmental importance (to name only the same examples) via semantic information of analytical signals [1, 2]. This process, usually relatively simple from the chemical point of view (selection of a proper receptor, recording of the signal, and its interpretation) is a nontrivial computational task from the point of view of information theory, as sensing can be regarded as information encoding process, which is an essential part of computation. Classical sensing protocols are based on a linear information flow: from the analyzed sample (or environment) to the observer. Moreover, classical sensing is a one-time event, or a question-answer-like situation – classical sensing is an instantaneous process, at least from the point of view of information theory [3]. In a more practical approach the time scale of chemical sensing is limited by the reaction kinetics, but most of the practical systems operate on the basis of fast ionic reaction, so time constants for the sensing processes are not relevant. In more complex cases, especially in the case of sensor arrays, advanced postprocessing, e.g. principal factor analysis or other chemometric methods, may be time and computational resource-consuming actions [4-15].

In this review, we want to show another possibility of application of advanced information processing for chemical sensing: reservoir computing (RC) [16]. This idea has been theoretically addressed pointing out that applications of the principles of reservoir computing in chemical sensing (i) are possible and (ii) may bring a significant improvement in sensor performance.[17, 18] Despite the great computing power of RC systems there are only TWO published reports on experimental integration of the principles of RC with chemosensors: application of RC algorithms in post-processing of gas sensors, reported by Fonolosa et al. [19] and RC computing device with sensing element as an active node in reservoir, as reported by Przyczyna et al. [20] The main reason for this lack of popularity may originate from apparent stochasticity and randomness of RC systems, as well as from technical difficulties in constructing good physical reservoirs [16, 21-24].

The current paper presents a brief overview of the principles of reservoir computing with a special focus on implementations, which create the possibility of direct applications in chemical sensing, followed by an overview of vavarious photoelectrochemical sensing protocols. It will be further demonstrated that, among all sensing techniques, photoelectrochemical sensing, because of its hybrid character, is naturally the most suited for integration with RC systems.

## 2. Principles of reservoir computing

Reservoir computing is one of the fields in unconventional and neuromorphic computing domain. It originates from the concept of artificial recurrent neural networks, especially their application in simulation of dynamic systems. In contrast to artificial neural networks, RC systems can be relatively easily (at least from theoretical point of view) implemented in systems other than networks: in time- and/or space-continuous systems with appropriate internal dynamics. RC systems, like any other computational systems, must exhibit computational primitives: the transmission, storage, and modification of input data [25]. In the case of RC systems the transmission is ensured by coupling input and output by a transfer function (cf. Eq.2), information storage is embodied in the fading memory feature and modified by the echo state property.

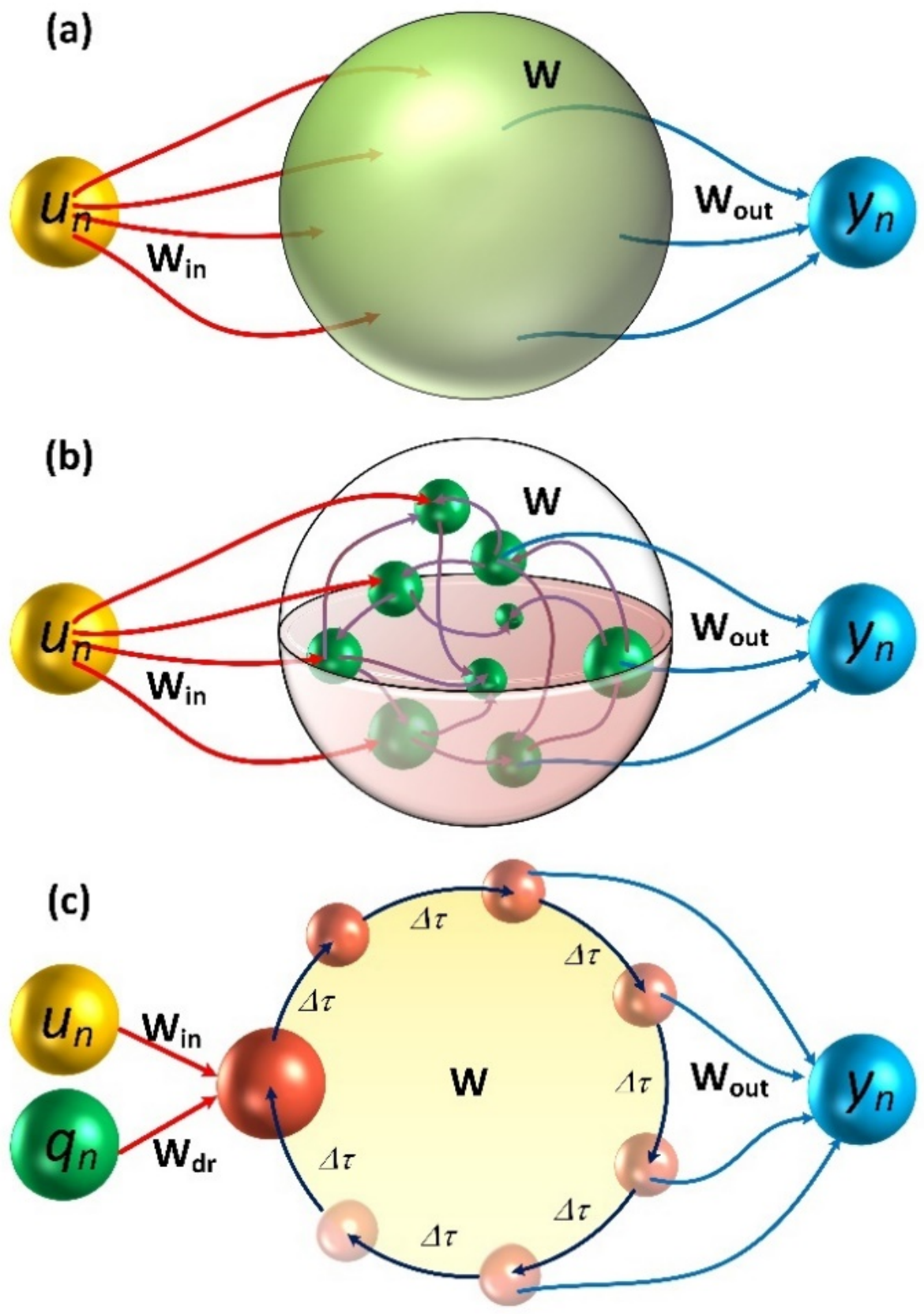

*Figure 1. Basic types of reservoir computing systems: liquid-state machine (a), echo-state network (b), and delay feedback loop with an external drive (c).*

The key notion in RC is a reservoir: any dynamical system that undergoes evolution in time and responds to external inputs. Furthermore, a system to be regarded as a reservoir must be characterized by a couple of features: (i) separability, (ii) the echo state property, and (iii) fading memory. A complete RC system must thus comprise three elements: input (an interface that couples the reservoir with the input or just its environment), reservoir itself, and a readout layer, a simple yet trainable neutral-network-like linear system. Unlike artificial neural networks in RC systems, the readout layer is the only trainable part of the system.

The basic rule of the RC approach is to harness the internal (spontaneous or stimulated) dynamics of the system for information processing purposes. It is a challenging task to find systems of suitable complexity for a given computational task. Trial-and-error searching is the most common method, which, however, cannot guarantee success [22, 26].

There are three most common implementation of RC systems: liquid state machines, echo-state networks, and (single node) delayed feedback systems (Figure 1) [27, 28]. Regardless of the implementation, the formal description and principles of operation remain the same; however, the first two systems can be regarded as collective state computing systems, whereas the last (at least in most physical implementations) is a sequential state computing system [29]. The first and the last one can be easily implemented in physical systems, whereas the second is usually implemented in software.

In physical implementations, the liquid state machine is usually based on a sample of macroscopically structureless material, *e.g.* container with liquid or a properly shaped piece of polymer (Figure 1a) [21, 30, 31]. The input interface $\mathbf{W}_{\mathbf{in}}$ provides appropriate coupling with the environment or data input, whereas the trainable output layer $\mathbf{W}_{\mathbf{out}}$ (in the simplest case a linear, feedforward neural network) yields the results of computation to the information recipient. Echo state networks (Figure 1b) are based on a finite number of nonlinear nodes with unknown connection topology and unknown but stable synaptic weights [32]. In physical systems, they can be implemented in polycrystalline layers of semiconducting materials, where each crystallite may serve as an individual node [33, 34]. The delayed feedback systems are based on a single, usually highly nonlinear node, and a feedback loop providing proper dynamics (Figure 1c) [23, 24, 35]. In this case, the output layer takes states of the node at different points in time as individual internal states of the reservoir, i.e., output is time multiplexeds and not space multiplexed, as in either types of reservoir computers. In other words, this approach takes advantage of the feedback-stimulated input signal evolution. Regardless of the architecture of the reservoir, its main task is the geometric transformation of the phase space representing the input data (Ω) into a different phase space representing the result of the computation (Ω'), i.e. (1):

$$\mathfrak{F} : \Omega \rightarrow \Omega' \tag{1}$$

As the dimensionalities of the spaces Ω and Ω' may be different, the transformation $\mathfrak{F}$ is neither homeomorphism nor projection (since the dimensionality of Ω' may be higher than Ω) (Figure 2). In the case where the data are represented as time series, the RC system can be considered as a filter. For simplicity, we will specify a reservoir based on a feedback loop that operates in discrete time steps $\Delta\tau$. In this case, the state of the reservoir at a given time $t$ can be labelled as $x_n$ and at time $t + \Delta\tau$ as $x_{n+1}$. With this approach, and considering regular time intervals, the update of the internal state of the reservoir can be seen as (2):

$$x_n = (1-\alpha)x_{n-1} + \alpha f\left(\beta \mathbf{W_{in}} u_n + \gamma \mathbf{W_{dr}} q_n + \delta \mathbf{W} x_{n-1}\right) \tag{2}$$

where $f$ is the non-linear activation function and $x_n$ is the $m$-dimensional vector $x_n = \left[x_{1,n}, x_{2,n}, x_{2,n}, \ldots, x_{m,n}\right]$ denoting the state of the reservoir in a given disctrete time moment $n$. Continouos time representation of the reservoir evolution is also possible, but the simplest representations of echo state machines (cf. Figure 1c) are usually presented in discretised time domain [36]. The most common activation functions in the software-based reservoir are: Heaviside (3), logistic (4) and hyperbolic tangent (5) functions:

$$f(x) = \begin{cases} 1, x > 0 \\ 0, x \leq 0 \end{cases} \tag{3}$$

$$f(x) = \frac{1}{1+e^{-x}} \tag{4}$$

$$f(x) = \frac{e^x - e^{-x}}{e^x + e^{-x}} \tag{5}$$

$\mathbf{W_{in}}$, $\mathbf{W}$, and $\mathbf{W_{dr}}$ are weight matrices for input, internal states of the reservoir, and coupling with the drive, respectively, and $\alpha$, $\beta$, $\gamma$ and $\delta$ are the global scaling parameters controlling the performance of the reservoir. The first one, $\alpha$, controls the leak rate: the mismatch between the input and the reservoir dynamics, or the mixing between the input and the past state of the reservoir; when $\alpha = 1$ the previous state does not leak into the current state and the system does not have explicit memory. The input scaling $\beta$ controls the non-linear response of the reservoir on the current input, whereas the drive scaling parameter $\gamma$ does the same as the response to the drive signal. The internal scaling factor $\delta$ controls stability and responsiveness to external signals: low values dampen internal activity and increase response to inputs, while high values lead to self-sustained chaotic activity [26]. Thus, the state of the reservoir in the short form (cf. Eq. 2) can be described as follows:

$$x_n = \mathfrak{F}\left(x_{n-1}, u_n, q_n\right) \tag{6}$$

Any reservoir computing system for satisfactory operation should exhibit two principal attributes: to achieve universal computing power for time-series-type inputs: (i) the pointwise separation/generalization property (fulfilled when different inputs produce appreciably different states, while similar inputs should map to closely related reservoir states) and (ii) the approximation property (the readout layer can assign the internal state of the reservoir to the required output state with the predefined accuracy) [37]. These concepts are further discussed in detail in recent review articles [22, 38] and specialized books [39-41].

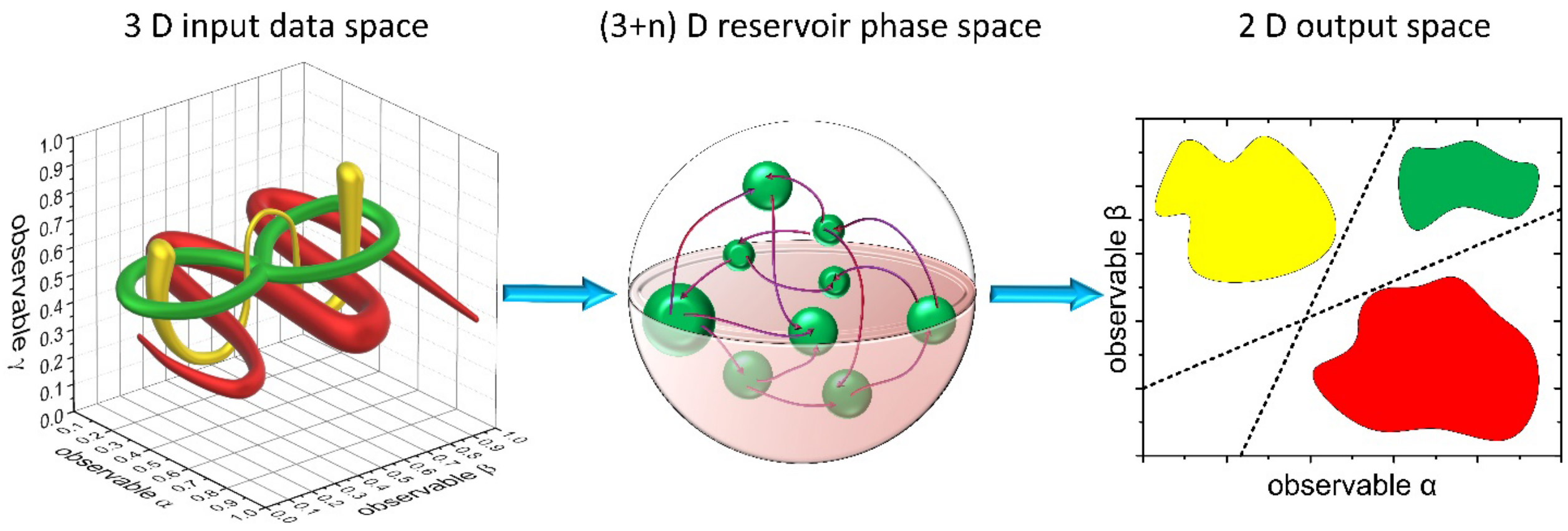

*Figure 2. Graphical representation of reservoir computer action on a complex shape in three-dimensional (3D) phase space resulting in a simple two-dimensional (2D) patterns, easily described by two observables. Dashed lines indicated possible separators calculated by a linear perceptron (output layer).*

The essential and congenial features of reservoir computing systems are: (i) fading memory and (ii) echo state property [42, 43]. Fading memory feature ensures (through the increasingly weaker, generation after generation, the influence of the primordial input on the present state of the reservoir) that the operation of the reservoir does not run into chaotic oscillatory mode. Formally, it is controlled by the $\delta$ scaling factor (cf. Eq. 2) It can be demonstrated that reservoir systems, alike any other dynamic systems used for computation, perform most efficiently at the edge of chaos, i.e. in the region of the phase space where the transition from ordered to disordered behaviour can be encontered [25, 44-46]. This provides the highest sensitivity to even a subtle change in the input signal, however, may lead to over-separation of similar inputs and generalization property will be lost. The RC system in the chaotic mode does not meet the generalization criterion: even infinitesimally small changes in the input may result in complete separation of outputs in the output phase space. In this case, no classification can be performed. However, as the name implies, the echo stat properties depend on the influence of the input signal on the dynamics of the reservoir, which in turn is reflected in the future state. In other words, the trajectory of the reservoir in the phase space depends, to some extent, on all previous inputs. It can be formally defined that the dynamic system has an echo state property if there exists a function $\mathfrak{S}$ formulated as (7) :[35]

$$x_n = \mathfrak{S}\left(u_{n-1}, u_{n-2}, u_{n-3}, \ldots, u_{-\infty}\right) \tag{7}$$

In this approximation, the dynamics of the reservoir does not depend on its initial stat, but only on the history of inputs (dependence on drive input is omitted for the sake of clarity, but formally any drive signal may be considered as an input, which is especially important in sensing applications). For arbitrarily chosen systems, the existence of the $\mathfrak{S}$ function cannot be guaranteed. Finite-time probing of the echo states on the basis of equations (2) and (7) gives a recurrent recipe for the state of the reservoir (8):

$$x_n = \mathfrak{H}^{(h)}\left(u_{n-1}, u_{n-2}, u_{n-3}, \ldots, u_{n-h+1}, x_{n-h}\right) \tag{8}$$

For a small number of steps, the recipes for obtaining the final state of the reservoir are relatively simple; examples for $h = 2$ and $h = 3$ are shown (9-10):

$$\mathfrak{H}^{(2)} = \mathfrak{F}\left(\mathfrak{F}\left(x_{n-2}, u_{n-1}\right), u_n\right) \tag{9}$$

$$\mathfrak{H}^{(3)} = \mathfrak{F}\left(\mathfrak{F}\left(\mathfrak{F}\left(x_{n-3}, u_{n-2}\right), u_{n-1}\right), u_n\right) \tag{10}$$

Although the existence of the $\mathfrak{S}$ function for an arbitrarily chosen dynamic system cannot be guaranteed, the limit (11) can be safely taken for systems with an echo state property.

$$\mathfrak{S} = \lim_{h\to\infty} \mathfrak{H}^{(h)} \tag{11}$$

The fading memory feature allows us to simplify for a finite number of time steps (12):

$$x_n = \tilde{\mathfrak{S}}^{(h)}\left(u_n, u_{n-1}, u_{n-2}, \ldots, u_{n-h+1}\right) \tag{12}$$

A more detailed analysis of reservoir dynamics can be found in works of Jaeger and Konkoli [17, 47-53].

Finally, the trainable readout layer should give the desired output of computation $\mathscr{D}$, constructed in such a way that it will preferably be a function of the most recent input and drive (13):

$$y_n = \mathbf{W}_{\mathbf{out}} x_n = \mathscr{D}\left(u_{n-1}, q_{n-1}\right) \tag{13}$$

Memory features are retained as much as possible, at least with respect to the input of the analyte, *u*. However, the success of this operation depends on the appropriate design of the drive signal [17, 18].

The quality of the reservoir computing system can be quantified as the ability to separate and generalize the input data: good systems should efficiently separate *significantly different* data sets and, at the same time, categorize *similar* data sets into the same category. The meanings of 'significantly different' and 'similar' are quite arbitrary here and depend on particular computational (or sensing) problem. Separability can be quantified as a number of different binary features that can be extracted by hyperplanes that bisect the phase space of the reservoir. Usually, the richer the dynamics of the reservoir or the higher the rank of matrix **W**, the better the separability. However, linear separation of the phase space into the largest possible number of fragments is not always a desired behaviour. Similar objects should be gathered in the same category, because most computational problems require some degree of generalization, *e.g.* classification of objects or patterns according to a defined set of features, whereas other features (and noise) should be neglected. In this case, the rank of the **W** matrix should be much smaller than for maximal separability.[54, 55] Furthermore, the reservoir's memory capacity should be high,[56] however for numerous applications, including prospective applications in sensing the entire depth of the reservoir's memory are not explicitly used.

Another measure of reservoir performance is the stability of its internal dynamics. This can be quantified using the spectral radius, that is, the largest eigenvalue) of the internal weight matrix **W**, which can be considered as the global scaling parameter (cf. Eq. 2). This parameter globally controls the weights of the internal connectivity of the reservoir and may move the system between different dynamic regimes, thus modulating its nonlinearity and complexity of the response. Effectively, the scaling parameter alters the internal timescales of the system, therefore, inflecting the echo state property. For operation reservoirs, the value of the spectral radius is usually centred around a value of one; smaller values are attributed to a stable regime (a fading response to input stimuli); whereas values significantly larger than one render reservoir highly unstable [57].

In principle any physical system with internal dynamics and any kind of hysteresis may work as core of the reservoir computing system [43]. The most common implementations of artificial neural networks are software-based [58, 59], whereas reservoir computing systems are in most cases heterotic ones [60] – a physical reservoir joint with efficient software-based readout layer [16, 21, 22]. The diversity of approaches is tremendous: electronics, photonics, spintronics, quantum dynamics, mechanics, and fluidics. The variety of materials for physical impementations of reservoirs is equally diverse: glasses and optical materials [61-64], nanomaterials [65-69], superconductors [70-73], magnetic materials [43], bulk and thin layer semiconductors and other inorganic materials [33, 34, 74-77], electrolytes and ionic liquids [78-82], conducting polymers [83, 84], analog and digital electronic circuits [85-87], as well as biomaterials and living tissues [88]. The only common feature for all these systems is their rich internal dynamics and the fact, that they have not been intentionally made or designed for computation, but can be creatively exploited for information processing tasks. For example, buckets are design to carry water of other liquid goods, but a bucket of water, equipped with a proper interface, can perform complex computational tasks [89]. Even without this dynamic behaviour (any mechanical stimulation of water in a bucket results in formation of waves, and complexity of wave physics helps to develop complex readouts [90]) if the stimulus results in a significant change in material's shape or morphology, the behaviour of such device can be understood in terms of morphological computing [21] and morphogenesis [91-93].

Each physical substrate shows different dynamics and responsiveness, therefore, for each material its utility for reservoir computing in particular task should be thoroughly examined. In the next step, an appropriate interface must be designed in order to maximize the coupling between the system's internal dynamics,  data input layer and the readout layer. Finally, a dictionary translating dynamic behavior patterns of the reservoir on stimulation, bearing syntactic and semantic information, into more pragmatic data formats. The latter is done by the trainable output layer, which first must be developed [94, 95]. All these aspects make the reservoir computing concepts hardly accessible, as there is a lot of serendipity required for the design of the functional system.

One on the most common approaches towards reservoir computing in based on application of thin layer memristive structures. Individual memristors can be incorporated into delayed feedback loops [33, 34, 96] or used as a dynamically reconfigurable matrix [97-100]. The former approach is ideally suited for time-series analysis (e.g. speech recognition), whereas the later one - for spatially-distributed data (e.g. image recognition).

Among all these approaches, the one based on photoelectrochemical processes is a unique one. In this approach the reservoir system based of photoelectrode can react to three different information channels with minimal cross-talk: (i) species in the electrolyte, (ii) potential applied to the photoelectrode and (iii) incident light intensity and wavelength. This situation is quite different than in other approaches, and can be exploded not only for computation, but for various sensing scenatios.

In most of the sensing protocols, an analytical target is bound with the receptor (as specific and selective as possible), which in turn affects the transduced that generates the analytical signal. Irrespective of the detection technique (spectroscopic or electrochemical), the operation of the sensor requires an appropriate external stimulus: voltage applied to the working electrode in the case of electrochemical sensors [101], current driven through chemiresistive sensors [102] or electromagnetic radiation in the case of optical / luminescent / NMR sensors [9, 103, 104]. Only electrochemical sensors with transistor-like architectures [105, 106] and photoelectrochemical sensors require two independent stimuli to operate: source-drain and gate-drain voltages in the case of transistor-like devices and combination of electrical potential and illumination in the case of photoelectrochemical ones. Therefore, these two cases seem to be suitable for RC integration, one of those stimuli can serve as a drive channel.

In this review we focus mostly on photoelectrochemical sensors based on different mechanisms, all of them enabling the use of the concept of reservoir computation to improve their functionality. It should be noted, however, that the reservoir computing approach, owing to its universality, computational power, and robustness, can also be used to improve the performance of other types of sensors, based on other physical phenomena. All these types of sensors are used in specific sensing device architectures and utilise different mechanisms for the detection and monitoring of the concentrations of various chemical entities. Reservoir computing (RC) based on large networks of randomly connected nonlinear elements being a reservoir is an approach which may significantly improve their sensing effectiveness through proper classification of resulted data.

In recent years, especially attention has been paid among others to the development of sensing methods based on chemoresistors [107], field effect transistors (FET) [108, 109], solid-state electrochemical sensors (SSES) [110, 111], and quartz crystal microbalance (QCM) [112]. Among them, chemoresistors and field effect transistors are the most promising in the frame utilisation of the reservoir computing concept. Chemoresistors are based on the effect of local chemical environment-dependent electrical resistivity. The changes in resistivity result from direct interaction between sensing material and chemical molecules. The field effect transistor is based on monitoring the current flow at a given voltage applied to the gate, which in turn affects the conductivity between the source and drain.

The excellent gas sensing properties of FETs and chemosensors made them widely investigated in the development of artificial olfactory sensing devices also known as electronic noses [113-116] being a great alternative to traditional gas chromatography measurements, which have many limitations related to the costs, lack of mobility, equipment size, and many others. The electronic nose is an electronic system that mimics the structure of the human nose that generates an output signal based on interactions between the sensor array and volatile compounds [116]. E-noses are useful, among others, in a wide spectrum of applications, e. g., disease diagnosis, food processes monitoring, environmental monitoring, agriculture, and explosives detection [114]. Optimization of artificial olfactory sensing systems behind the design of new sensing materials, sensor architectures, and sensing mechanisms including also the development of all-feature olfactory algorithms which can effectively improve their functional properties [117]. It should be noted that FET-based sensors have been commercialized toward pH-sensing and are considered for soil pH detection and agricultural applications [118, 119].

The sensing devices based on FET and chemoresistors creates also new opportunities due to the possibilities of their applications in wearable flexible sensors [115, 118, 120]. Sensors are mostly investigated towards healthcare applications aimed at monitoring key physical parameters such as blood pressure, body temperature, heart beat rate, blood oxygen concentration, respiration time, and body's daily motion [109, 121]. Organic field-effect transistors (OFETs), due to the development of industrial printing technology on flexible substrates and their sensing properties for a wide range of volatile analytes, ions, and light, are considered as great solution for OFET-based flexible biosensors [120]. Another interesting application of flexible sensors is ion sensing. Flexible ion sensors are mainly based on chemoresistors being active materials that interact with the analysed ion. The big advantage of chemosensors is the easy fabrication of sensing devices which makes them widely used for this purpose [120]. The effects of the combination of multitasking flexible sensors with reservoir computing were reported by Wakabayashi et al. Results indicated that reservoir computing creating possibilities of reduction of sensor integration complexity and limit power consumption for sensor and signal processing as well [122].

The reservoir computing approach in FET and chemoresistors-based systems is involved, among others, for the increase of the dynamics of the system response and increase of their sensitivity and selectivity. Seik et al. successfully utilised reservoir algorithms for continuous prediction in the case of a metal oxide (MOX) gas sensor used for the monitoring of ethanol, ethylene, and acetone [123]. The improvement of continuously exposed the dynamics of chemosensor arrays to fast variations of gas phase composition by utilization of the RC concept was proved by Fonollosa et al. [124]. The dynamics of the sensing system may be crucial from the point of view of many applications related to online gas composition monitoring. The dynamics of the sensor may also provide parameters useful in the discrimination of volatile compounds [125]. There are many ways to create possibilities for enhancement of the functionality of chemosensors. A promising approach in the context of future development of chemosensors is the application of a time delay neural network for the estimation of gas concentrations [126] and the acceleration of response processing by finding correlations between transient characteristics and steady-state resistance [127].

Wang et al. using commercial gas sensors, reservoir computing system based on volatile memristive devices, and nonvolotaile memristive device-based classifiers designed olfactory inference system with a precision of identification of 95% [128]. This demonstrates unprecedented possibilities of utilisation of the reservoir computing concept in chemoresistors and FET-based sensing systems. In the literature, there are many works concerning the application of RC concept in FETs or chemosensor-based sensing devices showing the importance of this scientific problem.

Although the manuscript focusses mainly on materials, sensing mechanisms, sensors architectures, sensor-analyte interactions, and signal generation it should be noted that the development of RC-based sensing systems is focused not only on the design of new materials of unique properties and signal generation mechanisms, but also on the development of novel classifiers creating possibilities of more accurate, faster, and predictable analysis of the sensor-environment interactions [129].

Apart from chemical sensing reservoir computing has found numerous applications in signal processing, time series analysis and other branches of information processing.

## 3. Photoelectrochemical sensing

The aim of photoelectrochemical sensors is to accurately detect and analyze the concentrations of important chemicals that significantly affect human life and environmental aspects. In recent years, the sensitivity of biosensors and the detection of toxic chemicals that can cause environmental damage have been widely studied [130]. The development of photoelectrochemical sensors is possible because of Becquerel, who in 1839 discovered the photoelectric effect [131]. Since then, photoelectrochemical phenomena have been used in photocatalysis, photovoltaics, sensing applications, and in a broad field of medical research [132-135]. For photoelectrochemical sensing applications, semiconducting nanostructures can be combined with various molecular receptors, including enzymes, for example, they can detect abnormalities in protein kinase, which is applicable for monitoring diseases such as Alzheimer, diabetes, or cancer. Many nanoparticles (NPs), such as ZnO, CdS, or CdTe, have also been shown to work as biosensors by binding to glucose oxidase, resulting in photocurrent generation [136].

The crucial factors responsible for the proper performance of photoelectrochemical sensors and biosensors are the operating conditions, sensitivity, detection limit, dynamics, and stability [137]. These factors are strongly correlated with the properties of the sensing material, among others: the band gap value ($E_g$), the energy corresponding to the Fermi level ($E_F$), the location of the flat band potential ($E_{fb}$), the efficiency of photocurrent generation, and the operational range. Photoelectrochemical sensors are unique kinds of sensor that exhibit a simultaneous sensing ability towards electrochemical and light-driven processes. Compared to other detection techniques, such as spectrophotometry, chemiluminescence, or Raman spectroscopy, photoelectrochemical sensors have the merit of miniaturization while maintaining high selectivity and sensitivity. The lack of expensive instruments and quick response opened up a broad spectrum of applications for them.

In general, the photoelectrochemical process consists of absorption of incident photons, electron-hole excitation, and charge separation followed by charge carrier collection and interfacial electron transfer processes. The interactions between the sensing material and light result in photocurrent/photovoltage generation. Photoelectrochemical sensors are designed most often for work in standard three-electrode systems [138]. The photoelectric effect takes place on the working electrode, which is coupled with the counter and reference electrodes. Charge carriers photogenerated in a semiconductor-based electrode can generate anodic or cathodic photocurrents by moving toward a conductive substrate or toward the counter electrode, respectively. Charge carriers pass through the electrolyte, where the donor/acceptor species undergo reduction or oxidation reactions. The analytes present in electrolytes can also participate in charge transfer by changing the charge transfer path. Depending on the unique properties of the molecule, the observed current/voltage signal may be disturbed in several ways. The photocurrent/photovoltage signal may be amplified, attenuated, or even changes in the polarization direction are possible. The overall process consists of partial charge transfers at the analyte/sensing material and then at the sensing material/conducting substrate interfaces. Thus, the only complex characterization of sensing material and analyte properties, knowledge of environment properties, and understanding of all processes occurring at several interfaces allow the design of sensors of proper photoelectrochemical performance. Three main mechanisms can be distinguished as operating bases of photoelectrochemical sensing: charge transfer, energy transfer and reactant determinant mechanisms, as shown in Figure 3.

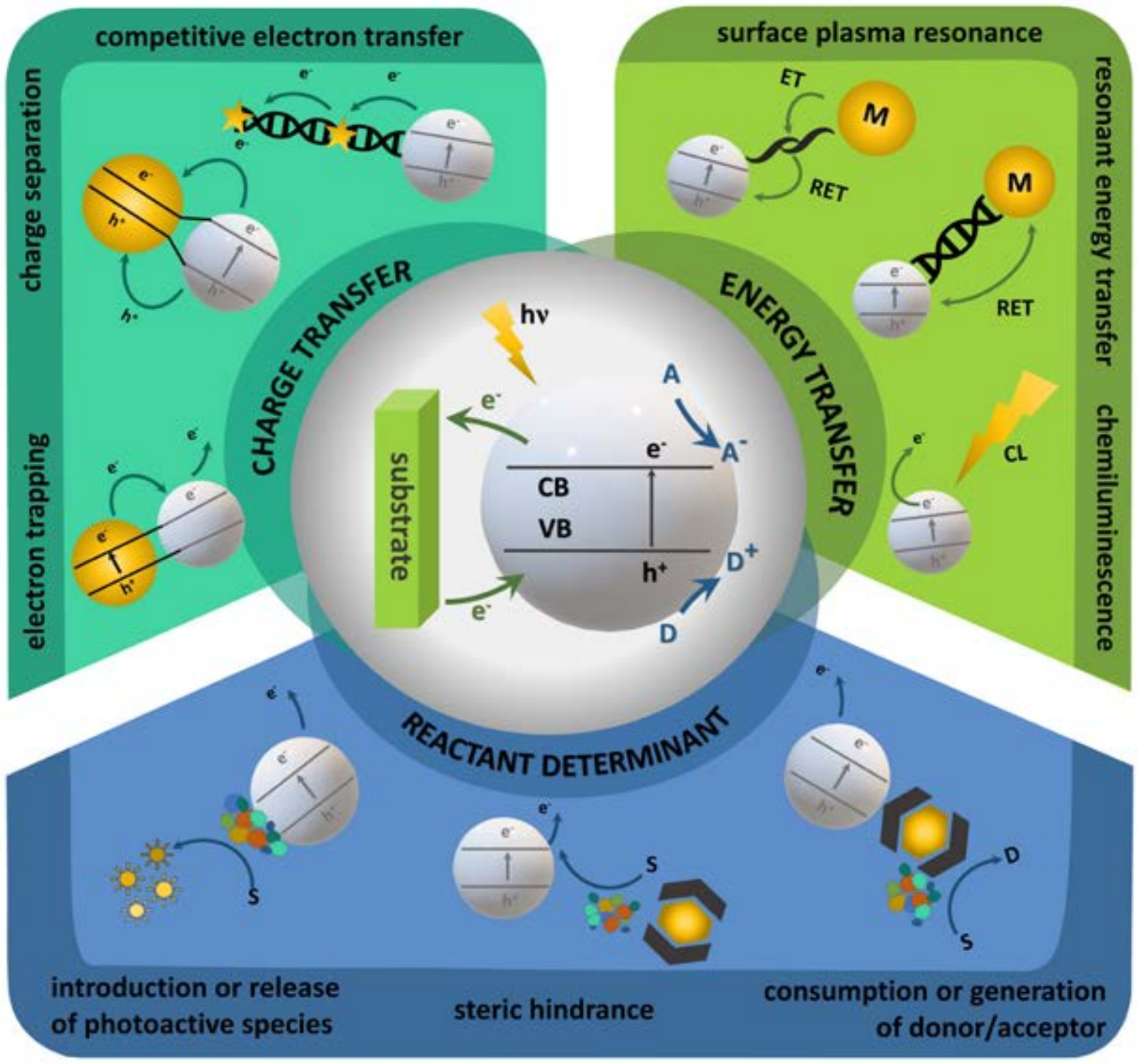

*Figure 3. Strategies for creating photoelectrochemical sensors.*

Semiconductor-semiconductor heterojunctions constitute very popular systems for the designing of advanced photoactive materials for different applications: molecular logic devices, memristors, transistors, or photoelectrochemical sensors. Specific charge transfer can be achieved by tuning the properties of the working electrode. In that case, *p*- or *n*-type semiconductors with a certain polarization can be used to generate photocurrents in competitive electron transfer. The transport and separation of charge carriers in the interface region may lead to an improvement in the efficiency of devices and their detection limits [139, 140]. Control of the photocurrent flow direction can be obtained by introducing additional reactants. One way to increase selectivity is by introducing surface trapping centers that can be removed by detected analytes. Induced surface states may also act as recombination centers. This process has been used to mimic natural photosynthesis in the Z-scheme technique.

Energy transfer approaches are often used to modulate the intensity of the photocurrent [141]. Under illumination, electrons from metal NPs can undergo collective surface oscillations, which may enhance the output signal of the sensor. This type of energy transfer is known as surface plasmon resonance (SPR). Other transfers, based on resonance (FRET, PIRET), can occur between acceptor-donor pairs through dipole-dipole interactions and plasmon-induced resonance, respectively. Energy transfer can also be controlled by the addition of luminous material and coreagent, which affect the excitation process of acceptors.

Another category of the principle of photoelectrochemical signaling for biosensors proposed by Zang *et al.* [142] is the determinant mechanism of the reactant. In this type of sensing, photoactive species and electron donors/acceptors are the reagents that determine the correct course of the photoelectrochemical reaction. Photoactive species can be used as a signal modifier; the change between the signal label and the electrode can lead to attenuation or enhancement of the photocurrent signal. A similar effect can be achieved by introducing a redox reaction between the working electrode and the electrolyte. In that case, electrons from the conduction band (CB) of the conductor can reduce the amount of agents in the analyte, or holes from the valence band (VB) can oxidize it. Often used in immunosensors is also the effect of steric hinderance, where electron donors/acceptors are directed to suppress the diffusion of surface photoactive species surfaces [143].

There are two main features that make most photoelectrochemical sensors suitable for integration with RC systems: (i) generation of analytic signal (photovoltage or photocurrent) requires two inputs: analyte and light and (ii) the dependence of photovoltage/photocurrent amplitude versus light intensity is given by a nonlinear function. While the first feature is obvious, the second requires a bit more detailed analysis.

The photovoltage versus light intensity dependence is given by [144]:

$$E_{ph} = -\frac{kT}{e}\ln\left(1+\beta J_0\right) \quad (14),$$

where $J_0$ is the light flux and $\beta$ is the proportionality factor related to doping density and diffusion length of charge carriers. In the first approximation, for highly doped material and low light intensities, the photocurrent depends linearly on the light intensity [144]:

$$i_{ph} = eJ_0\left(1-\frac{e^{-\alpha L_{SC}}}{1+\alpha L_P}\right) \quad (15),$$

where $J_0$ is the light flux, $\alpha$ is the absorption coefficient, $L_{SC}$ is the thickness of the space charge layer, and $L_P$ is the diffusion length of holes.

However, in a non-ideal case, various parasitic processes accompanying charge carrier generation and collection should be taken into account: inhomogeneity of the semiconductor electrode, surface and bulk recombination losses, diffusion of electrochemically active species to the electrode surface, local thermal effects, and other processes [145, 146]. This leads to the power-law dependence of photocurrent intensity versus light flux [147]:

$$i_{ph} = i^0 \left(1 - e^{-\xi J_0}\right) \quad (16),$$

where $i^0$ is the limiting current and $\xi$ is the proportionality factor associated with the kinetic parameters of the process [147]. With a few competitive, or even just parallel processes at the photoelectrode, the generalized photocurrent response is a sum of all contributions from all processes (17):

$$i_{ph} = \sum_n i_n^0 \left(1 - e^{-\xi_n J_0}\right) \quad (17).$$

Taking into account that processes responsible for generation of anodic ($i_n^0 < 0$) and cathodic ($i_n^0 > 0$) photocurrents may be present at the same time, simple light intensity modulation may result in reversal of photocurrent polarity, the phenomenon known as Light Intensity-Induced Photocurrent Switching (LIIPS) may occur [147]. This, along with Photoelectrochemical Photocurrent Switching (PEPS) effect [148-165] and photoinduced neuromimetic phenomena [166-168] renders photoelectrochemistry of wide band gap semiconductors a perfect playground for various sensing and unconventional computing studies.

It may seem counterintuitive that nonlinearity is a desired feature in sensing applications, but nonlinear flux vs. photocurrent dependence allows generation of time-domain analytical signals, *e.g.* high harmonic distortion allows detection of the second harmonic of the light modulation frequency with much increased signal-to-noise ratio.

The very general image of photoelectrochemical sensors suggests its compatibility with the principles of RC: The analytical signal can be generated only in the presence of both analyte (input, $u_n$) and light (drive, $q_n$). The response to at least one of the stimuli is non-linear (cf. Eq. 4 and Eq. 16) and shows internal dynamics due to the high capacitance of the electrolyte-semiconductor junction, finite doping density, and charge carrier mobility. In the simplest case and in the absence of significant charge carrier recombination, the time dependence of photocurrent vs. time in the case of pulsed illumination is given by:

$$i_{ph} = i^0 \left(1 - e^{-t/RC}\right) \quad (18)$$

where $i_0$ is the steady-state photocurrent, $R$ is the resistance of the photoelectrochemical cell and $C$ is the joint capacitance of the electrical double layer and the space charge layer. In can be further extended to:

$$i_{ph} = \sum_n i_n^0 \left(1 - e^{-t/R_n C_n}\right) \qquad (19)$$

if a series of charge trapping processes are involved.

There are, however, open questions: (i) Is it really possible to apply the RC approach in photoelectrochemical sensing and (ii) can it significantly improve the performance of the sensor. Although the integration of photoelectrochemical sensors with the RC platform should be feasible, and some preliminary attempts have already been reported [169], the second question, especially in light of the 'no free lunch theorem', is much more difficult to address [170]. As already mentioned in the Introduction, chemical sensing is a class of information processing tasks, so the overall cost (in terms of computational effort) should not depend on a specific approach. On the other hand, exploitation of the internal dynamics of physical systems for information processing purposes may significantly decrease the computing cost [171].

In this review, we focus on photoelectrochemical sensing strategies that may be a starting point for the construction of reservoir computing-powered analytical devices. Recent developments in the construction of photoelectrochemical sensors are presented, with emphasis given to different sensing mechanisms. Finally, all of these mechanistic approaches will be screened for their compatibility with the RC principles.

## 4. Charge transfer photoelectrochemical sensing

### 4.1 Competitive electron transfer

Light-induced charge transfer is a crucial factor in the design of the group of charge-transfer-based photoelectrochemical sensors. Semiconducting sensing systems based on the charge-transfer effect create possibilities for the detection of many species of acceptor or donor character. A wide range of available semiconducting *n*- or *p*-type materials of different band structures and, on the other hand, possibilities of the proper optimization of analysis conditions through the adjustment of electrode potential, wavelength, and incident light flux intensity make this method applicable for the detection and quantitative analysis of a wide group of organic and inorganic species. The possible mechanisms of photocurrent generation in *n*- and *p*-type semiconducting materials in the presence of acceptor/donor character species on polarized sensing electrodes at different potentials are presented in Figure 4. Knowledge of the properties of semiconducting material and analyte along with recorded photocurrent action maps allows for the precise determination of the conditions corresponding to the highest sensitivity and appropriate detection level of the sensing material.

Figure 4 presents the example intensities of the output signal resulting from the interaction of the single semiconductor *(n*- and *p*-type) sensing electrode with an analyte of donor or acceptor character in several potential regimes under dark conditions and light irradiation. Visible output signals allow the prediction of how interactions of the analyte with the sensing materials affect the output current in several potential regimes. It may be concluded that the *n*-type semiconducting sensor is the most suitable for analysis of the donor character, while the chemical entities of the *p*-type are the most suitable for analysis of the acceptor character.

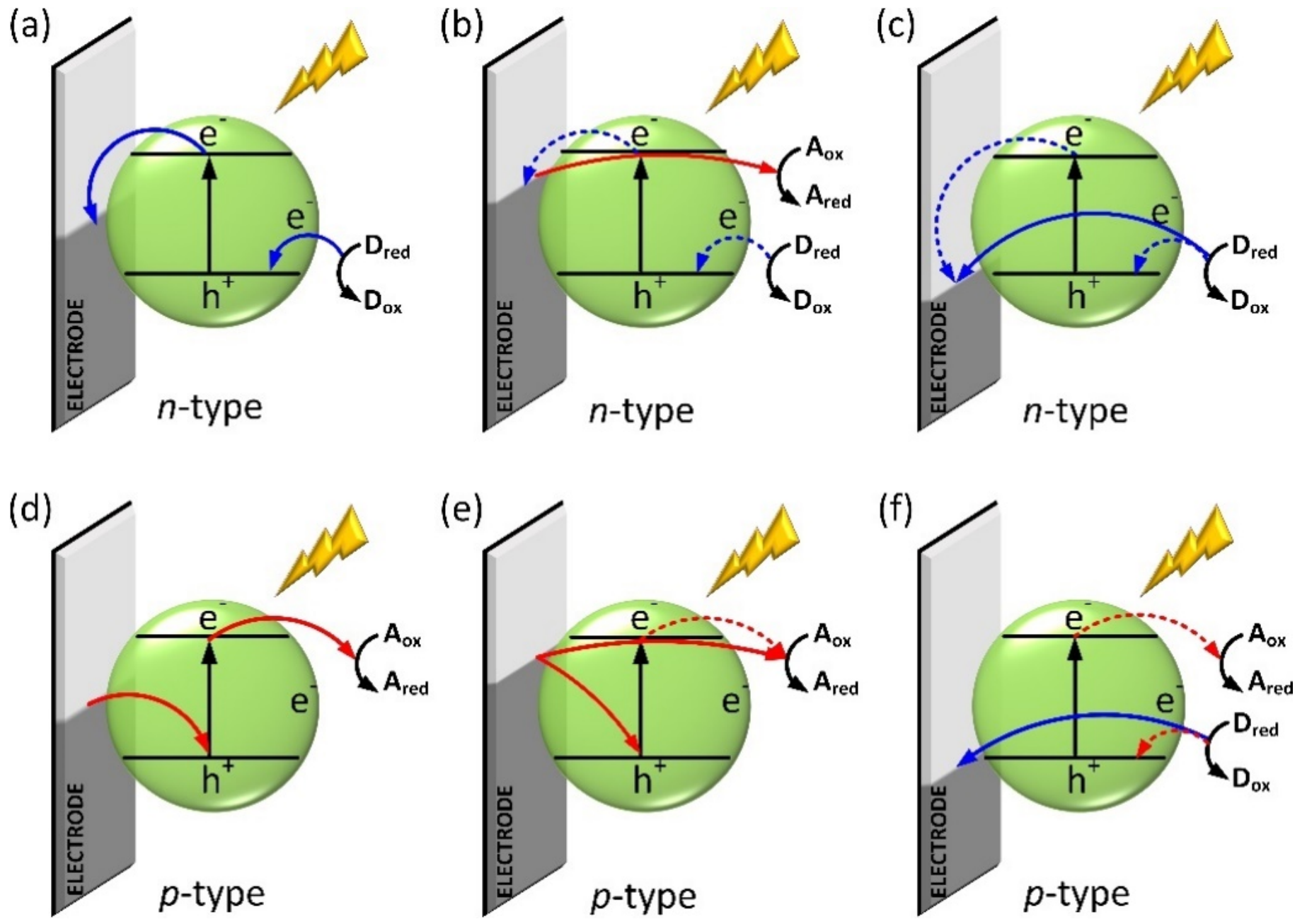

*Figure 4. Possible mechanisms of photocurrent generation in n- and p-Type Semiconductor-Based Sensing Electrodes. a and d) Photocurrents generated under light irradiation at open circuit potential; b and e) photocurrents generated under light irradiation at a negatively polarized electrode; c and f) photocurrents generated under light irradiation at a positively polarized electrode. (blue lines, generation of anodic current; red lines – generation of cathodic current; solid line, predominant paths of charge flow; dashed lines – secondary paths of charge flow; the upper edge of the grey bar, location of the electrode potential).*

The polarization of the sensing electrode, depending on the type of conductivity toward a more electronegative or electropositive direction, as well as the light application, allows output current amplification, providing conditions for the best sensitivity and lowest detection limit. Sensing electrodes based on charge transfer aiming their composition generally may be classified into a few groups: (i) single semiconductor, (ii) cocatalyst coupled semiconductors (carbon nanomaterial coupled semiconductors, metal NP coupled semiconductors), (iii) semiconductor – semiconductor heterojunctions (conventional heterojunctions, Z-scheme heterojunctions *p-n* heterojunctions, and multicomponent heterojunctions) [172]. Single semiconductor-based sensing electrodes are used for some specific applications, but due to greater flexibility towards the limitation of the photogenerated charge recombination, improved sensitivity, detection limit, stability, and proper adjustment of overall band structure enabling application of sensor for detection of specific analytes, many attentions is devoted to multicomponent sensing electrodes. Taking into account that the current part has to be focused mainly on electron-transfer-based sensing processes, we distinguished a few subgroups within which the mechanism of output signal generation that will be discussed in the context of practical applications.

### 4.2 Photoelectrochemical sensing of chemical entities that are electron acceptors

Photoelectrochemical sensing of electron acceptors is widely investigated mainly in the frame of the detection of metal ions, which are a very important factor in the influence of living organisms and the environment, as well as other chemical entities of acceptor character. Co, Zn, Se, Ni, Fe, Cu, Cr, Mn, Mg and Mo in proper low amounts are essential for various biochemical and physical functions and are considered beneficial [173] while in excess they are toxic.[174] Therefore, fast, reliable and precise detection and monitoring of their concentrations in different media are of great importance.[175]

The detection of acceptor-character analytes is a more complicated issue than that of donor character. The higher complexity of the system results from the possibilities of precipitation of reduced species onto the surface of the sensing electrode, as well as in situ formation of heterojunctions. Thus, not only the kinetics of the analyte reduction process but also the interactions of reduction products with sensing materials, which influence the charge-transfer path, recombination processes, and photocurrent generation, have to be considered in the design of the sensor.

The sensing of the acceptor character analyte based on the application of *p*-type semiconductor consists of the analysis of changes of cathodic photocurrent intensity. The photoexcited electrons from the CB of the *p*-type sensing material are transferred to the acceptor-type analyte, which undergoes reduction, while holes from the VB are transferred to the substrate. Dependently on the localization of $E^{o}_{A_{ox}/A_{red}}$ sensing material may be also polarized for proper alignment of energy levels, and thus optimization of its sensitivity. Du et al. designed a sensor based on *p*-type graphitic carbon nitrides (*g*-$C_3N_4$) for the detection of $Cu^{2+}$ ions. They used the mechanism described in Figure 4e – a negatively polarized electrode under light illumination generates cathodic photocurrents that increase with increasing concentration of $Cu^{2+}$ ions in the electrolyte. The authors postulated that photoexcited charge carriers would be captured by $Cu^{2+}$ being electron acceptors, enhancing charge separation, which in turn increases the output cathodic photocurrent. The authors underlined that the high selectivity and sensitivity of the *g*-$C_3N_4$ based sensor resulting from proper localization of the $Cu^{2+}$ redox potential and its preferential strong adsorption properties towards $Cu^{2+}$ ions proved by the results of calculations based on density functional theory [176].

The application of n-type sensing material in properly optimized conditions and environment may also allow the determination of analytes of acceptor character. Wang *et al*. invented a mechanism based on damping of the anodic photocurrent produced by illuminated CdS quantum dots (QDs) to detect $Cu^{2+}$ ions (Figure 5) [177].

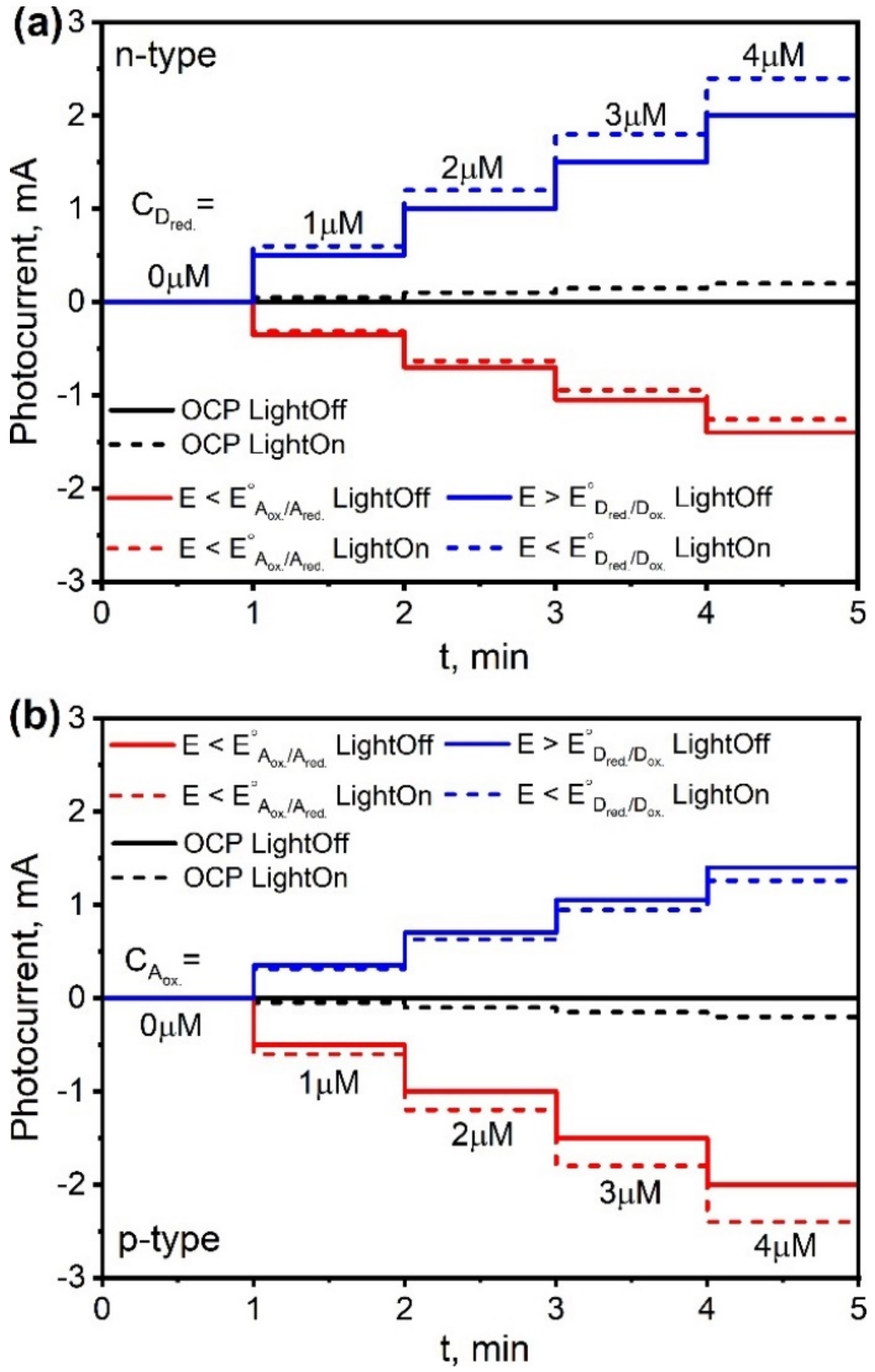

*Figure 5. Scheme of the output current signal generated at a) n-type and b) p-type semiconductor-based sensing electrodes.*

The mechanism is based on the interplay between the electrode surface of CdS QDs and $Cu^{2+}$ present in triethanolamine (TEA), resulting in the formation of $Cu_xS$-doped CdS QDs. This, in turn, disrupts the CdS electron transfer from the CB to the ITO, resulting in a decrease in photocurrents. They reported that the $Cu^{2+}$ ions present in the solution can bind to the surface of the semiconductor, then under illumination, the $Cu^{2+}$ ions may be reduced to $Cu^{+}$, which in turn may result in the formation of $Cu_xS$ on the surface of CdS. $Cu_xS$ formation is possible due to the chemical displacement of $Cd^{2+}$. The products of $Cu^{2+}$ reduction present on the surface of CdS create additional energy levels located below CB of CdS, ensuring the right conditions for charge carrier recombination. Therefore, the reduction of $Cu^{2+}$ itself and the improvement of recombination result in a decrease in anodic photocurrents [178].

An *in situ* electrodeposition approach is very interesting in the context of sensing acceptor analytes that may result in the formation of semiconducting compounds. On the surface of the polarized electrode, the analyte undergoes a reduction, resulting in the formation of the compound. Simultaneous illumination of the electrode surface allows for analysis of the generated photocurrents which are proportional to the quantity of the compound at the electrode surface. The quantity of electrochemically synthesized semiconductors is strongly related to the kinetics of electrodeposition, which is mainly dependent on the applied potential and the concentration of analyte in the electrolyte. Thus, depending on the type of conductivity, an amplification or damping of the photocurrents for a cathodically polarized electrode allows for the determination of analyte concentration.

The idea was successfully used by Qiu et al. for sensing $Cu^{+}$ ions through the analysis of the photoelectrochemical response of *p*-type $Cu_2O$. In this case, the mechanism of photocurrent generation is relatively simple. Photogenerated holes are transferred from $Cu_2O$ VB through ITO to the anode to play a role in water oxidation. Electrons from CB are transferred to the semiconductor/electrolyte interface to reduce water molecules. The separation of charge carriers is also supported by the polarization of the sensing electrode at -0.2 V (versus Ag / AgCl) [179].

Luo et al. also developed a sensing method that uses in situ electrodeposition of PbS NPs on $TiO_2$ nanotubes (NTs) useful for the determination of the $Pb^{2+}$ concentration. Observed anodic photocurrents are amplified by the increase in the concentration of $Pb^{2+}$ in the electrolyte due to the increase in the quantity of PbS particles [180]. Unfortunately, the authors did not describe the mechanism of photocurrent generation. The probable explanation for the observed effect can be found in the work of Zhang et al. focused on the application of $TiO_2$/PbS heterojunctions of the same structure in the purification of 4-chlorobenzoic acid. Illumination of the $TiO_2$ NTs results in the formation of electron-hole couples, but due to the localization of CB and VB of PbS present on its surface, some charge carriers are transferred to PbS particles which act as electron acceptors. This in turn may limit the recombination of charge carriers in an $TiO_2$ resulting in increased anodic photocurrent response [181].

The interactions between the ZnO nanorod (NR) sensor and CdS formed by in situ electrodeposition from electrolyte containing $S_2O_3^{2-}$ and ethylenediaminetetraacetic acid were utilized by Wang et al. for $Cd^{2+}$sensing. The authors described that sensing of $Cd^{2+}$ is possible due to the enhanced separation of the charge photogenerated in ZnO caused by CdS. The authors reported that in the presence of visible (VIS) light irradiation, just a small number of photoexcited electrons from the ZnO CB were transferred to ITO, which in a turn resulted in low-intensity photocurrent. Electrons photoexcited to CB of electrodeposited

CdS of high light harvesting efficiency are transferred to CB of ZnO and then transferred to ITO, resulting in significantly higher anodic photocurrent intensity [182]. Liang et al. utilized the same mechanism of photocurrent generation in the case of a sensor based on $TiO_2$. They analyzed the concentration of $Cd^{2+}$ based on the output anodic photocurrent amplified by the combination of $TiO_2$ and in situ electrodeposited CdSe of narrow band gap (such as CdS) [183].

Wang *et al*. designed an in-direct highly sensitive and selective method of dopamine (DA) concertation analysis based on CdS QDs utilizing the effect of damping the anodic photocurrents in the presence of polydopamine. PolyDA has abundant benzoquinone groups responsible for its acceptor properties. The method requires prior oxidation of the DA chemically in a weakly basic solution in presence of oxygen or electrochemical oxidation. The mechanism of output photocurrents is as follows: In the absence of polyDA upon light illumination, electrons from CB of CdS are transferred to ITO, generating anodic photocurrents. The presence of polyDA results in limitations of electron transfer to ITO which, in turn, results in a decrease in the intensity of photocurrents generated photocurrents [177].

There are many possibilities of sensing analytes of acceptor character, but independently of the sensing material applied and products formed as the result of the reduction process, knowledge of the charge-transfer mechanisms is crucial for proper and reliable detection and determination of analyte concentration.

### 4.3 Photoelectrochemical sensing of chemical entities that are electron donors

One of the simplest and most widely studied charge transfer-based sensor applications is the detection of glucose [184-189]. The process is based on the non-enzymatic electrochemical or photoelectrochemical oxidation of glucose to gluconic acid. Photoelectrochemical glucose detection was studied with the use of different single semiconductor materials such as $TiO_2$NR array [184], $TiO_2$ATO (anodic titanium oxide),[189] α-$Fe_2O_3$,[185] $Co_3O_4$,[186] and $BiVO_4$.[187] Application of the photoelectrochemical method led to a decrease in the overpotential for glucose oxidation reaction, which limits the kinetics of interference reactions compared to the electrochemical method [184]. Liu *et al.* reported that the sensing process in the case of $Co_3O_4$@FTO (fluorine-doped tin oxide) except for direct reduction of glucose is also accompanied by conversion between Co (III)/Co (IV) couples supporting the generation of current response [186]. They reported that sensing in the case of the electrode polarized at 0.6 V (versus Ag / AgCl) under illumination significantly increases the sensitivity (even up to 7753 $\mu A \cdot mmol^{-1}\ cm^{-2}$) and shifts the lower limit of the detection range to 0.01 µM. In the case of separate electrochemical and photoelectrochemical sensing, they observed a higher limit of the detection range of 0.1 and 0.2 µM, respectively. The electrons in the case of the remaining reported glucose sensors based on the $TiO_2$NR matrix [184], $TiO_2$ATO [189], α-$Fe_2O_3$ [185], and $BiVO_4$ [187] of different morphologies are not involved in the oxidation/reduction of chemical entities other than glucose. The mechanism of output current generation for a single *n*-type semiconductor-based sensor polarized toward an anodic direction after illumination was presented in Figure 1c. Syrek *et al.* revealed that the linear detection range, as well as sensitivity, in the case of the hybrid electrochemical/photoelectrochemical method, can be optimized by modulating the sensing electrode polarization potential and in the case of the $TiO_2$ATO-based sensor. The change in the polarization potential from 0.2 to 1.0 V (versus SCE) results in an increase in sensitivity from 7.3 µA $mmol^{-1}{\cdot}cm^{-2}$ even to 237 µA $mmol^{-1}$ $cm^{-2}$ [189]. The mechanism of anodic photocurrent generation was also utilized for ultrasensitive sensing of DA as an electron donor by Hun et al. [190]. In the case of the $MoS_2$ modified gold electrode, they observed anodic photocurrents for the electrode polarized at anodic potentials. The generation results from the transfer of photoexcited electrons from the $MoS_2$ VB to the gold electrode. On the other hand, DA is oxidized to dopamine quinone by holes that simultaneously the recombination process Alves et al. used the same photocurrent generation to detect DA based on $WO_3$ [191].

The mechanism was also used for the detection of ciprofloxacin, an antibiotic that can be toxic to humans. For this purpose, Yan et al. designed a photoelectrochemical sensor based on a 2D thin sheet structure of nitrogen-deficient graphitic carbon nitride. Nitrogen vacancies act as charge traps that may effectively limit the charge recombination effect. The oxidation of ciprofloxacin by photogenerated holes (Figure 6) results in amplification of the observed anodic photocurrents. The mechanism led to a very sensitive analysis of the concentration of the analyte in a wide range of 60 ng/L to approximately. 19 µg/L with the low detection limit of 20 ng/L. The sensor sensitivity is also improved by the 2D structure that significantly affects the separation and transfer of charge carriers, dramatically improving its performance [192].

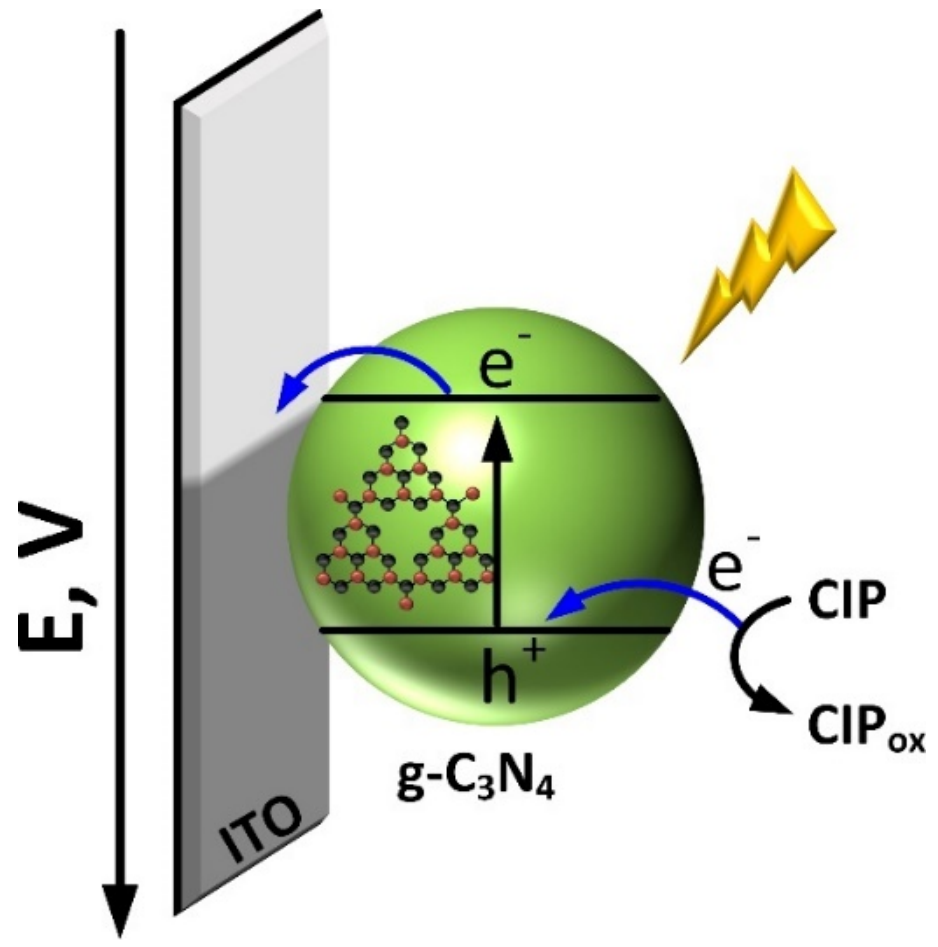

*Figure 6. Schematic mechanism of nitrogen-deficient g-$C_3N_4$ based ciprofloxacin (CIP) detection* [192].

### 4.4 Charge separation-based sensors

Semiconductor heterojunctions are the most effective due to the separation and transfer of light-induced electrons and holes in the contact area of various semiconductors [193]. In semiconductor-semiconductor heterojunctions, we can distinguish 3 main groups: conventional heterojunction, *p-n* heterojunction, and Z-scheme heterojunction. In all of these systems, photoelectrochemical detection is based on electron transfer, but the path of electron transfer is different [193, 194]. The charge-transfer path may be adjusted by a composition of two or more semiconductor materials with the appropriate arrangement of the band edges. The main advantage of semiconductor heterojunction is the ability to increase photocurrent generation by improving the light absorption range and decreasing the recombination of light-induced carriers. From the point of view of photoelectrochemical sensors, we achieve a better signal and can determine a lower concentration of the analyte.

### 4.5 Heterojunction of conventional semiconductors

The basic issue in the construction of conventional semiconductor heterojunctions is to match the energy levels of the individual components of the junction. The heterojunction of two or more different energy band-gap semiconductors has attracted attention because the co-sensitized structure has a wide detection range. Furthermore, the use of several different semiconducting materials effectively prevents charge recombination and also improves the light absorption capacity, resulting in a high photoelectric conversion efficiency. This type of heterojunction is observed mainly for *n*-type semiconductors in which the positions of the CBs and VBs have different energy (Figure 7). The internal field appears and improves the electron-hole separation efficiency, charge carrier lifetime, and electron transfer rate. Under the influence of light, in the first stage, an electron from the valence band is transferred to the conduction band (in one or both semiconductors, depending on the wavelength), and the generated carriers are separated, then they can can undergo recombination processes or be transferred to the conduction band of another semiconductor. At the same time, there is also a transfer of holes from the valence band between the two semiconductors. The transport of photoinduced charges occurs spontaneously, always from a higher to a lower level of the valence or conduction band. As a result of such a downhill energy scheme, the generated anodic photocurrent is enhanced and reaches a higher value compared to one semiconductor. The next step after the contact of the sensor with the analyte results in decrease of the photocurrent. The advantage of this type of system is the increase in the quantification limit of the analyte in the tested sample [193, 194].

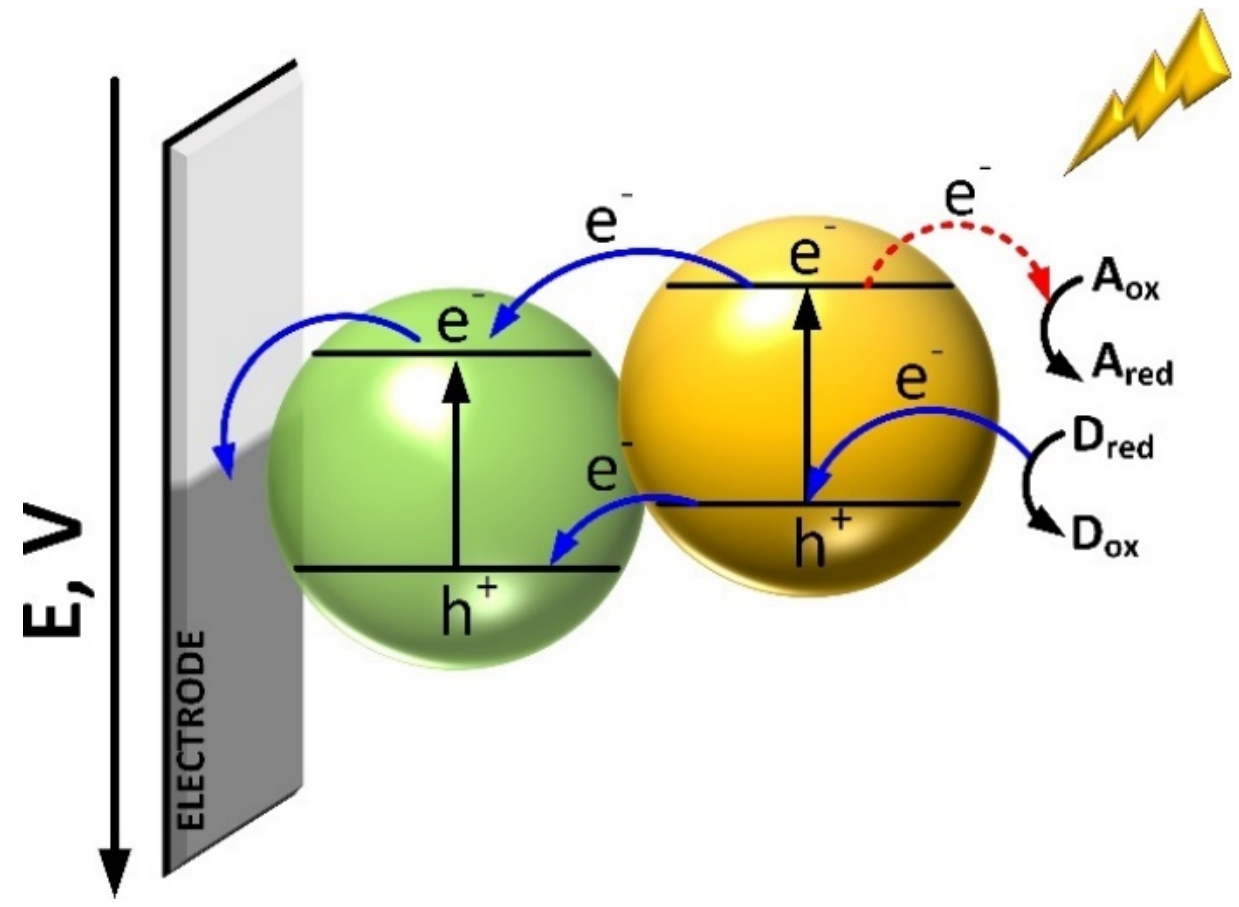

*Figure 7. Mechanism of photocurrent generation utilizing the charge separation effect in two semiconductor heterojunctions: photocurrents generated under light irradiation at open circuit potential (blue lines – generation of anodic current, red lines - generation of cathodic current, solid line – predominant paths of charge flow, dashed lines – secondary paths of charge flow).*

One of the examples of conventional heterojunction with two *n*-type inorganic semiconductors is the ZnO/CdS nanospheres that are used for the $Cu^{2+}$ ions sensing. The electrode prepared from ZnO/CdS was immersed in copper ions solution and a decrease in photocurrent intensity was observed (Figure 8). It was caused by the synthesis of $Cu_xS$ on the surface of cadmium sulfide and generated a lower CB energy level. Electron transfer from the CB of CdS to the CB of $Cu_xS$ was more favorable than to ZnO [195]. An analogous example of the determination of copper ions was based on heterojunctions of $SnO_2$/CdS [196] and $WO_3$/CdS [197] heterojunctions. All of these systems are very promising tools to monitor copper pollution in the environment.

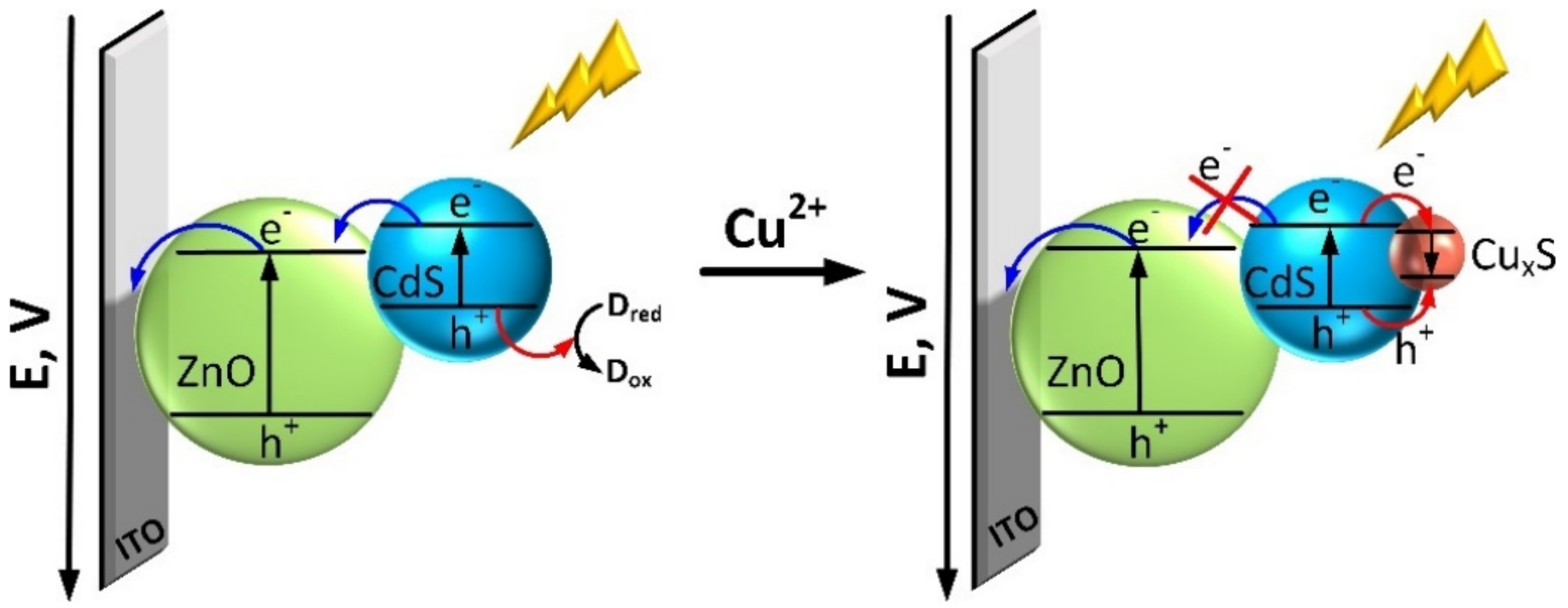

*Figure 8. The possible mechanism of photoelectrochemical detection of $Cu^{2+}$* [195].

Another example of a heterojunction-based photoelectrochemical sensor is the detection of dopamine (DA) by the $BiVO_4$/FeOOH system. Generation of the photocurrent in this system is possible because the water oxidation process. In this case, very important is the thickness of FeOOH present on the surface of $BiVO_4$ because it serves as the active centre and cannot be too thin, while too thick a layer would reduce the absorption of light by the system. The presence of DA in the solution caused that electrons from the CB of $BiVO_4$ can participate in the polymerization reaction and polyDA formation. Furthermore, a large number of polyDA benzoquinone groups accept electrons very efficiently, and this process is more favourable than the electron transfer from $BiVO_4$/FeOOH to the ITO electrode. As a result, a lower intensity of photocurrent was observed [198].

Photoelectrochemical sensor based on *p-p* heterojunction is very seldom observed. An example of such a system is the $Cu_2O$/BiOI composite for the detection of $H_2O_2$ [199]. The mechanism of generating photocurrent is analogous to the *n-n* heterojunctions, but in this case the cathodic photocurrent is observed. As a result of intense illumination, electrons are transferred from BiOI to $Cu_2O$, where on the surface they take part in the reaction with $H_2O_2$, thus increasing the intensity of the generated photocurrent [199].

### *4.6 p-n* heterojunctions

The charge transfer path may also be adjusted by a combination of two types with semiconductors with proper band-edge arrangement. The general idea assumes that both the VB and the CB of *n*-type semiconductors are more positive than those of the p-type, and thus the electrostatic field in such an *n-p* heterojunction forces the movement of photoexcited electrons from the CB of the *p*-type semiconductor to the CB of the *n*-type one. The electrons are then transferred to a conductive substrate. Limited recombination improves the efficiency of hole transport, providing better conditions for analyte oxidation and, thus, for efficient photocurrent generation.

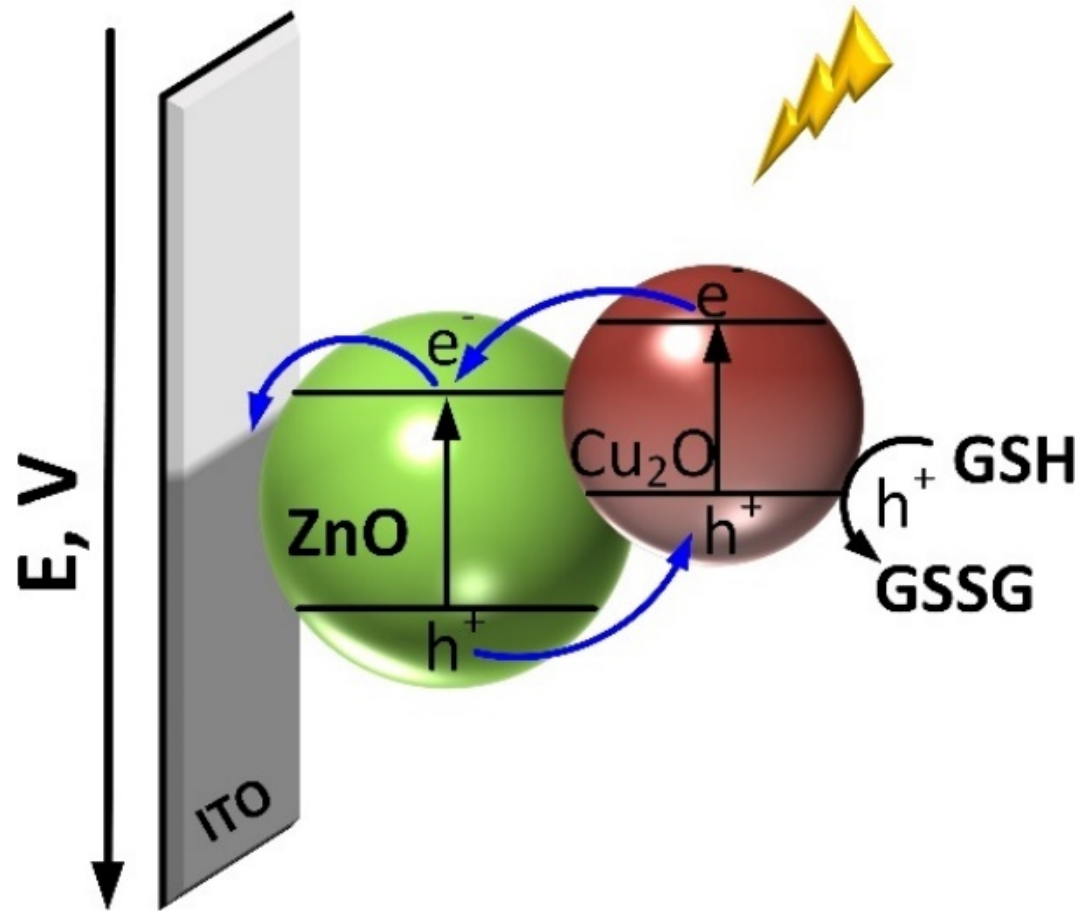

*Figure 9. Mechanism of photoelectrochemical detection of GSH in the $Cu_2O$/ZnO heterojunction-based photoanode* [200].

Wang et al. designed the $Cu_2O$/ZnO *p-n* heterojunction that exhibits sensing properties for glutathione (GSH). The combination of $Cu_2O$ and ZnO exhibits a synergistic effect resulting from the modification of the charge transfer path. In system light, excited electrons from $Cu_2O$ CB are transferred to the CB of ZnO and then to ITO, resulting in a significant increase of observed photocurrent. Simultaneously, positive charge migrate towards the sensor/electrolyte interface to oxidize GSH (Figure 9) [200]. Yan et al. utilized the same effect of synergic interaction between *n*-$BiPO_4$ and *p*-BiOCl in the detection of 4-chlorophenol. The combination of both materials ensured proper conditions for efficient electron-hole charge-carrier separation and thus amplification of the generated anodic photocurrents. The holes generated in the VB of $BiPO_4$ are transferred to the VB of BiOCl, resulting in more efficient kinetics of 4-chlorophenol oxidation [201].

The important group of heterojunction materials is semiconductors modified with other materials with semiconducting properties, such as polymers, porphyrins, and *g*-$C_3N_4$. In addition to adjustment of the potential and light wavelength and/or intensity, the increase in the anodic response of *n*-type sensing material can also be achieved through the functionalization of the surface of semiconducting material, *i.e*. molecularly imprinted polymers [194, 202]. Li et al. functionalized $TiO_2$ NPs with poly(3-hexylthiophene) (P3HT) to modify the electron transfer path. Electrons excited in P3HT under light irradiation are transferred to the $TiO_2$ CB and then to the glassy carbon electrode (GCE) while holes migrate from $TiO_2$ through the VB of P3HT to generate hydroxyl radicals that allow the conversion of chlorpyrifos (*O*,*O*-diethyl-*O*-(3,5,6-trichloropyridin-2-yl) phosphorothioate) into chlorpyrifos radical, in turn amplifying the current intensity of the sensor [203].

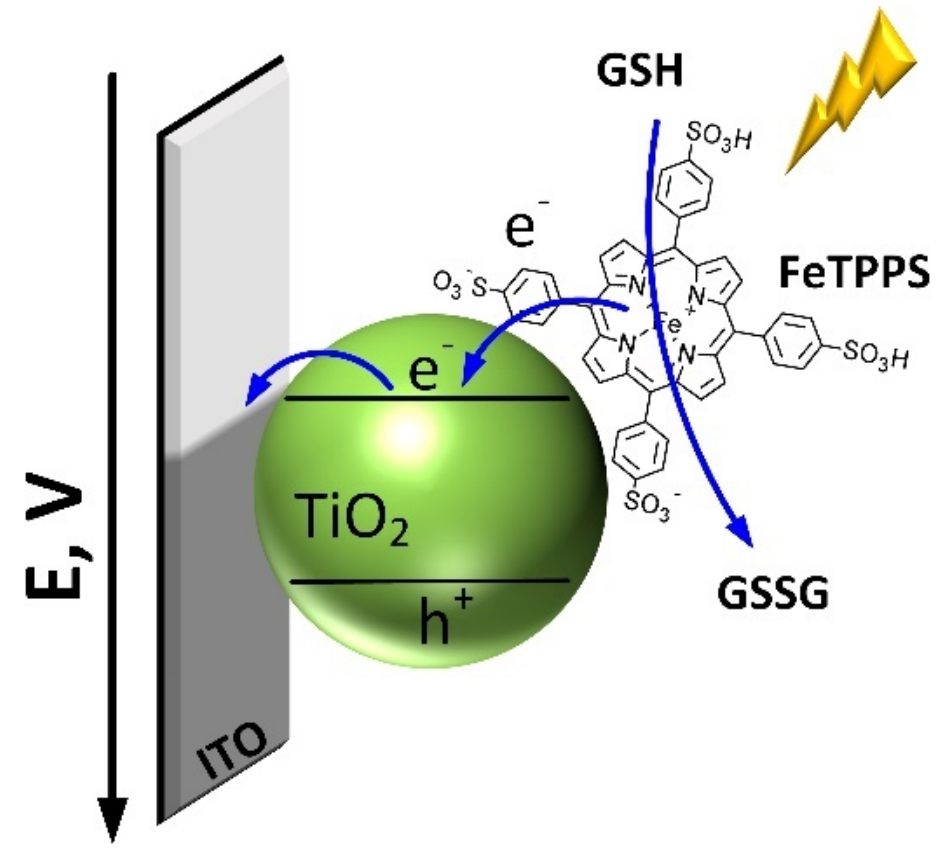

*Figure 10. Diagram showing the operation of the FeTPPS-$TiO_2$-based sensor for GSH detection* [204].

As an example of a well-functioning sensor made of a semiconductor and porphyrin, we can indicate $TiO_2$, the surface of which has been modified with FeTPPS porphyrin particles. As a result of this modification, the generated anode photocurrent was significantly enhanced in comparison to that of the single components. This is due to the strong electronic coupling between the excited state of the porphyrin and the conduction band of the semiconductor. This system was used to build a sensor designed to detect GSH, which is oxidized at the electrode as a result of the reaction with holes coming from $TiO_2$ valence band, what in turn results in amplification of the photocurrent generated in the system (Figure 10) [204].

### 4.7 Sensors based on the Z-scheme mechanism

Hybrids of photoactive semiconductors represent three different types of valence band maximum ($VB_{max}$) and conduction band minimum ($CB_{min}$) alignments, named type I, type II, and Z-scheme. The last category consists of a short path for electron transfer from the conduction band of a photocatalyst (PC I) to the valence bands of the second active photocatalyst (PC II) in the band energy diagram, under light irradiation (Figure 11). This results in a reduction of the recombination rate of excitons (electron-hole pairs), and as a consequence the photocurrent intensity in the cell increases. Based on the electron transfer pathway and types of reactive participants, three types of Z-scheme photocatalysts have been introduced: (1) direct contact of two photocatalysts (Figure 11 a), (2) plasmonic centres (eg metallic nanoparticles (NPs)) being solid-state mediators between two semiconductors (Figure 11 b), and (3) traditional Z-scheme with a pair of the acceptor (A) and donor (D) as ionic photoactive mediators (Figure 11 c) [205].

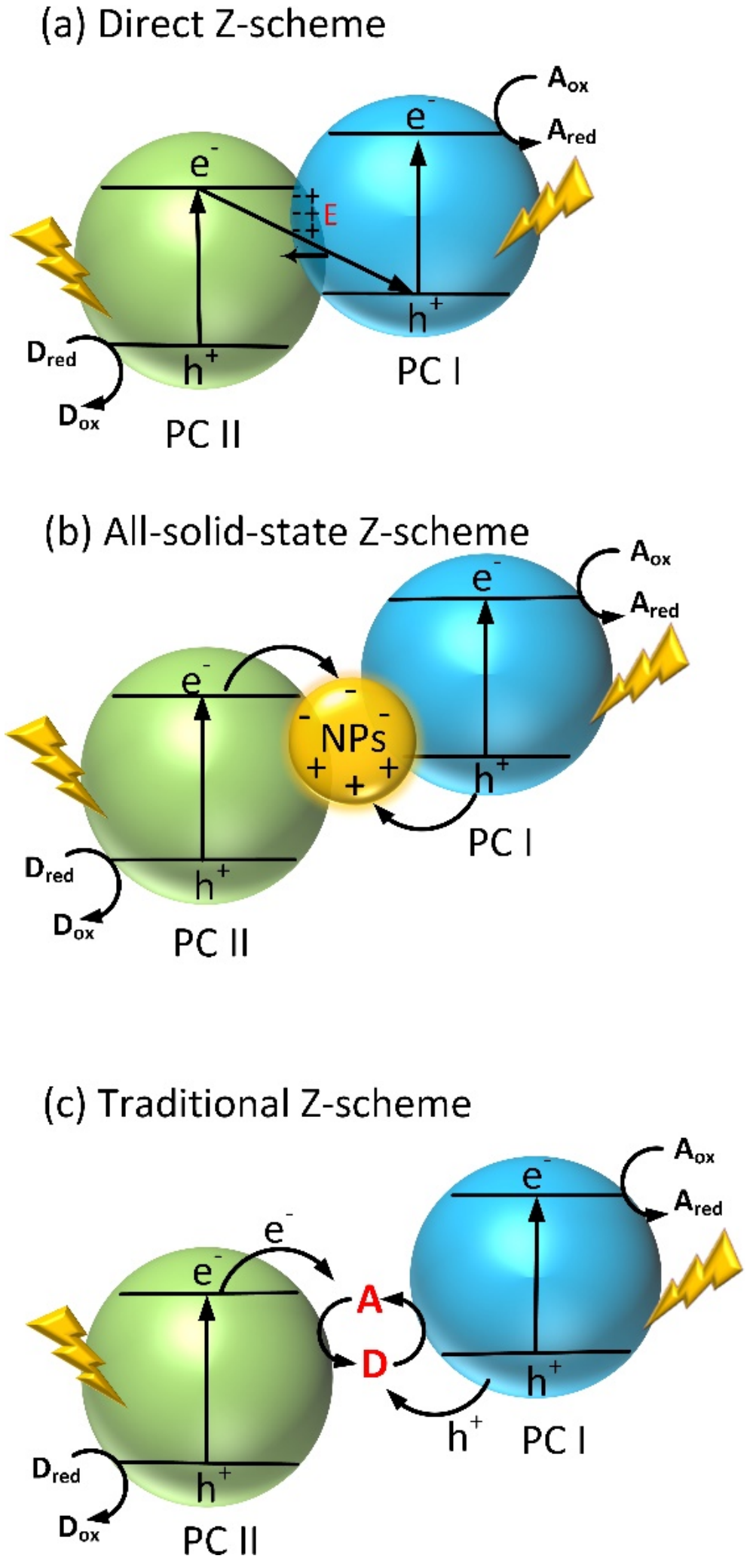

*Figure 11. Schematic illustration of a) direct Z-scheme and b) NPs as solid- state electron mediators in Z-scheme, and c) donor/acceptor pair in the traditional Z-scheme. PCI and PCII stand for photocatalysts I and II, respectively, and E represents the direction of the internal electric field.*

To demonstrate the importance of band energy alignments in combinations of photocatalysts, a hybrid of bismuth oxyiodide and cadmium sulfide (BiOI-CdS) and nitrogen-doped graphene QDs (N-GQD)/I-BiOCl, as two examples of direct Z-scheme, were fabricated and employed as photocathodic sensors for detection of $Cu^{2+}$ and chlorpyrifos, respectively, by Wang et al. [206-209]. The resulting Z-scheme composed of BiOI-CdS sensors showed a remarkable photocurrent response after exposure to 470 nm light at a -0.1 V bias potential. However, when adding $Cu^{2+}$ analyte to the medium, the intensity of the photocurrent decreased, due to the formation of a new centres on the surface of the heterostructure that accelerated the electron-hole recombination However, when adding $Cu^{2+}$ analyte to the medium, the intensity of the photocurrent decreased, due to forming a new center on the surface of the heterostructure that accelerated the electron-hole recombination [209].

Advanced investigations demonstrated that mediators have influential roles in Z-scheme structures, which affect photogenerated electron transfer and promote the accuracy of the measurement. For example, the effects of ionic pairs including $Fe^{3+}/Fe^{2+}$ and $IO_3^-/I$ on the enhancement of photocurrent intensity were investigated in photoredox reactions in fluids [210]. Furthermore, in this framework, solid-state mediators such as reduced graphene oxide, AuNPs, and silver NPs (AgNPs) have been used in photoelectrochemical Z-scheme systems recently [210, 211]. Zhao et al.[210] applied AuNPs in heterostructures made of CdS QDs and $BiVO_4$ to produce a Z-scheme photosensor for the detection of prostate specific antigen (PSA). The recently reported Z-scheme in photoelectrochemical sensing applications is summarized in Table 1. [176-180, 182-187, 189-192, 195-201, 203, 204, 208-219]

**4.8 Sensors based on the electron trapping processes**

Mechanisms based on the electron-trapping process are also very usefulpossible in the photoelectrochemical detection of analytes. This case mechanism is observed when the analyte forms an intermediate species that captures electrons and interrupts the transfer of electrons from the semiconductor to the electrode (Figure 12) [212]. The meso-2,3-dimercaptosuccinate-modified CdTe QDs immobilized on the electrode were used as $Cu^{2+}$ sensors. CdTe is a *p*-type semiconductor, which under illumination generates a cathodic photocurrent. In the presence of $Cu^{2+}$, in which the formed $Cu_xS$ competes for electrons from the CB of the QDs with the electrode, a decrease in cathodic photocurrent was observed. Unlike the mechanism of anode photoelectrochemistry, in this case $Cu_xS$ was not an electron acceptor, but it was working acted as the trapping centre, causing a decrease in photocurrent intensity [212].

One step further in photoelectrochemical sensors that use electron trapping is the so-called ‘on-off-super on’ system based on porous carbon nitride. The ‘on’ state is a photoanode made of carbon nitride that exhibits anodic photocurrent. Modification of the surface of g-$C_3N_4$ surface with copper ions causes a decrease in electron transfer rates and is defined as the ‘off’ state. This material could be an efficient sensor for hydrogen sulfide because the $S^{2-}$ ions could abolish the influence of Cu-induced surface exciton trapping and amplify the photoelectrochemical response by the formation of g-$C_3N_4$/$Cu_2S$ heterojunction.

In this situation, the photoinduced electron from the CB of CuS is transferred to carbon nitride and improves photocurrent generation, the ‘super on’ state [213].

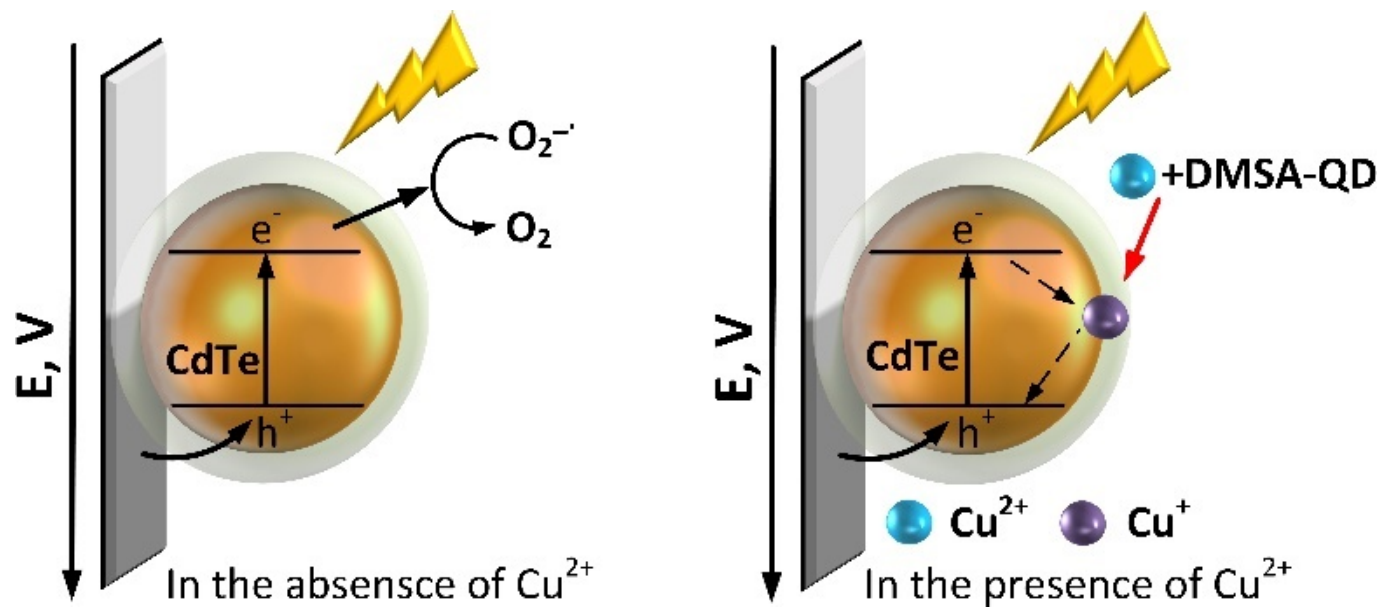

*Figure 12. Mechanisms of photocurrent generation utilizing electron trapping effect in two semiconductor heterojunction, photocurrents generated under light irradiation at open circuit potential* [212].

## 4.9 Competitive analyte-mediated electron transfer sensing

This mechanism is associated with abrupt changes in the photocurrent of the photoelectrochemical system due to the collection of charge carriers by species in the electrolyte medium. This requires the incorporation of redox reactants either for the analyte or just as an additional reactant (*e.g.* as quenching factor). Directional electron flow upon photoelectrochemical sensor operations can be realized with sole analyte and joint controllable charges relaying through the building electrode composite material. The overall effect should be to sweep charge carriers (electrons) in the direction from the semiconductor. Examples include systems where an analyte (or any type of analyte complex) acts as a photocurrent quenching factor or where the photocurrent flow direction depends on the analyte (analyte complex) or simply where external stimuli define the type of generated photocurrent – either cathodic or anodic type. The only requirement in these types of photoelectrochemical sensing systems (as well as the ones working in the opposite direction, *i.e.* toward the electrode) is that the photocurrent signal should be controlled proportionally to the analyte concentration.

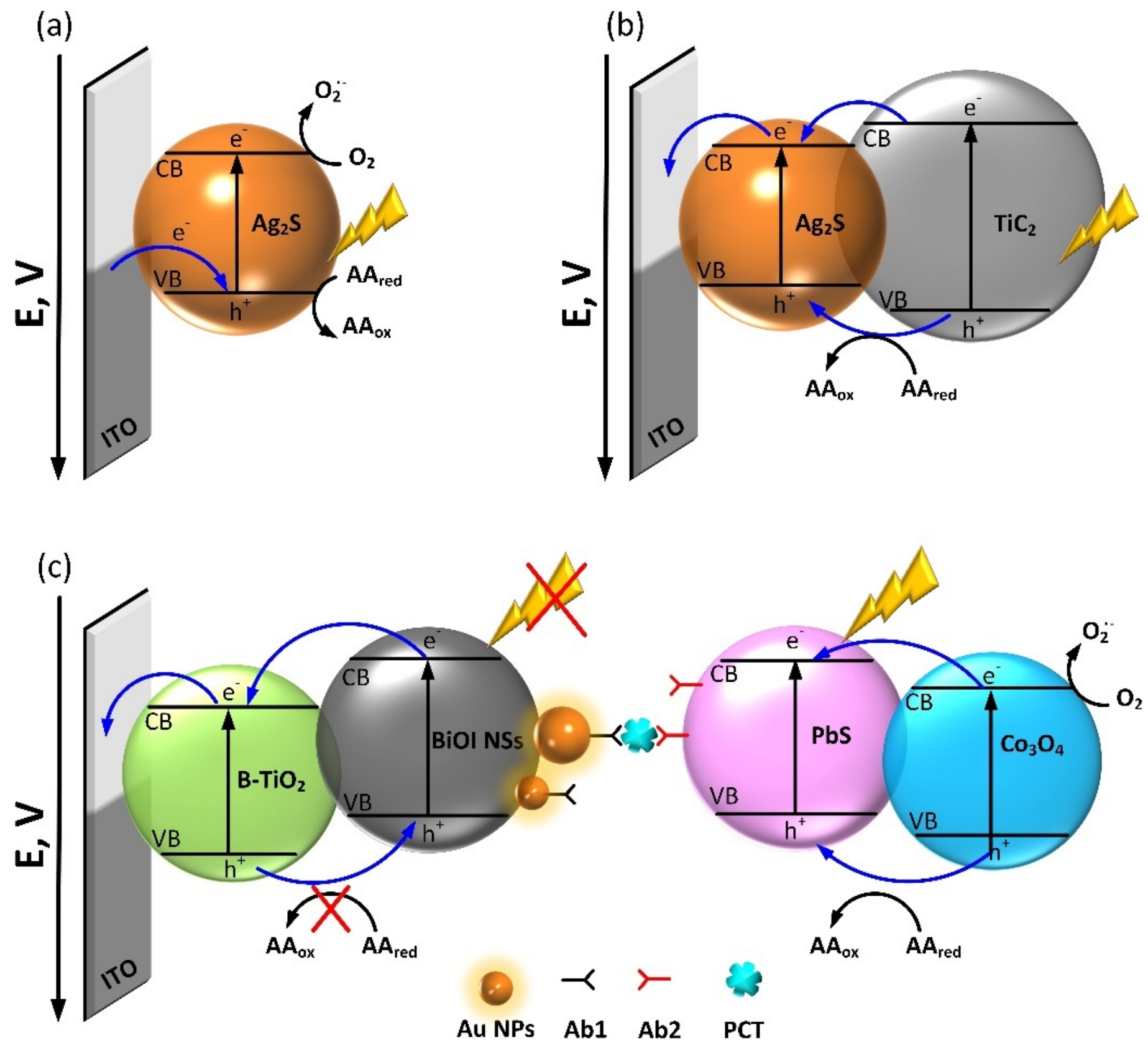

*Figure 13. Photoelectrochemical system design, based on electron transfer (type I), where the electrolyte medium is involved in the collection of charge carriers. (A) High-selectivity and-sensitivity biosensor: $Ag_2S$ NP immobilized with ITO with Ab1 antibodies detects binding to the analyte (sCD146) – by photocurrent drop. Upon the addition of the type of $Ti_2C$-Ab2 structure, the photocurrent changes from cathodic to anodic [218]. (B) Another strategy relies on signal-off phenomena. The base sensing matrix, built from $TiO_2$ NP, BiOI and functionalized Au NPs, generates enhanced photocurrents. Competing and also functionalized toward analyte bonding, $PbS/Co_3O_4$ system suppresses the electric signal upon detection – due to both the steric obstruction of one of the reagents (AA) to the sensing matrix and competing absorption of light [217]. (C) Photocurrent flow can be modified – both in direction and amplitude - by external stimuli. Exemplary performance of the Au/$CeO_2$ core/shell NP photoelectrochemical sensor performance can be tuned by external voltage. This effect does not exist for the bare $CeO_2$ NP system* [219].

The above strategy allows the exploration of new architecture types for photoelectrochemical sensors. A good example is the Au NP sensitization structure/BiOI nanosheets/black $TiO_2$ NP (Au NP/BiOI NS/B-$TiO_2$ NP), immobilized on the ITO surface as a photoactive matrix [217]. Under light irradiation, an enhanced photocurrent was measured – due to increased conductivity, capture more light, and in turn leading to improved photoelectric conversion efficiency for the photoelectrochemical immunosensor. Then, PbS/$Co_3O_4$ system was utilized as a signal label. The collective effects of steric impedance and competition for light and electron donor make this system a signal-off photoelectrochemical immunosensor. The latter was used for the detection of procalcitonin in the electrolyte containing ascorbic acid (AA). In the sandwiched system, electrode and label materials were equipped with Ab1 and Ab2 immobilized antibodies, respectively.

The underlying mechanism was based on the signal-off phenomenon. A barbed photoactive matrix had the potential for photocurrent enhancement owing to the favorable band structure of the constituting moieties and the effect surface plaque resonance of the AuNPs. The additional signal label varied the photocurrent for the photoelectrochemical immunosensor. The PbS/$Co_3O_4$ had been decreasing photocurrent - the PbS QDs absorbed full wavelength light and competed with the matrix for absorption of irradiation, and the steric resistance of $Co_3O_4$ obstructed the charge transfer between matrix and AA. Furthermore, the $Co_3O_4$ and PbS QDs experienced photogenerated electron transfer from the CB of $Co_3O_4$ to the CB of PbS QDs. The electron reacted with the dissolved oxygen present in the electrolyte, forming the superoxide anion radicals. These in turn could react with AA. In total, both factors were responsible for signal suppression from the photoactive matrix.

Recently, a photocurrent polarity switching strategy was designed to construct photoelectrochemical biosensors with high selectivity and sensitivity, capable of removing false positive or false negative interferences [218]. It was presented as an alternative to enzyme-linked immunosorbent assay (ELISA) tests for the soluble cell adhesion molecule sCD146, which is a biomarker in the diagnosis, therapy, and prognosis of lung cancer. The sensor uses one of the $Ti_2C$ MXenes, representative of the class of two-dimensional inorganic materials, from the novel family of 2D materials. It is made of the $Ag_2S$/ITO electrode, modified capture antibodies (Ab1). Upon binding of sCD146, the cathodic photocurrent intensity drops. The addition of $Ti_2C$ MXenes labeled with a detection antibody ($Ti_2C$-Ab2) creates a sandwich immune recognition structure. This in turn results in signal switching for the whole photoelectrochemical system, from cathodic to anodic photocurrent. The mechanism is based on the efficient transmission of photogenerated electrons/holes through the matched energy levels between $Ag_2S$

QDs and $Ti_2C$ MXenes. Changes in photocurrent flow, due to competitive electron transfer, can also be induced by external stimuli – such as electric potential. This was done, in conjunction with plasmonic visible light photocurrent enhancement, in hybrid Au/$CeO_2$ core/shell NPs [219]. The described system was used as a photoelectrochemical $H_2O_2$ sensor, with high selectivity (with the presence of ascorbic acid, uric acid, dopamine, glucose, NaCl and KCl) and good stability (several days). Superior photocurrent generation from hybrid Au/$CeO_2$ core/shell NPs occurs also for only Au core excitation, meaning that for this sensor, ultraviolet (UV) excitation can be avoided, increasing the number of potential applications. Regarding photocurrent switching under different applied potentials, the inspiration for such a mechanism came from previous literature examples reported for CdS [220] or CdSe/ZnS [118, 119]. The magnitude depends on the type of NP and is approximately 20 times higher for the hybrid Au/$CeO_2$ (at 500 mV vs Ag/AgCl), but for different NPs the photocurrent switch from cathodic to anodic direction usually occurs at a potential of about 200 mV vs Ag/AgCl.

The main factors of the photoelectrochemical sensors are presented in Figure 13 for each of the systems.

## 5. Photoelectrochemical sensing of energy transfer

### 5.1 Surface Plasmon Resonance

By definition, surface plasmon resonance is based on collective surface oscillations of electrons induced by incident light. It is usually observed in metal NPs at the metal-dielectric interface. Nanoparticles are frequently present in photoelectrochemical systems, which can be designed, according to Figure 14 – a transparent electrode, made of ITO or FTO, is being covered by a thin semiconductor thin film, with optional addition of photoactive molecules. In these photoelectrochemical systems, NPs usually make up most of the outer layer. Under irradiation, the generation of surface plasmons by NPs may be the reason for the altered photocurrent in joint semiconductors, both organic and inorganic. For these heterostructures, NPs usually contribute to final photocurrents in a way that amplifies the output signal. This is done in multiple ways: firstly, the hot electrons in the nanostructure are transferred by the means of surface plasmon resonance to the orbital or CB of the adjacent photoactive moiety, molecules, or semiconductors and further towards the electrode. Second, the semiconductor itself generates excitons, but their recombination is prevented because electrons can be transported to the nanostructure, contributing to increased photocurrent value. The SPR-generated electric field, depending on the NP energy and physical positions within the sensing system layer, can increase or decrease the photocurrent response. Another type of photocurrent modification is the formed Schottky junction which can suppress the backflow of charge carriers. Sensors based on photoelectrochemical techniques usually exhibit detection limits than electrochemical ones, as their response is enhanced. However, additional mechanisms of energy transfer presented in the following sections can work in an opposite way, contributing to the diminishing of the photocurrent. This is also utilized for sensing applications and can be incorporated, with more restrictions, toward information processing, at best, in reservoir systems.

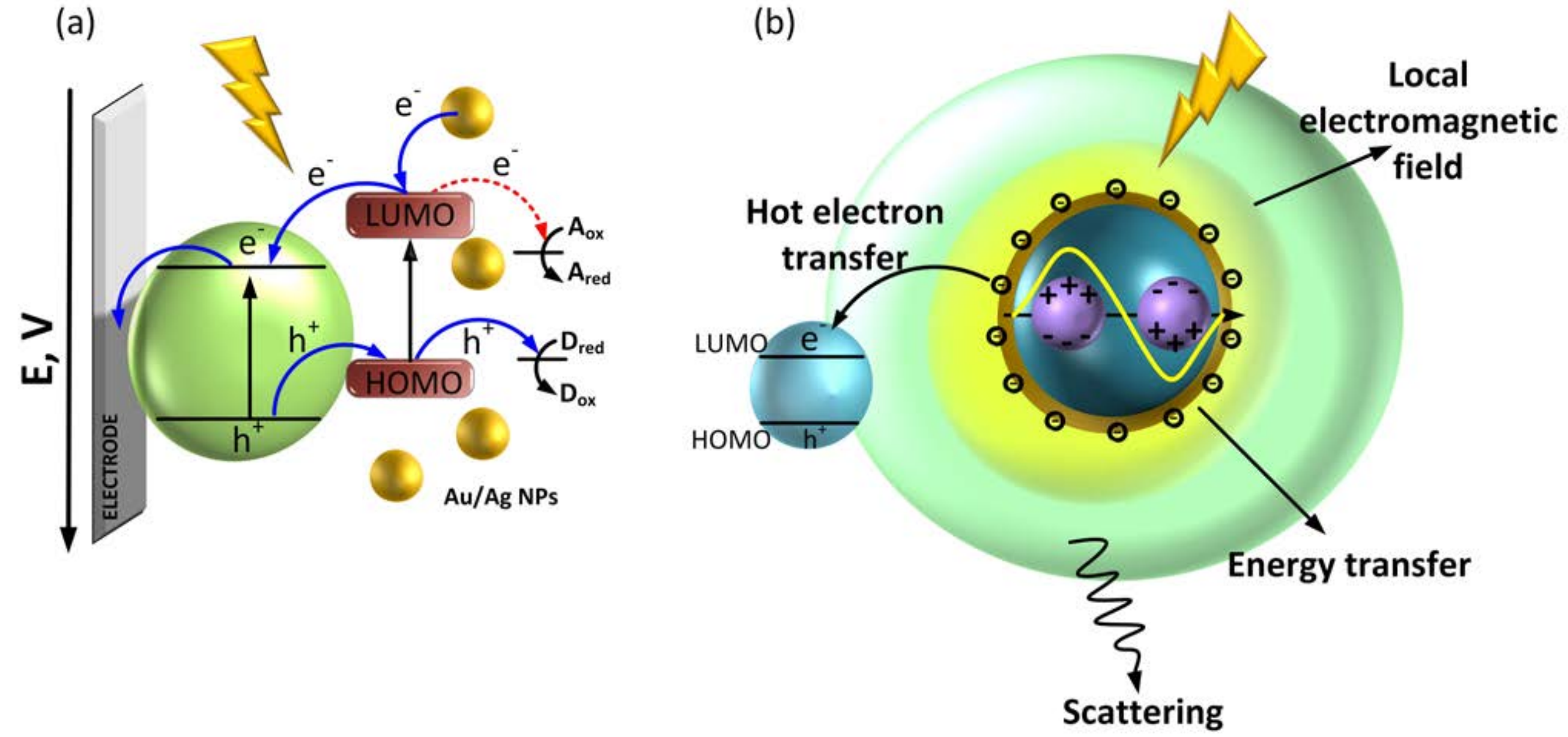

*Figure 14. Typical photoelectrochemical sensor system design, based on nanoparticle incorporation. Gold nanoparticle-based photoelectrochemical sensor systems have abundant ideas for enhanced photocurrents. They are added to different types of semiconductor-based photoelectrodes, usually contributing to photocurrent amplification. Processes are shown from the perspective of semiconductor nanoparticle (a) and plasmonic metal nanoparticle (b) [222].*

For information processing, in case of no external signal amplifier, the photocurrent decrease should not be significant, because then after only a few signal generations, the response will decay into noise. If one of the constituting materials for the photoelectrochemical sensors is memristive, then the underlying resistive switching mechanism (such as interfacial charge carrier trapping) can incorporate the necessary delay in the reservoir system, implementing the internally reservoir concept. If not, the delay as well as the amplification can be implemented by additional hardware pieces. When the photocurrent response, i.e. signal, is passed numerous times through the same processing center, the response can carry more complexity than during only one instance of read-out sequence. So far, the utilization of photoelectrochemical sensors in reservoir systems is scarce, yet potential candidates are presented within the range of current review. For reservoir computing to be feasible, you should implement them as processing/storage units in the center of processing systems.

The gold NPs can be incorporated into the structure of the $Bi_4NbO_8Cl$ perovskite, by means of a solid-state reaction, and the electrode is made by a layer-by-layer technique in ITO (Figure 15a) [221]. The mechanism of photocurrent enhancement by NPs is based on the 'hot electrons' (near the Fermi level) being excited to the surface plasmon state and, due to the correct alignment of the band, transferred to the electric circuit. The Au NPs decrease the charge trapping on the surface states and thus facilitate fast charge separation and concomitant transfer. Furthermore, the holes in $Bi_4NbO_8Cl$ and Au NP would simultaneously oxidize cysteine biomolecules.

Plasmonic NPs can cause an increase in the photocurrent response due to several effects, such as electrical field amplification, enhanced light absorbance, and interfacial charge transfer in the semiconductor/metal interface. In this way, the photocatalytic process can be enhanced, but the generated photocurrent can also serve as a detection signal. The introduction of gold NPs to g-$C_3N_4$ resulted in a decrease in the recombination process of the $e^-/h^+$ pair, leading to a decreased photoluminescence intensity for such an altered system, but generating a stronger photocurrent response (Figure 15b) [223]. An interface was observed to form between AuNPs and g-$C_3N_4$ – which strengthened electron transfer effects through the composite structure. After irradiation, electrons were introduced to the CB of both AuNPs and g-$C_3N_4$. These electrons are then instantly transferred to the electrode, contributing to a higher photocurrent. This system was used as a photoelectrochemical biomolecule detection system and did not require the irradiation from ultraviolet range as both moieties absorbed within the visible range.

Another example is the $BiVO_4$-based photoanode constructed on Mo:$BiVO_4$ and Au nanoparticles, which also experiences higher generation and less bulk recombination of charge carriers [224]. The multifunctional nature of $BiVO_4$-based electrode systems was recently reported,[225] where Ag-$BiVO_4$ nanocomposite needs an additional layer of reduced graphene oxide. This time, the system was also introduced as a sensor for nitrites. The role was to determine faster charge separation and also to electrochemically oxidize $NaNO_2$. Once again, potential usage in neuromorphic computing is replaced in the above publications by an example used not only in sensing but also in photoelectrochemical water splitting. It is worth mentioning that even the bare $BiVO_4$ nanoparticle thin film system is known for its good memristive features and can be used in neuromorphic computing [226].

The deposition of nano-sized plasmonic Au particles was associated with subsequent hydrogenation of the $TiO_2$ semiconductor, leading to Au@H-$TiO_2$ (Figure 15c) [227]. The introduction of Au nanoparticles induced a plasmonic effect, which allowed light absorption to move from the UV range to the VIS range. The hydrogenation process improved the oxidation activity. The system was used under visible light irradiation for selective detection of different types of sugars. This included glucose, fructose, sucrose, and lactose in the presence of the aromatic compound potassium hydrogen phthalate, without the need for separation. The plasmonic Au NP allowed sugars (*e.g.* glucose) to be oxidized by hot-hole transfer, leaving the aromatic compound unreacted. Under UV irradiation, the sensor detected the overall amount of organic compounds by nonselective photodegradation of these compounds (without AuNP involvement in the process).

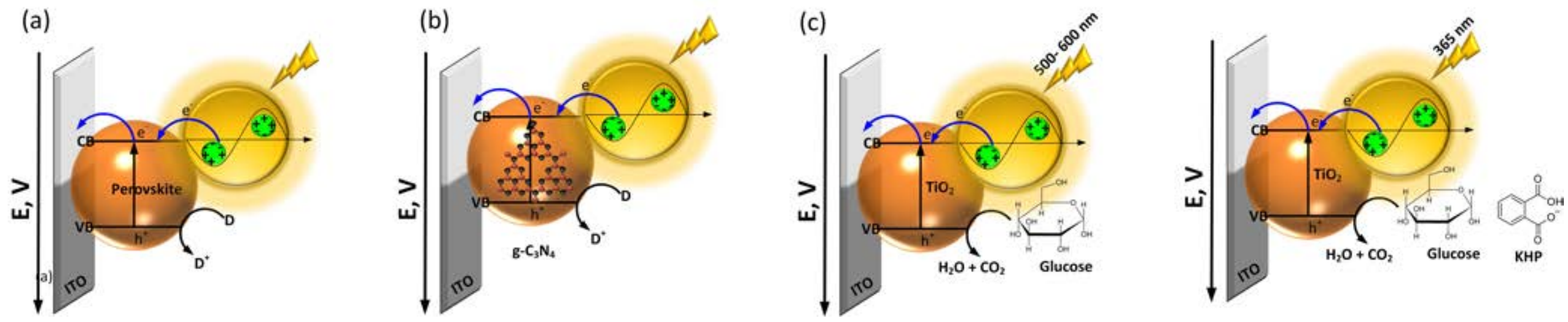

*Figure 15. A variety of SPR-based photoelectrochemical systems: (a) AuNPs incorporated into the layer of $Bi_4NbO_8Cl$ perovskite* [221]. *(b) AuNPs joint with g-$C_3N_4$ – decrease in recombination rate of the $e^-$/$h^+$the pair (lower photoluminscence intensity), increase in photocurrent* [223]. *(c) Au NPs with hydrogenated $TiO_2$. The photoelectrochemical oxidation activity of this system was selective for different sugar types, and other aromatic compounds unreacted* [227].

Another type of SPR-enhanced sensor is presented as a $TiO_2$-based sensor, where $TiO_2$ nanowires were grown directly in FTO glass and then decorated with AuNP (Figure 16) [228]. Next, Au NPs were functionalized with ganglioside GM1 – cell membrane receptor for binding of cholera toxin subunit B. The proximity of Au NP to $TiO_2$ provides strong SPR effects, resulting in ~100% increase in the photocurrent response. In addition, sensitivity got improved, as the receptors are bound directly to the Au NP surface, minimizing the electromagnetic field coupling efficiency. The SPR effects in this system include the electrical field amplification effect and the injection of hot electrons into the CB of $TiO_2$. The first phenomenon involves coherent oscillations of free electrons in the CB of Au NPs and electromagnetic field formation, which can span through hundreds of nanometers in depth and sensitivity to its surroundings. Binding of the target molecule interferes with the electrical-field coupling efficiency and causes a photocurrent change. This effect is even stronger (stronger attenuation of the electromagnetic field coupling) than for direct binding of the molecule on the $TiO_2$ surface. Redox reactions with proteins are not also present. the second of the mechanisms presented allows for transport of SPR-generated hot electrons into the CB of $TiO_2$, and thus increases the photocurrent.

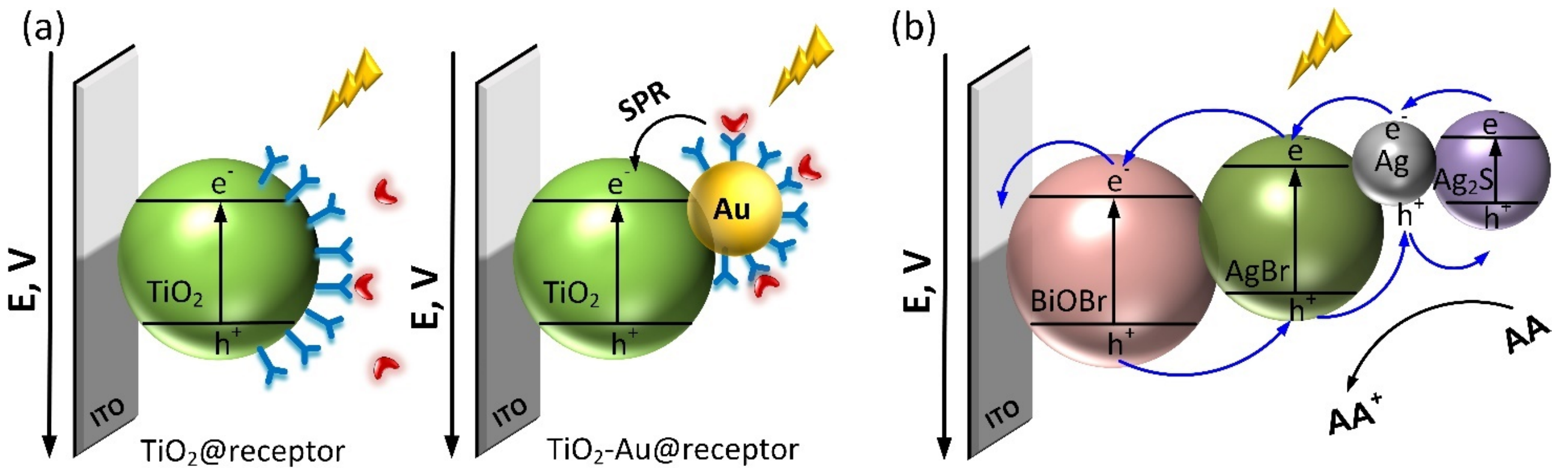

*Figure 16. A variety of SPR-based photoelectrochemical systems: (a) AuNP on $TiO_2$ nanowires. AuNPs functionalized with receptors increase the overall photocurrent response of the system* [228]. *(b) Ag/AgBr/BiOBr heterojunction for PSA sensors. The AgNPs suppress charge carrier backflow and strengthen electron acceptor behavior* [229].

Plasmonic NP can be used in a way that does not require their incorporation into the sensor itself – as in [230], where self-assembled thiol-terminated biotin molecules bound to the Au nanoislands /$TiO_2$ photoelectrode were successfully used to modify the photocurrent response to the analyte (Figure 17). Gold NPs were modified with STA Au NPs and acted as analytes for the in situ real-time measurement of biotin-streptavidin binding kinetics. The binding of thiol-terminated biotin to Au nanoislands was due to the extremely strong Au-S bond, which prevented interactions with other functional groups and provided good resistance to acid, alkali, and external forces for the sensor system. After the interaction occurred, the distance between the Au nanoislands and the AuNPs was only a few nanometers, allowing for interparticle interactions, yielding intensification of the hot electron transfer process. Photocurrent enhancement was reported as a result of SPR effects, and the response was sensitive to different concentrations of STA-AuNP while being irradiated with 600-nm light.

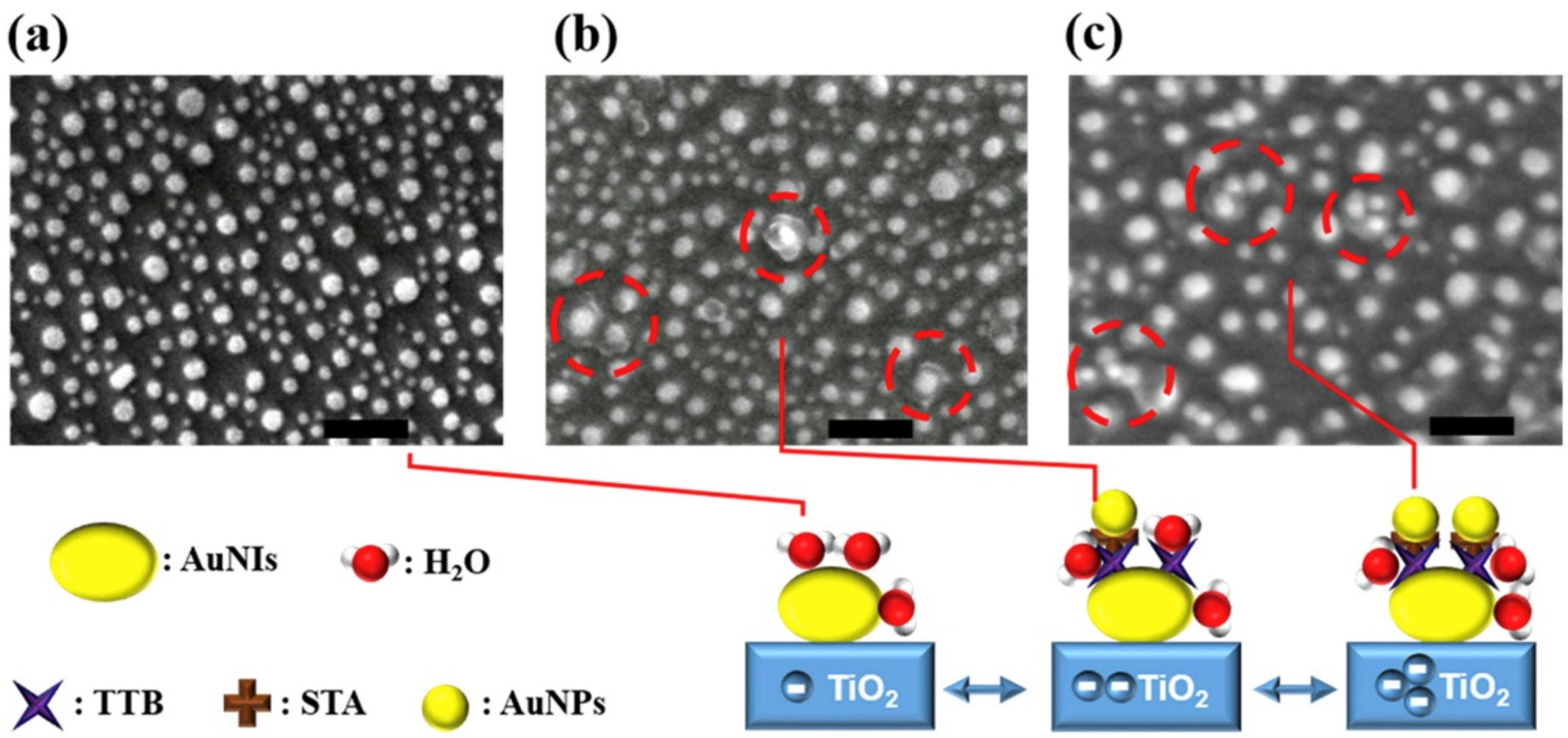

*Figure 17. Modified AuNP (STA-AuNP) allowing for SPR-based signal enhancement on the $TiO_2$ electrode. Reprinted from Ref.* [230] *under Creative Commons CC-BY.*

Other types of plasmonic NPs are AgNPs, which contribute in a manner similar to enhanced photocatalytic activity for the base material while being part of the composite materials. In addition, Ag-NPs suppress charge carrier backflow and strengthen electron acceptor behavior. Ag-NPs and nanostructured BiOBr composites are mostly used in photocatalysis, yet, with slight modifications, they can be applied to detection, as for the exemplary Ag/AgBr/BiOBr heterojunction for prostate-specific antigen sensors (Figure 16c) [229].The sensor had a wider linear range, lower detection limit, and better specific response, compared to similar PSA detecting systems.

The Au@Ag core@shell structures combined with the $TiO_2$ nanosheet film were reported by Zhang et al. as a mercury ion photoelectrochemical sensor [231]. The photoelectrochemical sensor was used in $Hg^{2+}$ detection: It had a significant increase in the photocurrent response, while being irradiated at 430 nm. With the addition of $Hg^{2+}$ ions, the photocurrents were selectively and quantitatively decreased. The use of two noble metals contributed to better NP stability, but also allowed tuning of SPR peaks – pure Au NPs would not absorb below 500 nm, whereas the addition of Ag would move the absorption to 400 nm. This in turn leads to desired plasmon-induced charge separation connected with hot electron injection to the CB of the base semiconductor over the Schottky barrier at the interface.

PEC systems with metal NPs can be fine-tuned – usually by variations of metal core sizes, but also, in some cases – by variations of protective layer thickness. To make strategies toward signal amplification more organized, fine control of the NP size and density in the thin layer is usually implemented. This allows tuning of the surface plasmon waves at the surface of the metal to optimize the response of the sensors.

The underlying need to execute such optimization lies in the fact that most common systems use metals and semiconductors that stay in direct contact. This in turn favors nonradiative energy transfer from the photoexcited semiconductor to metal NPs and may, in some cases, contribute to decrease the photocurrent. At the same time, the SPR-mediated electric field increases light absorption of the semiconductor, and light scattering improves light harvesting efficiency throughout the whole process. The latter effects occur even when the NP is separated by an insulating layer of nanometer thickness. That is why the Ag@$SiO_2$ core-shell system with additional gold nanoclusters (Au nanoclusters-Ag@$SiO_2$) was studied [222]. In the thickness of the example, the $SiO_2$ layer was changed and influenced the photoelectrochemical sensor for alkaline phosphataseactivity, with AA as an electron donor. The optimum distance between Au nanoclusters and Ag NPs was found, 15 nm, where the transfer of hot electrons, the local electric field, and the light scattering effects were maximized to improve photoelectric performance.

Plasmonic NPs do not necessarily need sophisticated semiconductors- amplification can occur solely on a simple ITO electrode (Figure 18) – this time used as a photoelectrochemical sensor towards the hydroquinone [232]. The signal amplification mechanism is described to be based on the hot-electron injection process resulting from plasmonic excitation. Additionally, in the presence of an analyte, the photocurrent increases, due to subsequently oxidation of hydroquinone to benzoquinone by transferred (from ITO) photogenerated holes. Modification of the ITO electrode by Au NPs enhanced the current response to hydroquinone 9 times, whereas additional irradiation 4.3 times.

In the current example, the photoelectrochemical sensing platform for the detection of 4-chlorophenol was made of Au/g-$C_3N_4$ compounds [233]. In this system, the Au-NP SPR improved light adsorption and enhanced the photoelectric conversion efficiency of pure g-$C_3N_4$. The composite sensor achieved a photoresponse 3.4 times stronger than the g-$C_3N_4$ sensor.

The underlying mechanism is presented as being based on the local electromagnetic field enhancement, where the NP acts as a charge carrier mediator, separating photogenerated electron-hole pairs. First, though, the SPR needs to occur upon irradiation leaving photogenerated electrons. Then additionally at Au/*g*-$C_3N_4$ interface the electrons from CB of *g*-$C_3N_4$ can travel through the SPR effect of Au, remaining on the surface of Au NP. Together with other electrons accumulated in the CB of the *g*-$C_3N_4$ composite parts, they contribute to the overall increase in photocurrent. When exposed to the 4-chlorophenol analyte, holes in the *g*-$C_3N_4$ are used in the oxidation process. In summary, enhanced charge separation suppresses electron-hole recombination, increasing the photocurrent intensity. Noble-metal NPs act as electron mediators, improving the light absorption and photoelectric conversion efficiency of a joint semiconductor.

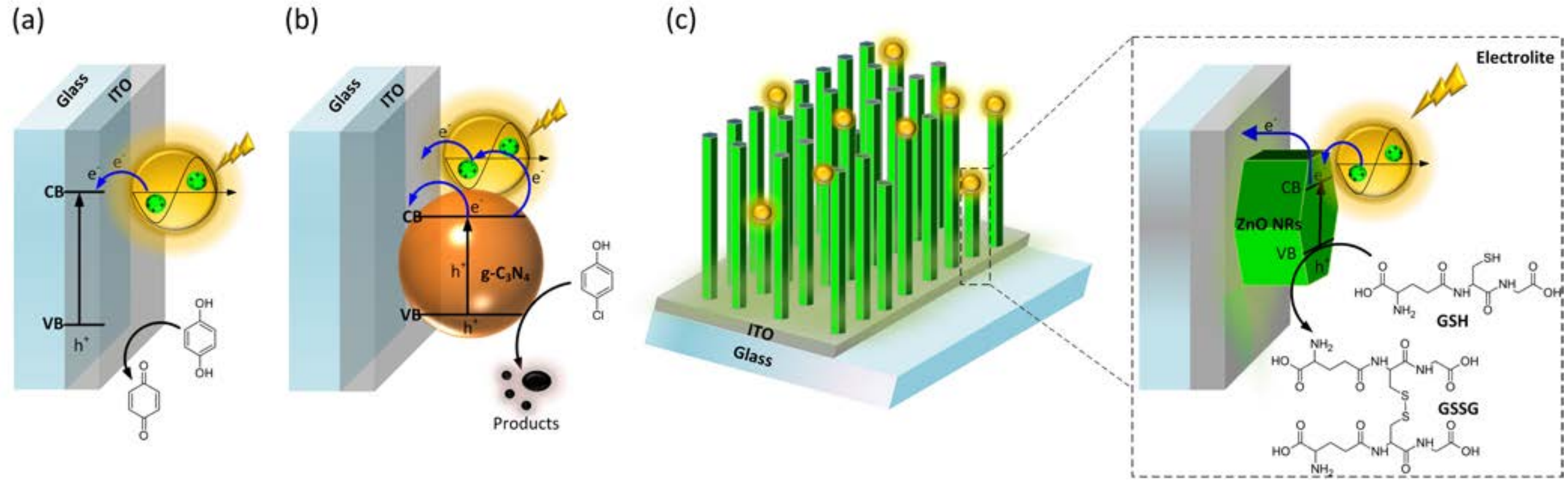

*Figure 18. A variety of SPR-based photoelectrochemical systems: (a) AuNPs on bare ITO photoelectrode. The photocurrent increases and the response to hydroquinone sensing is enhanced 9 times, whereas additional irradiation results in another enhancement 4 times* [232]*. (b) AuNP with g-$C_3N_4$NPs on ITO electrode. The SPR of AuNP enhances light adsorption and the composite sensor achieves 3.4 times stronger photoresponse than the pure g-$C_3N_4$ sensor* [233]*. (c) In the Au/ZnO NR system, hot electrons from Au NPs are injected into ZnO and then driven toward the photocathode. The composite system utilizes the visible-light region, generating much higher photocurrent response* [234].

As was written in the Introduction, SPR-related energy transfer phenomena do not necessarily contribute to electrode photocurrent in an incremental way. A sensor based on AuNPs connected to poly[(9,9-dioctylfluorenyl-2,7-diyl)-co-(1,4-benzo-thiadazole)] (PFBT) polymeric dots (Pdots) by chains with an incorporated telomerase starter is one of these examples [235]. Local electric fields originating from SPR effectively quench the photoresponse from the Pdots electrode. The sensor systems are built in such a way that upon telomerase activity in cell extracts, AuNPs are released, removing the Au SPR-induced electric field, leading to the recovery of the signal from Pdots (photocurrent increase).

One of the classic examples of metal NP / semiconductor is Au/ZnO NRs [234] where hot electrons originating in Au NPs, due to Vis range light absorption and SPR, are injected into ZnO and then driven towards the photocathode. At the same time, photogenerated electrons from UV light absorption in ZnO are transported in the same direction. The ZnO-based photogenerated holes move from the surface of ZnO to the surface of ZnO and then take part in the oxidation of GSH to glutathione disulfide GSSG. The introduction of Au NPs with absorption maximum at 520 nm significantly enhanced the utilization of visible light - pristine ZnO exhibited absorption below 380 nm. For reservoir systems, the memristive character of ZnO [236] can be taken into account when making decisions about the materials of the information processing systems.

The general trend in the construction of MNP-enhanced electrodes is to significantly improve the photoelectric response of the sensor. These are mainly ascribed to SPR-generated hot electrons, near the Fermi level, which can be transported to the CB of the semiconductor. In addition, the SPR related local electric field can influence the absorption efficiency of the nearby semiconductor. The light-scattering effect on NPs can improve the overall light-harvesting efficiency. In Figure 18 the reader can find a summary of the designs used for SPR-based photoelectrochemical sensors. The following Table 2 presents the summary of sensors with a short description of the underlying mechanisms on the changes in the photocurrent of the photoelectrochemical sensor. Strategies for output signal alternation can include both increase and decrease in the final readout, making nanoparticle-based photoelectrochemical sensor modification methods the most universal and straightforward sensor tuning approaches.

### 5.2 Resonance energy transfer

Most recently, the photoelectrochemical process associated with the resonance energy transfer (RET) strategy has resulted in new opportunities to advance photoelectrochemical-based analysis and provide a remarkable pathway to investigate various analytes in photoelectrochemical sensors, especially bioassays [237]. Förster resonance energy transfer (FRET) is a dipole-dipole coupling process between two fluorophores; named donor (D) and acceptor (A). Photoexcitation leads to excitation and the creation of a dipole moment in the oscillating donor molecule ($\mu_D$), which is followed by an induced oscillating dipole moment ($\mu_A$) in the acceptor species. When the excited donor (*D) relaxes to the ground state and stops oscillating, the excited states in the acceptor (*A) are generated in the medium, and energy transfer occurs simultaneously. The energy transfer rate ($W_{RET}$) in dipole-dipole coupling by FRET is defined through Eq.18:

$$W_{RET} = \frac{9000\Phi_D \log \kappa^2}{128\pi^5 N_A n_r^4 \tau_D r^6} J(\lambda) \qquad (18)$$

where $\Phi_D$ is the donor fluorescence quantum yield, $\tau_D$ is the donor emission lifetime, $n_r$ is the refraction index of the host medium, $N_A$ is the Avogadro number, $\kappa$ is an orientation factor, and $J(\lambda)$ is the spectral overlap given by (19):

$$J(\lambda) = \int v^{-4} \varepsilon(v) I(v) dv, \quad (19)$$

where $v$ is the wavenumber, $\varepsilon(v)$ is the absorption spectrum profile of the acceptor, and $I(v)$ is the normalized emission spectrum of the donor. As shown in Figure 19, the donor must have a broad emission intensity to overlap ($J(\lambda)$) with the absorption spectrum of a particular acceptor to meet the requirement for energy transfer [238-240]. It is imperative that the distance between fluorophores be around 2 nm < r < 10 nm.

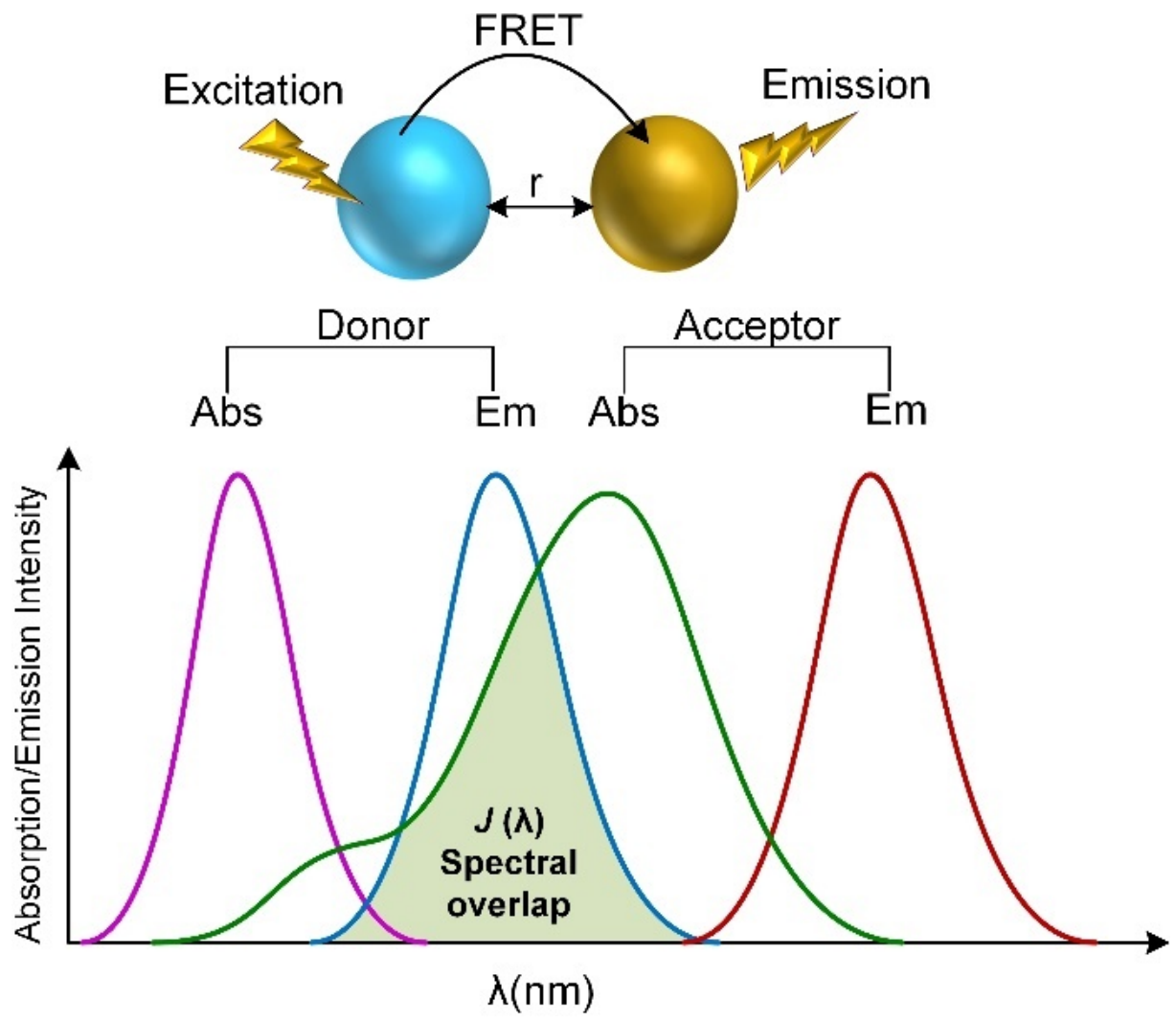

*Figure 19. Representation of the FRET process and the spectral overlap between the absorption and emission spectra of photoactive materials in FRET* [240].

### 5.3 Plasmon-induced resonance energy transfer

An approach that holds promise for enhancing the functionality of photoelectrochemical electrodes is the utilization of plasmonic metals along with semiconductors to modulate the photocurrent intensity. Direct electron transfer and light scattering have been thoroughly discussed in section 5.1 (Figure 14). Plasmon-induced resonance energy transfer (PIRET) is another probable phenomenon that occurs using light stimulation of semiconductors in the presence of plasmonic centers, which are separated by an insulating layer with thickness up to ~25 nm [241-244] (Figure 20). The overlap of absorption line of semiconductor and the resonance band of plasmonic metal as donor, as well as their separation distance, both have a significant impact on PIRET efficiency.

In contrast to FRET, which transferred energy has a longer wavelength and counts as a red-shift process, PIRET has a blue-shift energy-transfer mechanism. As a result, plasmonic nanostructures can absorb the near-infrared and visible range of sunlight and overcome the restrictions of single semiconductors with high band gap energy in photoelectrochemical cells and expand the application of devices [245].

Contact of metal NPs and metal oxide photocatalysts can build a Schottky barrier in the energy band diagram (work function semiconductor > work function metal) and prevent recombination by improving electron-hole separation, conductivity, and amplified photocurrent intensity.

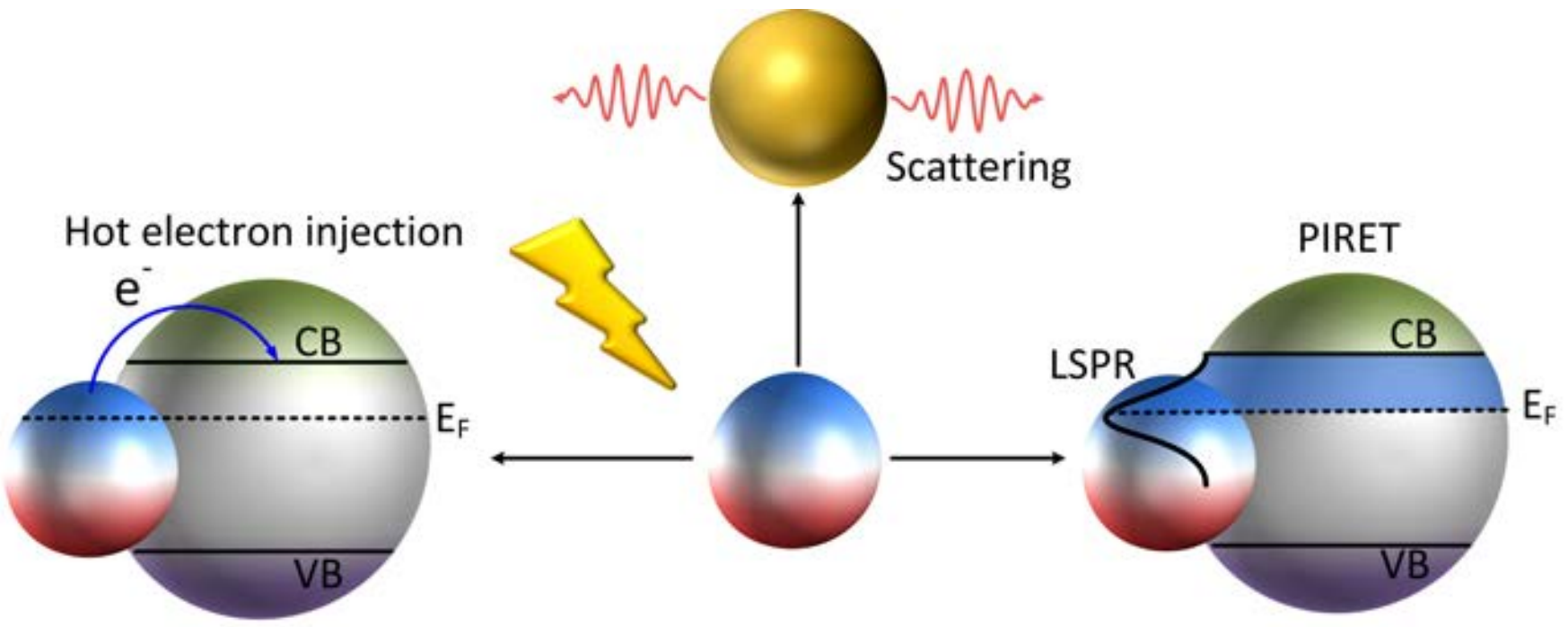

*Figure 20. Schematic models of possible mechanisms that affect photocurrent intensity after irradiation of the plasmonic center and semiconductor by light stimuli. (a) Hot electron injection; (b) Light scattering; (c) PIRET [244].*

Ma et al. [246] reported 9 plausible processes that participate in the anodic photocurrent of QDs (eg, CdS QDs), which we have represented in Figure 21. Upon exogenous light stimulation (process 1), electron-hole pairs (exciton) form inside the semiconductors (process 2). Moreover, holes can be consumed by hole scavengers in solution (process 3), and electrons transfer to the electrode surface (process 4) which can increase the photocurrent or recombination of electron holes through phonon decay (nonradiative) (process 5) and photon decay (radiative) (process 6) can happen. Comprehensive understanding of the nature of energy transfer process between plasmonic centers and photocatalysts is a challenging subject that needs to be explored to optimize utilization of the SPR-exciton systems and justifying photocurrent intensity in experimental findings. As shown in Figure 21, if emission spectra of semiconductors (process 7) overlap considerably with the absorption of metallic NPs (e.g., Au NPs), could in turn accelerate radiative decay of QDs through surface plasmon resonance energy transfer (process 8). Additionally, vicinal Au NPs can accelerate process 5 by exciton energy transfer (process 9). SPR and exciton energy transfer have a cooperative relationship in the photoelectrochemical system and compete with the electron transfer process, which decreases the overall photocurrent intensity.

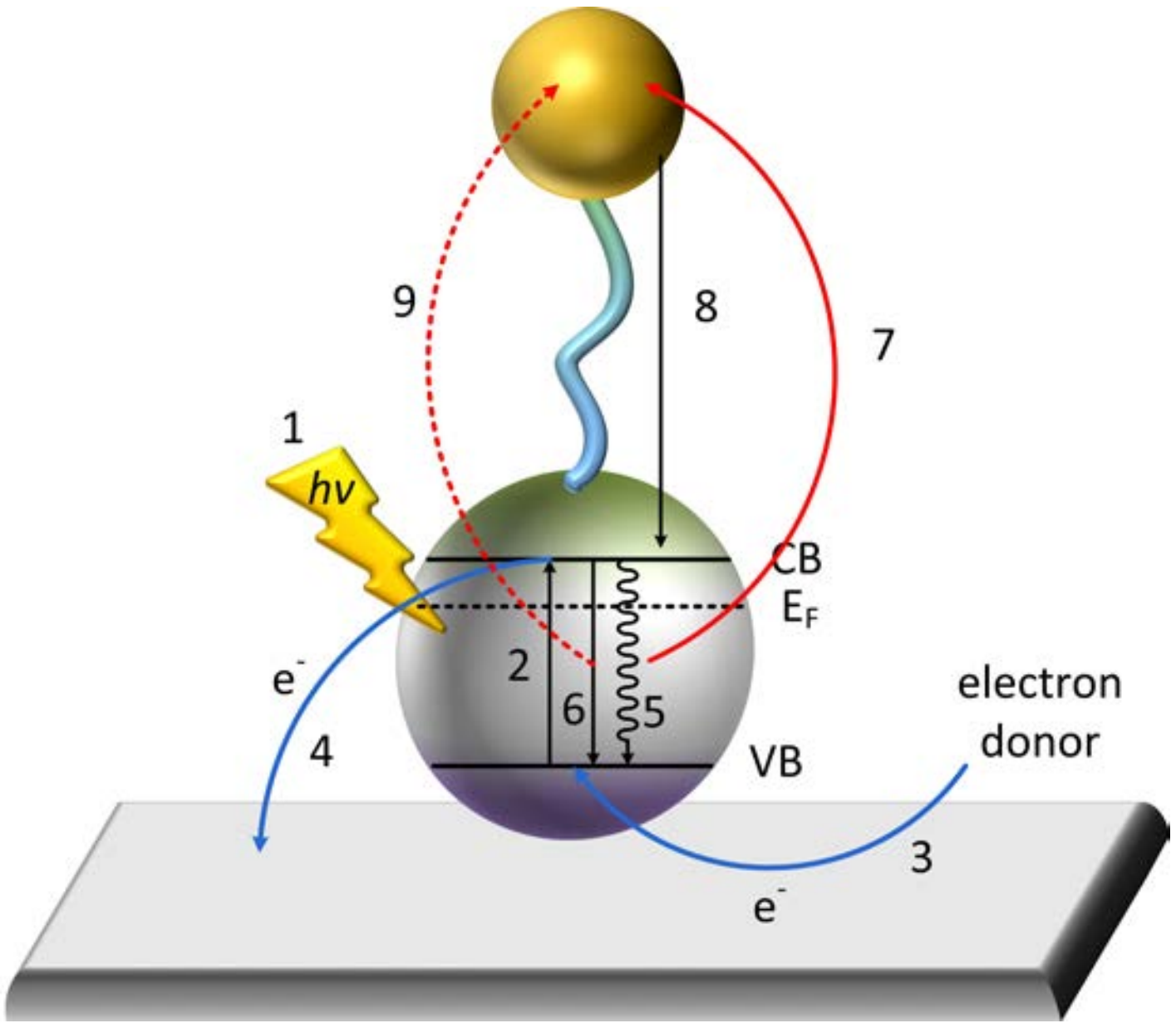

*Figure 21. Illustration of possible mechanisms in the operation of the photoelectrochemical device made of semiconductors and plasmonic metals separated by linkers in sensors* [246].

The remarkable photochemical stability, size-dependent emission, and compatible interactions with biomolecules have made quantum dots (QDs) ideal energy donors in photoelectrochemical processes (Table 2) [247]. Fan et al. investigated the effect of alloying QD and increasing the number of plasmonic metals in the field of RET sensing and increasing the sensitivity of aptasensors [237, 247, 248]. They showed that exciton-plasmon energy transfer leads to increased electron-hole recombination and decreased photocurrent intensity in CdSeTe QDs. After incubation of the aptasensor with thrombin and blocking AuNps, energy transfer was inhibited and the electron transfer process was activated, and as a consequence, the photocurrent intensity dramatically increased [247].

### 5.4 Chemiluminescence-based energy transfer.

The excitation light source is an essential part of photoelectrochemical detection. However, external light sources (xenon lamps, monochromators, etc.) restrict the application of sensing equipment. In recent years, chemiluminescence (CL) as an internal light source utilized to excite molecules or atoms has attracted considerable interest. In addition to the luminous material and coreactant (such as $H_2O_2$ or $K_2S_2O_8$), a suitable quencher or enhancer is required for the construction of a CL sensor. Metal NPs such as Pt NP, Au NP, and Ag NP, p-iodophenol, cyanine dye have been used to enhance or quench the CL intensity based on the chemiluminescence resonance energy transfer or surface plasmon resonance in sensors [249-251]. Such energy transfer can affect the excitation of acceptor and the probability of carriers, subsequently can change the photocurrent intensity [252]. The oxidation of luminol by enzyme in the presence of hydrogen peroxide is a known CL reaction, which has inspired researchers to use this concept to develop the sensitivity of photoelectrochemical biosensors [252]. Golub et al. [253] introduced a sensing system based on the CL process in hemin / G-quadruplex in the presence of luminol/$H_2O_2$, and the straight internal chemiluminescence acts as a light source that stimulates semiconductor excitation and generate the electron-hole pairs. Electron transfer from the conduction band (CB) to the electrode surface and concurrent scavenging of the holes by triethanolamine (as an electron donor) trigger an amplified photocurrent (Figure 22).

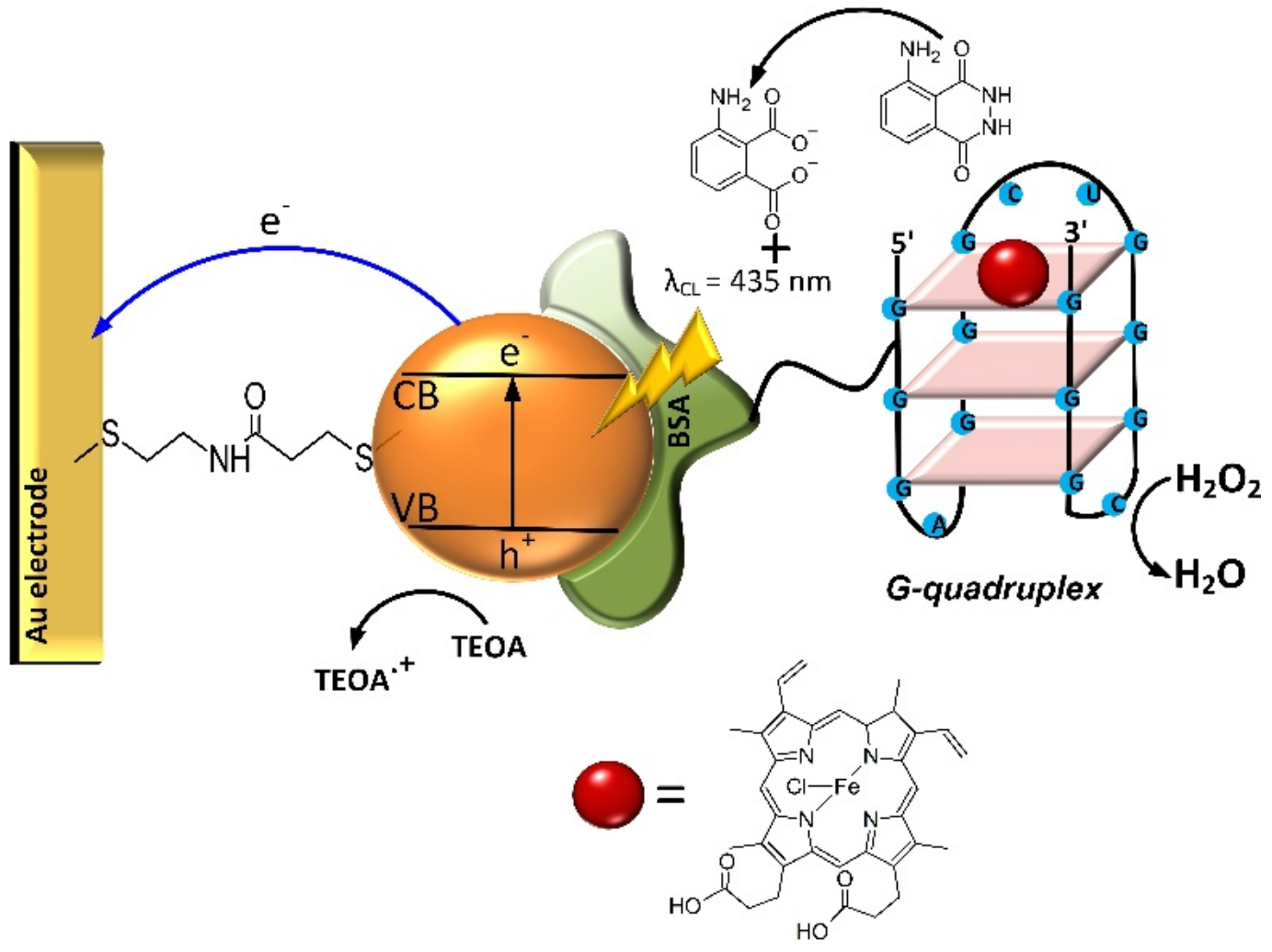

*Figure 22. CdS QDs connected to an Au electrode and functionalized by a hemin/G-quadruplex tethered to a layer of bovine serum albumin (BSA) produce photocurrents. The creation of the photocurrent and the excitation of QDs are caused by the chemiluminescence that results from the catalyzation of luminol $H_2O_2$ oxidation by hemin G-quadruplex [253].*

The attachment of Au NPs to glucose oxidase (GOx) and N-(aminobutyl)-N-(ethylisoluminol) in the microfluidic paper-based photoanalytical system reported by Wang et al., displayed increased sensitivity, as shown in Figure 23 [249]. Combining the CL process and electron transfer from excited CdS to $TiO_2$, surpass the electron-hole recombination and boosts the photoelectrochemical performance of the device.

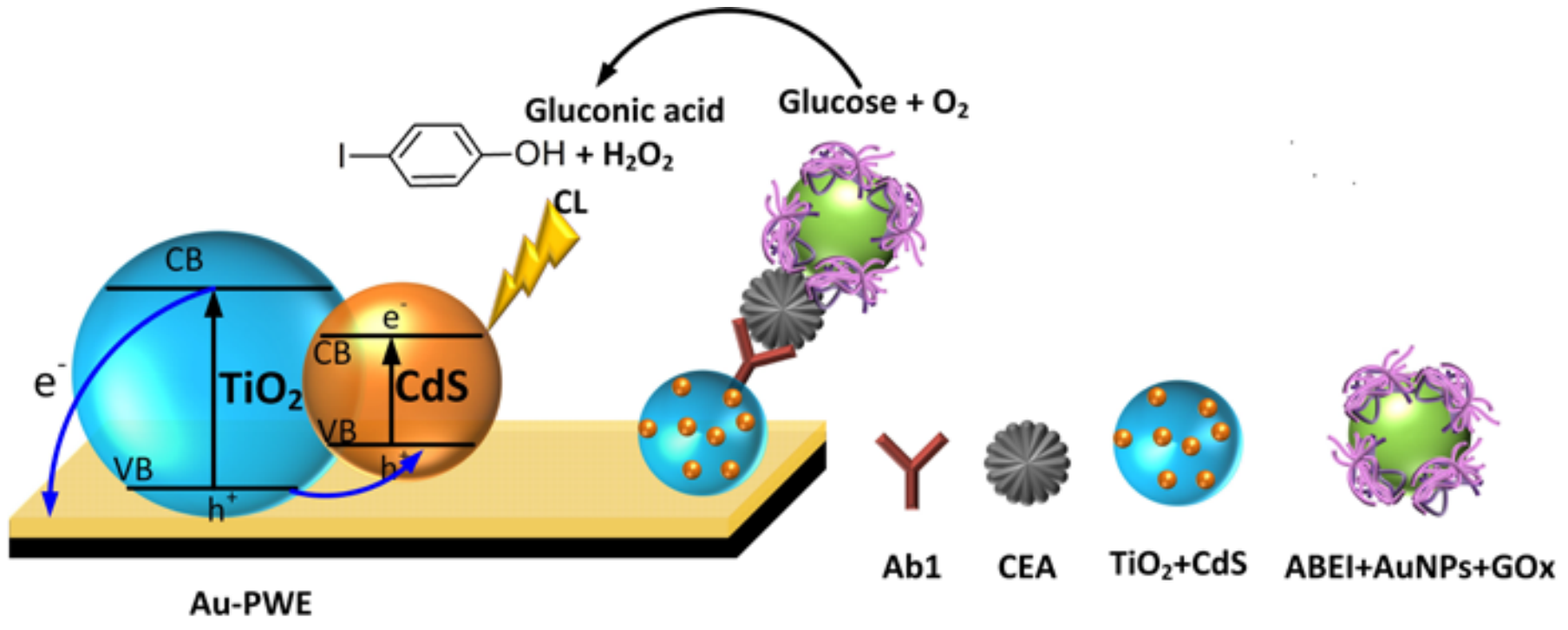

*Figure 23. Photocurrent generating mechanism in the modified Au paper working electrode (Au-PWE) under an internal CL light source used for detection of carcinoembryonic antigen (CEA). Screen-printed carbon working electrodes are referred to as SPCWE, SPCCE stands for screen-printed carbon counter electrodes, while ABEI for N-(aminobutyl)-N-(ethylisoluminol) [249].*

Recent methods developed based on the energy transfer mechanism are summarized in Table 2. [221-223, 227-235, 246, 247, 249, 252-260]

## 6. Reagent-Related Sensing Approaches

### 6.1 The induction/release of photoactive species

The induction or release of photoactive species in response to an analyte seems to be a convenient method for photoelectrochemical sensing. Several approaches may be involved in the use of these photoactive species as the signal modifier. One of them is to modify the distance between the signal label and the electrode. Zhang et al. [261] reported a photoelectrochemical biosensing strategy based on conformational changes in deoxyribonucleic acid (DNA). Figure 24 shows that this sensor consists of ITO modified with $SnO_2$ NPs, which are silanized with 3-mercaptopropyl-triethoxysilane and covered with Au NPs. Probe DNA is assembled on the electrode through the Au-S bond of the sulfhydryl group at the one end of DNA chain, while the $Ru(bpy)_2(dcbpy)^{2+}$) photosensitizer is at the other end. In the absence of target DNA, the probe DNA is in the hairpin form, so the photosensitizer is close to the ITO, giving a strong photocurrent. After hybridization with target DNA, the probe DNA hairpin opens to the extended rigid duplex form, which significantly extends electron transfer tunneling distance resulting in reduction of photocurrent. This approach seems to be interesting in the context of applications to reservoir systems as long as the photocurrent conditioning reaction is reversible.

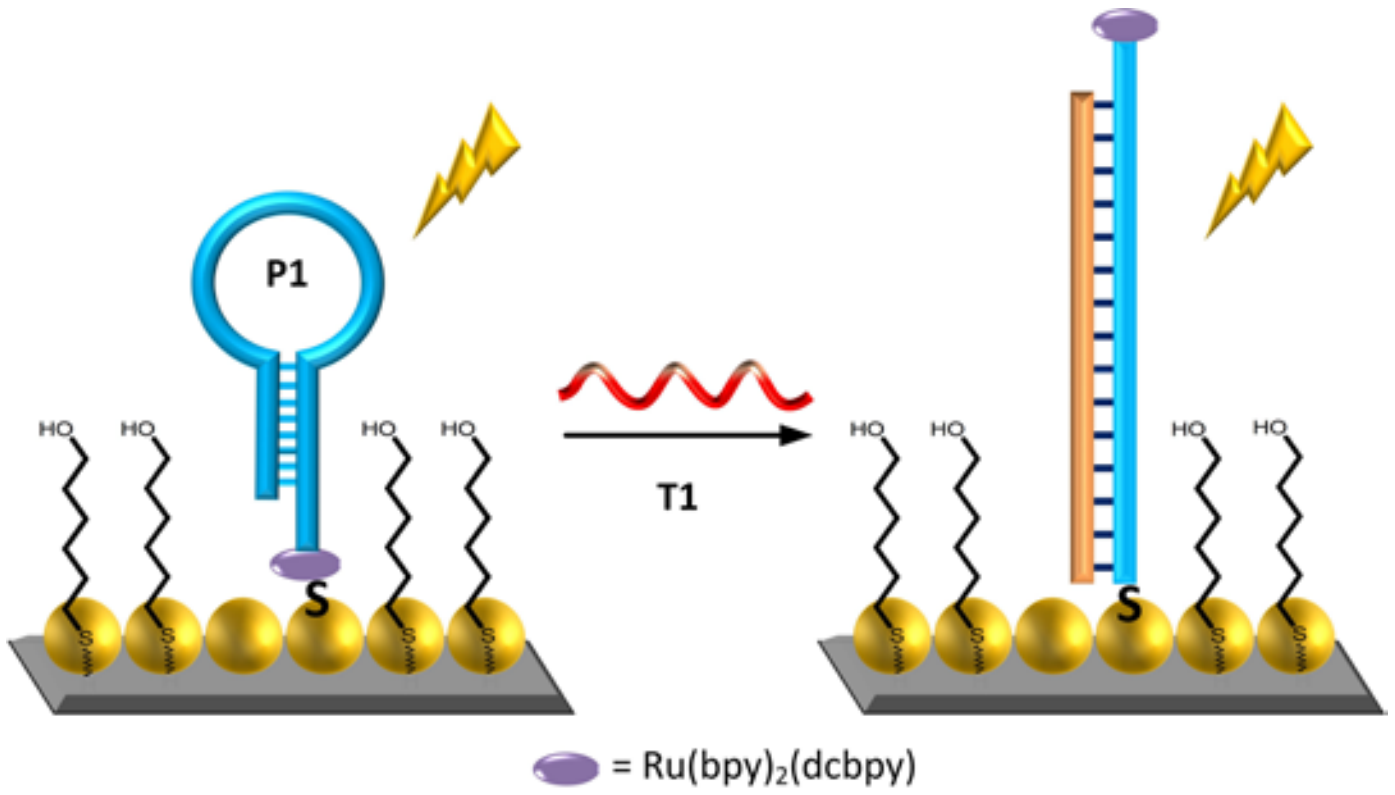

*Figure 24. Schematic diagram of the photoelectrochemical biosensor based on the conformation change of DNA* [261].

In a sense, Zhang adopted the opposite strategy in 2016 [262], where the photocurrent modifying agent is captured in the solution. Figure 25 shows the architecture: the ITO electrode is modified with CdS QDs, where molecular beacon probes are anchored. These probes have two segments that can hybridize with the target DNA sequence and with the DNA label that encapsulates Ag nanoclusters. After the molecular beacon probe is unfolded by the target DNA, the DNA-encapsulated Au nanoclusters are brought into close proximity to the CdS QD electrode surface. This results in a photocurrent quenching of CdS that originates from an energy-transfer process that originated from Ag nanoclusters. Thus, photoelectrochemical DNA bioanalysis is performed by monitoring the attenuation in the photocurrent signal.

Cao *et al.* [263] proposed a photoelectrochemical biosensor for the direct detection of micro-rybonucleic acid (miRNA), which is based on single-walled carbon nanotubes (SWCNTs-COOH) sensitized by DNA-CdS QDs acting as active species. A higher photocurrent is recorded in the absence of target miRNA as a result of the proximity of the CdS QDs and ITO/SWCNTs-COOH electrode. The hybridization chain reaction (HCR) of ssDNA with target miRNA results in the release of DNA-CdS into solution. DNA undergoes enzymatic hydrolysis (cleavage), while reducing the amount of CdS close to the electrode results in a decrease in photocurrent. In this case, detection is related to an irreversible process, which would make its application in reservoir systems difficult.

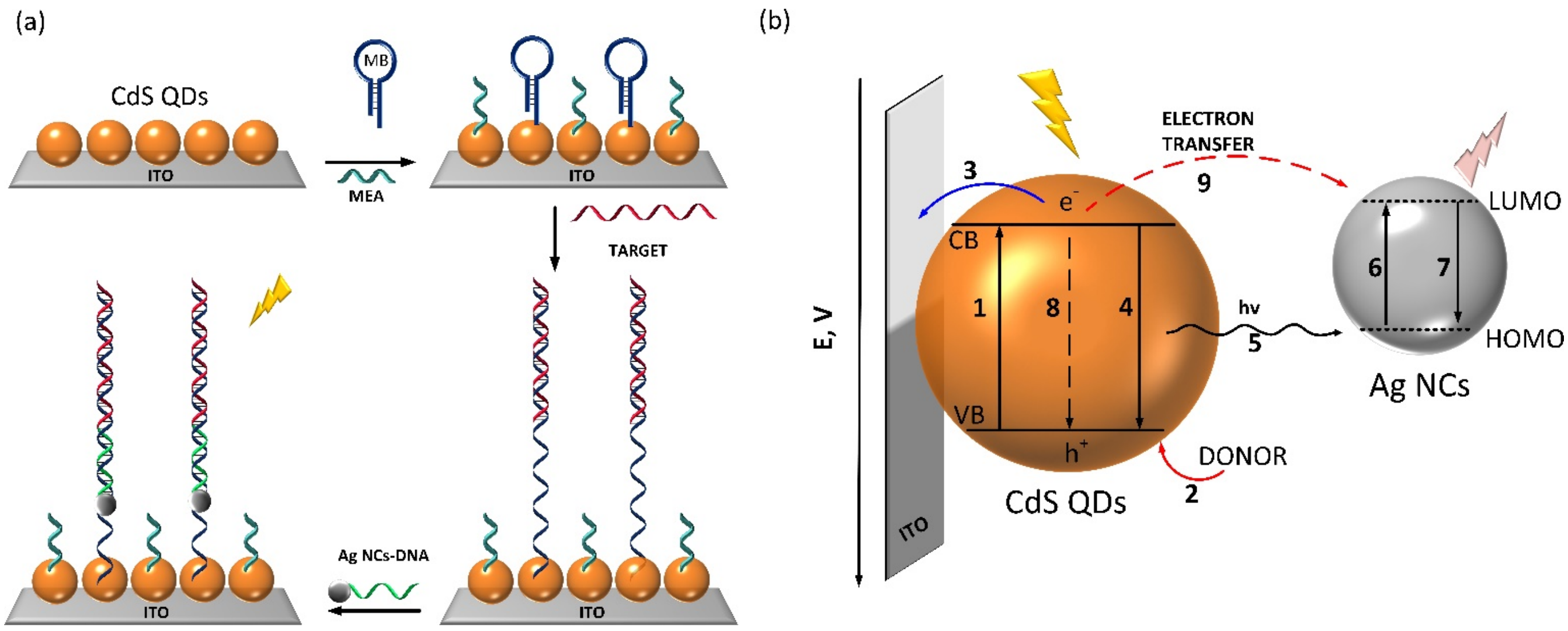

*Figure 25. (a) Schematic mechanism of photoelectrochemical DNA bioanalysis using the molecular beacon probe and Ag nanoclusters and (b) Illustration of the interparticle interaction between CdS QDs and Ag NPs (Process 1, photon absorption and electron transfer from the VB to the CB; Process 2, hole neutralization by electron donor; Process 3, electron ejection to the electrode for photocurrent generation; Process 4, radiative electron-hole recombination; Process 5, spontaneous emission originating from radiative decay; Process 6, electron transfer from the HOMO to the LUMO in AgNCs; Process 7, radiative electron-hole recombination of AgNCs; Process 8, non-Radiative electron-hole recombination in CdSQDs; Process 9, energy transfer(ET) from CdSQDs to AgNCs). Based on Ref.* [262].

Gao *et al.* [264] proposed an organic photoelectrochemical transistor for miRNA detection. Such a transistor uses photovoltage generated by the semiconductor to modulate the channel conductance. The gate is composed of Ti/$TiO_2$/CdS QDs/DNA, while the target miRNA opens the hairpin DNA that initiates HCR. This as-formed double-stranded DNA forms a non-conducting layer that modifies the gate capacitance and increases the potential drop over the gate. This drives the transfer curve to a higher gate voltage and leads to a change in drain current.

HCR amplification was also exploited in the work of Li et al. [265] reporting a DNA detection system employing the $[(ppy)_2Ir(dppz)]^+$ complex as an intercalated indicator. The sensing system consists of ITO/Au NPs/DNA, where a weak interaction between the Ir(III) complex and the short single-stranded DNA causes a low background photoelectrochemical signal. Single-stranded DNA undergoes HCR with the target DNA, resulting in the formation of long double-stranded DNA that permits intercalation of the Ir(III) complex that amplifies the photoelectrochemical signal.

Another approach for photoelectrochemical sensing is the in situ generation of the photosensitizer proposed by Barroso et al. [266]. In this work, CdS NPs are generated in situ in reaction with the analyte. Their amount is quantified by the graphite electrode modified with the Os complex, which transfers electrons from photocatalytic oxidation of thioglycerol driven by UV illumination and applied potential. A similar approach was reported by Zeng et al. [267], who used modified graphene oxide in ITO for the detection of carcinoembryonic antigen, or by Li et al. [268] who used $TiO_2$ on the ITO electrode and Cd functionalized titanium phosphate NPs for the detection of estradiol.

## 6.2 The constitution/generation of an electron donor/acceptor

The constitution or generation of an electron donor or acceptor present in the electrolyte is another type of reactant-determinant mechanism of photoelectrochemical sensing. The absorption of photons by semiconductor is followed by a redox reaction, i.e. electrons from the CB can reduce some agent in the electrolyte or, conversely, holes from VB can oxidize it. Tanne et al. [269] reported a photoelectrochemical sensor for glucose based on an Au / CdSe-ZnS electrode. Light-triggered oxygen reduction is observed at negative potentials, yielding a stable photocurrent. The glucose oxidase reaction consumes oxygen, leading to suppression of the signal of this oxygen-selective electrode. A similar reaction was utilized by Xu et al. [270] for the detection of alpha-fetoprotein using the ZnO photoelectrochemical electrode. Shu et al. [271] reported a photoelectrochemical immunoassay protocol for the detection of low abundance proteins. They utilized a porphyrin-sensitized $TiO_2$ electrode, where the photocurrent is amplified through enzymatic catalysis that accompanies the production of $H_2O_2$. *In situ* generated $H_2O_2$ acts as an electron donor, which is oxidized at low potential to assist signal amplification.

Zhao et al. developed photoelectrochemical immunoanalysis for prostate-specific antigen based on the in situ generation of electron donor ascorbic acid [272]. Figure 22 shows that the coupling of CdS QDs and $TiO_2$ NTs results in an enhanced photon-to-electron conversion efficiency, while Figure 26 shows schematic mechanism of photocurrent generation. Alkaline phosphatase is used for catalytic generation of AA for electron donation. The detection of a photoelectrochemical signal is due to the dependence of the photocurrent on the concentration of the electron donor. Ascorbic acid was also used as an electron donor by Sun *et al.* [273] for the detection of the avian leucosis virus using the ITO / $Bi_2S_3$ nanorod electrode and by Yin *et al.* [274] for the detection of protein kinase based on modified g-$C_3N_4$-Au NP. In recent work, Yang *et al.* [275] proposed a photoelectrochemical aptasensor for thrombin detection based on $C_{60}$@$C_3N_4$ quencher and modified Au NPs.

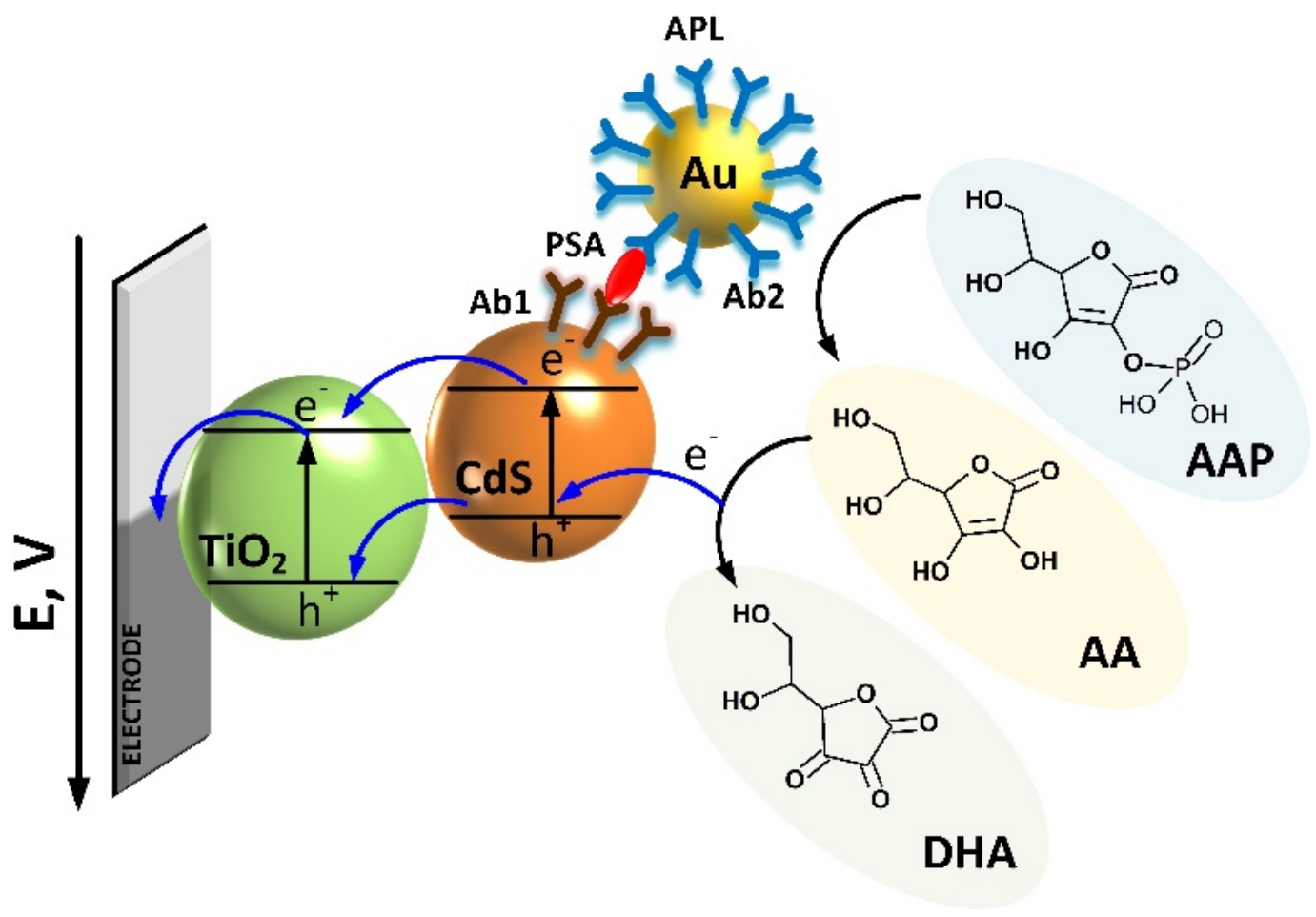

*Figure 26. Schematic mechanism of photocurrent generation involving as an electron donor the ascorbic acid (AA) obtain during the catalytic reaction from alkaline phosphatase (AAP)* [272].

Another approach was used by Zang *et al.* [276] who exploited the in situ generation of oxygen, serving as an electron acceptor, in the catalytic reaction of hemin toward $H_2O_2$. The photoelectrochemical platform is near-infrared CdTe QDs, decorated with a triple-helix molecular beacon on an ITO electrode. In this case, the electron acceptor of $O_2$ leads to an enhancement of the system's photocurrent response.

### 6.3 The effect of steric hindrance

The effect of steric hinderance is induced by molecular recognition reactions because electron donors/acceptors effectively suppress the diffusion of photoactive species surfaces, which is a common strategy in photoelectrochemical (PEC) immunosensor. The molecular recognition reaction is found in three main types [138, 142]. The most common type is molecular recognition within the DNA-analyte, here the PEC biosensor uses a visible light source to probe TATA sequence-binding protein in DNA-protein interactions (TATA is a transcription process that produces an RNA molecule from a DNA sequence) [277]. A strong DNA distortion appears when the DNA / protein complex has formed, preventing the diffusion of AA to the electrode surface, therefore attenuating the photocurrent signal [138, 277]. In addition, the development of many other PEC methods for the cytosensor and aptasensor were based on various recognition interactions [278]. The second type of molecular recognition reaction is the biotin-avidin interaction. The photocurrent response is decreased by the formation of an immune complex that leads to a blocked diffusion of oxidative quencher to the biotinylated ruthenium film, affecting the anti-cholera toxin antibodies [279]. The third type of molecular recognition reaction is the antibody-antigen interaction. For example, a PEC immunosensor based on a multiple hybrid CdSexTe1-x/$TiO_2$ ($0 < x \leq 1$) NTs-based could detect phenol using the antibody-antigen interaction; where the immunoreaction between the antibody and pentachlorophenol modified electrode greatly increases the quencher molecules diffusion around the electrode interface that causes a decrease in photocurrent [279]. Similarly, CdSe/anatase $TiO_2$ and Fe-$TiO_2$ sensitized Fe-$TiO_2$ modified ITO electrodes were used for the detection of Ochratoxin A and squamous cell carcinoma antigen, respectively [280, 281].

In addition, the formation of an insoluble product by using an enzymatic reaction is another strategy for signaling. For example, using peroxidase-induced biocatalytic precipitation of root vegetable, such as horseradish, to detect protein [223] Horseradish peroxide introduced leads to the oxidation of 4-chloro-1-naphthol by $H_2O_2$ to produce insoluble benzo-4-chlorohexadienone on the electrode surface, inhibiting AA diffusion and lowering the photocurrent [138, 282]. On the other hand, researchers show that the creation of a plasmonic nanohybrid-based PEC biosensor is a sensitive surveillance of polynucleotide kinase activity that depends on amplification of catalytic precipitation [223]. Furthermore, researchers developed immobilization-free biosensor that has a high detection of early cancer biomarker, depending on precipitation of mediated biocatalytic using enzyme-free cascaded quadratic amplification [283].

Due to the limitation of the metal-organic framework (MOF) shell of the 8 aperture of the zeolite imidazolate framework (ZIF-8) metal-organic framework (MOF) shell, the selectivity of the PEC sensor was built using semiconductor@MOF heterostructures. Where the template in the core-shell in ZnO@ZIF-8 NRs is made from ZnO NRs and ZIF-8 is formed from the $Zn^{2+}$ ions, as shown in Figure 27. The photocurrent responses depend on the different molecule sizes, so the addition of $H_2O_2$ or AA will give the opposite photocurrent response to the ZnO@ZIF-8 NR array response d. This PEC platform was successfully used to measure $H_2O_2$ in the serous buffer solution. In other words, a PEC sensor with molecule size can be achieved using semiconductor@MOF heterostructures [284].

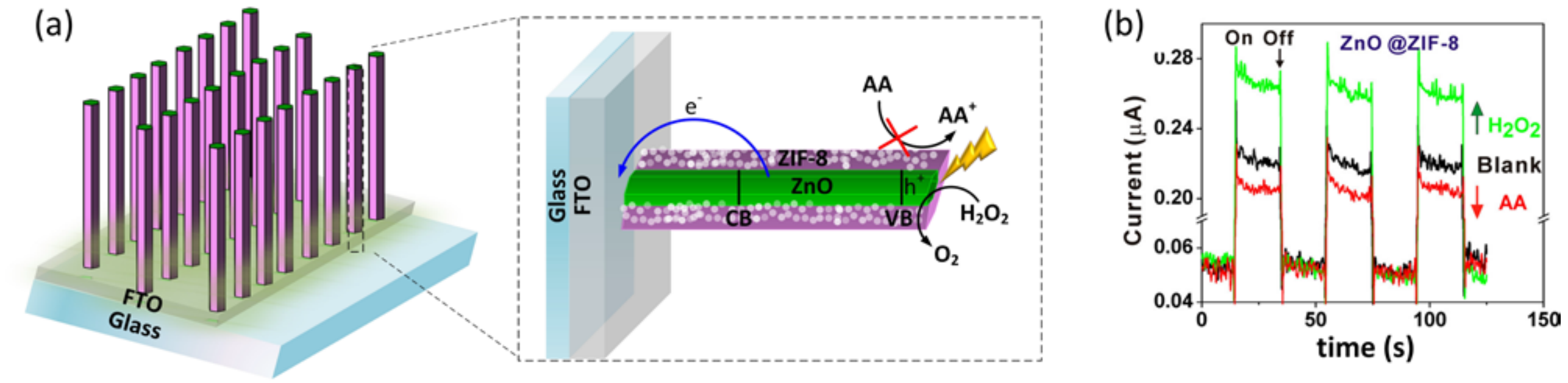

*Figure 27 Schematic diagram of the photoelectrochemical sensor (a) and distinct photocurrent response to $H_2O_2$ and AA (b). Partially reprinted with permission from Ref. [284]. Copyright 2013 American Chemical Society.*

The most up-to-date collection of reactant-determined photocurrent photoelectrochemical sensors is collated in Table 3. [223, 261-272, 274-276, 280-282, 285, 286]

## 7. Photoelectrochemical sensors in reservoir computing systems: Towards the future of sensing

As presented in the preceding sections, photoelectrochemical sensors are based on a plethora of various chemical, photochemical, and photophysical processes occurring upon illumination of appropriately modified semiconductor electrodes. Their common feature is the variation of photocurrent density as a function of a specific reagent, an analytical target. Is it enough to apply the principles of photoelectrochemical sensing by applying the principles of RC physically – integrating the sensing structure into the RC circuitry? In principle, yes, each photoelectrode can become part of an RC system, however, not each case will result in a ‘significant’ improvement of sensor performance.

Here we can envision two different architectures for reservoir computer integrated with the sensor: (i) sensor acts as a part of the reservoir, i.e. implementation of SWEET algorhitm or its variation [18] or (ii) sensor is implemented as an external source of signal then further passed to reservoir computing system. Flow diagrams for both approaches are presented in Figure 28. In each case the simplest echo state machine configuration has been used. These two approaches can be called for simplicity (i) direct and (ii) indirect reservoir sensing, respectively. Both architectures may offer significant improvements of sensor performance, however on different pathways and with different properties of the system being affected. The first, direct approach is definitely more difficult to achieve, but may also provide much much better improvement of the sensing performance. In this scenario the sensing element is an inseparable component of the reservoir. That way it becomes a fragment of much bigger dynamic system, leading to dynamic properties change during operation of the device. This feedback may be reflected in the improved sensitivity and selectivity of the device. Especially in the photoelectrochemical setup (cf. Figure 29) the output of the sensor dynamically controls stimulating light intensity, which leads to higher dimensionality of the phase space, and from there straight to better performance.

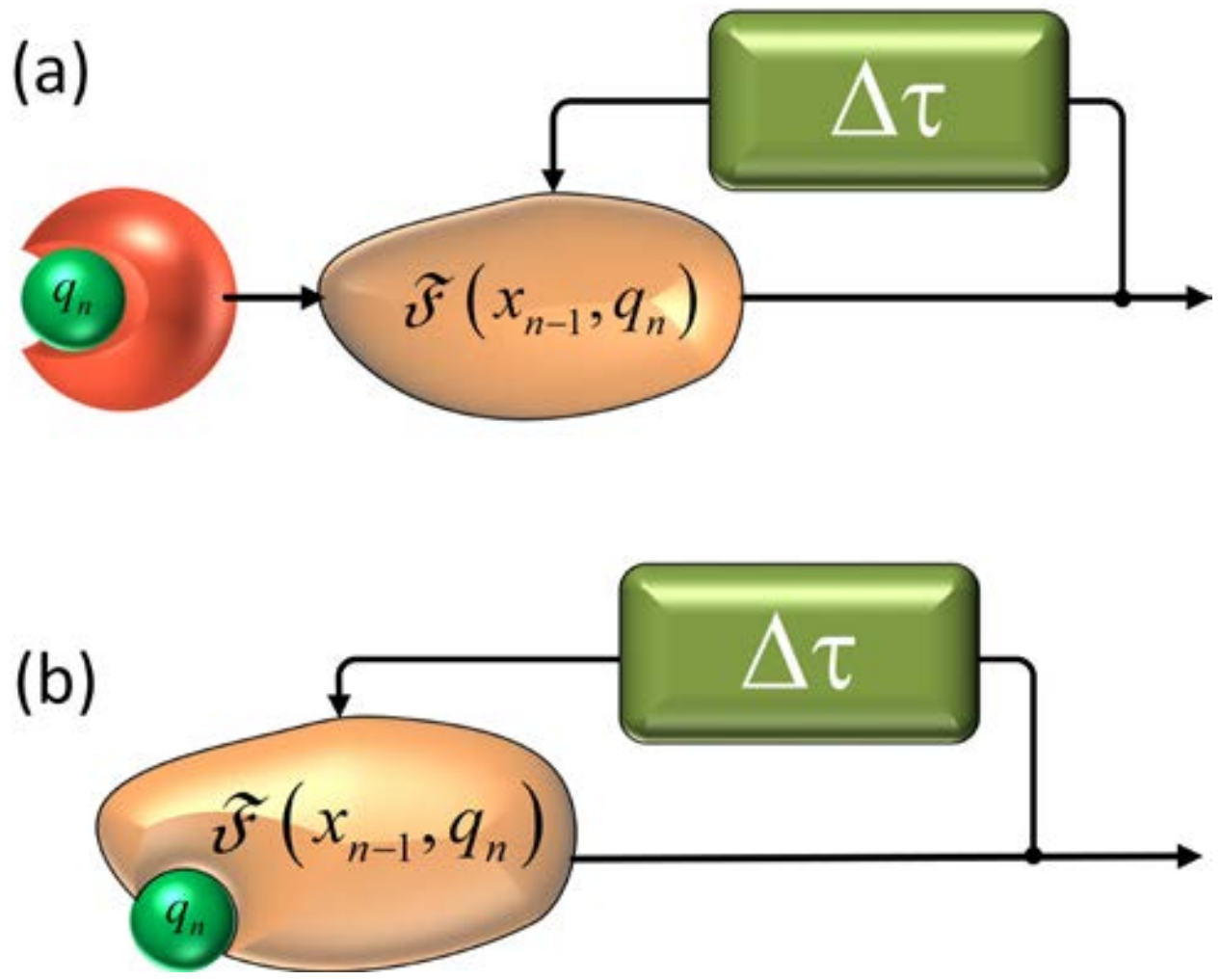

Figure 28. Two different architectures of reservoir computing for sensing: a sensor being a signal source for a reservoir (a) and a sensing reservoir (b).

The second approach, in turn, does not modify properties of the sensing element during operation of the sensor. Sensor response may be used for amplification of the analytical signal, noise reduction or preliminary decission making (e.g. input signal amplitude discrimination) [69, 124]. Such performances have been already demonstrated for similar feedback loop systems [33, 34].

The presented two architectures either contribute to tuning the properties of the sensor itself (direct reservoir sensing) or introduce separate auxilliary systems to sensing elements (indirect reservoir sensing). In the latter case numerous sensors (not necessarity photoelectrochemical) can be directly coupled with a single reservoir system and work in parallel, constituting even more complex computation system, fully capable of highly advanced decision-making performance [122, 287]. One of the most important aspects of this approach is sensory fusion and plurifunctionality, bot of which are very important in antonomous robotics. Management of data from various types of sensors that constitute robotic sensory systems, including e-skin (sense of touch), cameras and light detectors (sense of vision), microphones (sense of audition), torque and displacement (sense of relative motion, force, equilibrium), and other, task-orientated sensors, provides a huge stream of data that needs to be analysed in real time. The lack of synchronisation of events in the environment, various time characteristics of the stimuli, and the presence of disturbing, random stimuli makes navigation and decision-making even more complex, and hence energy consuming.

Reservoir computing offers a radically new vision of sensory information processing, sensory integration, and decision-making processes. Such way of development can be considered as an inspiration for prospective advancement of autonomous robotics. Here the direct inspiration can be taken from the world of insects. Due to their small dimensions and simplicity of their neural systems (at least when compared with higher organisms), they, as taken, are a perfect example of extremely efficient organisms. They easily navigate in 3D space, perform complex foraging tasks, efficiently look for food and partners, avoid predators, and possess social behaviour and learning abilities [288]. All these functionalities are realised in nervous systems consisting of ca. a hundred thousand neurons. This example perfectly illustrates the prevalence of neuromorphic systems versus digital computing implemented in von Neumann architectures. The energetic effectiveness (and therefore reduced carbon footprint) of neuromimetic computing is of key importance. Therefore, we claim that moving computation used for processing of sensory signal, proprioception, decision making and navigation to the simple dynamic memfractive systems will create novel energy-efficient technological platform for future generation of autonomous robots. Detailed discussion of these novel aspects of sensory information processing is, however out of the scope of current review. As the direct sensing approach is a much more challenging architecture to develop, it will be discussed more in the following section.

The basic implementation that should be considered as a platform for the implementation of an RC-based photoelectrochemical sensor is the single-node echo state machine with a drive signal (Figure 1c). It seems natural to use a light beam as a drive signal: in RC systems, drive adds extra complexity to the dynamics of the reservoir (or provides such dynamics) [22, 77] and in photoelectrochemical devices (not only sensors) light is a main source of energy [289]. This analogy suffices for successful application of RC principles in photoelectrochemical sensing.

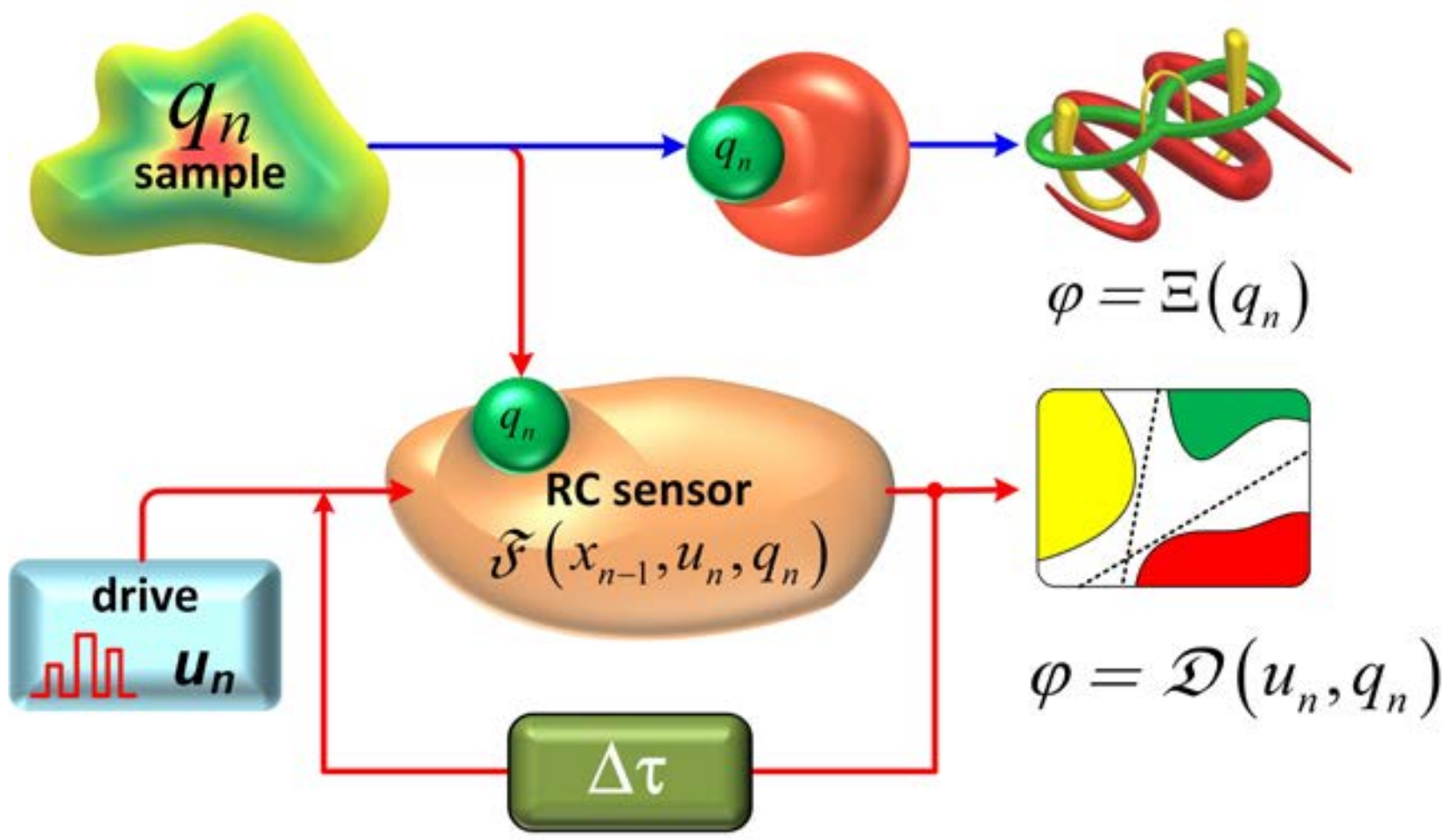

*Figure 29. Overview of the classical and RC-based sensor setups. The environment-sensitive reservoir is used as a sensing element to detect and quantify the analyte. The indirect sensing protocol is based on an appropriately selected drive signal $u_n$: in the case of photoelectrochemical sensor it is the initial sequence of light pulses. The delayed feedback Δτ sustains the dynamics of the system and increases the complexity of the configuration space, providing better performance of the RC system. On the contrary, the classical sensor directly samples the sensing target and gives a complex signal (i.e. analytical signal from the matrix) due to its low computational power [138].*

Let us consider a sample based on a complex matrix, which can be a source of numerous interferences [290]. One feasible approach will be a sensor matrix, which, despite the cross-sensitivity of individual sensing elements, can produce a reliable result upon application of a complex post-processing algorithm, eg, principal factor analysis or other chemometric methods.[291, 292] Then, either the sensor gives highly complex output $\Xi(q_n)$, containing signals from all disturbing species and species, or RC-based filtration of information is required, which will yield desired output $\mathcal{D}(u_n, q_n)$ as depicted in Figure 29. Other problems may be degradation of the sensor surface, e.g. by fouling or deposition of foreign material in the case of nondisposable devices. In numerous cases, with some properties of the electrode conserved, RC may render them usable even in the case of partial degradation without the need for replacement or recalibration.

The only cost visible at first glance is the time required for analysis (the cost of extra hardware is obvious and is not included in the analysis). In the case of classical sensors, the time required for analysis is related to the equilibration of the sensor with the analyte and the maxtrix. In RC sensors, however, this time can easily be an order of magnitude longer. This is the time (and also extra energy) associated with signal processing on the feedback loop (Figures 1c, 29 and 30).

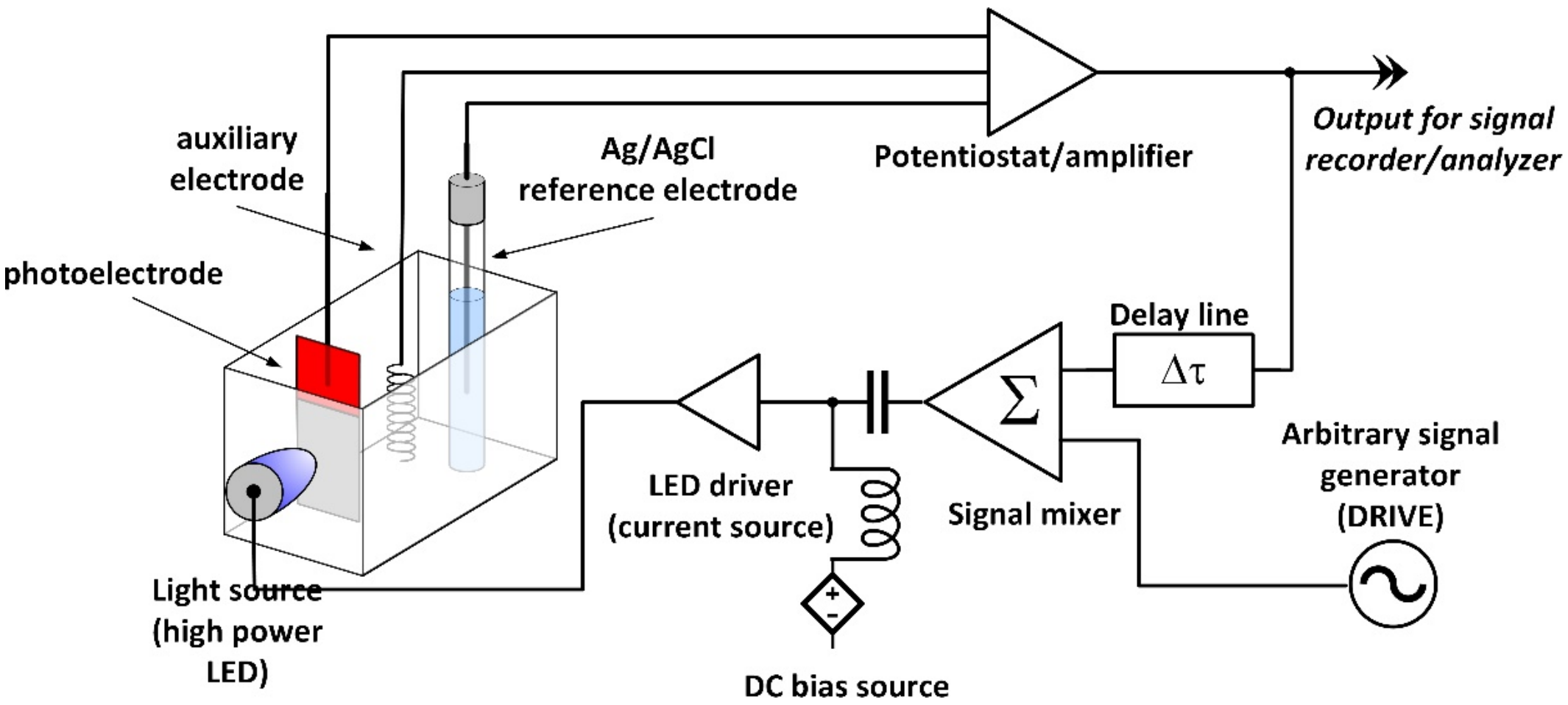

*Figure 30. Simplified circuitry of the RC-based sensing system. The light source is powered either from the initial drive generator or is controlled by the photocurrent via the delay line. The DC bias source along with the bias tee provides light intensity modulation. Adapted from Ref. [138].*

In the most obvious approach, the photoelectrochemical reservoir sensor is based on a light source (powered by a voltage-controlled current source), a potentiostat, a delay line, a mixed drive generator, and a signal (Figure 30). Some auxiliary elements, *e.g.* a bias tee, may be also required for proper operation of the system and tuning/modification of its performance. The system should operate as follows: A drive signal (sine wave, single pulse, train of pulses, etc.) is applied to the summing amplifier, which drives the light source. The photocurrent is generated in photoelectrochemical cells and recorded by a potentiostat (or current-voltage converter). This signal, in turn, is delayed and feeds back into the summing amplifier. This sequence repeats until the stable condition is achieved, (i) the signal is fully attenuated, and the system enters the initial silent mode, or (ii) the constant-amplitude persistent oscillation is achieved. During the process, the whole evolution of the signal should be recorded and subjected to (in most cases simple) postprocessing. Each epoch (the echo of the first pulse from the drive generator) can be regarded as an individual input to the perceptron or other (software-based) data processor. It may seem that such an approach does not lead to an improvement in sensor performance and leads to noise accumulation. Preliminary studies of the performance of such systems in the case of electrochemical sensors indicate the constant noise level throughout the experiment.[20] Moreover, this approach transforms the analytical signal into different domains, for example from voltage/current amplitudes into the time domain, which can be easily processed by the Fourier transform.

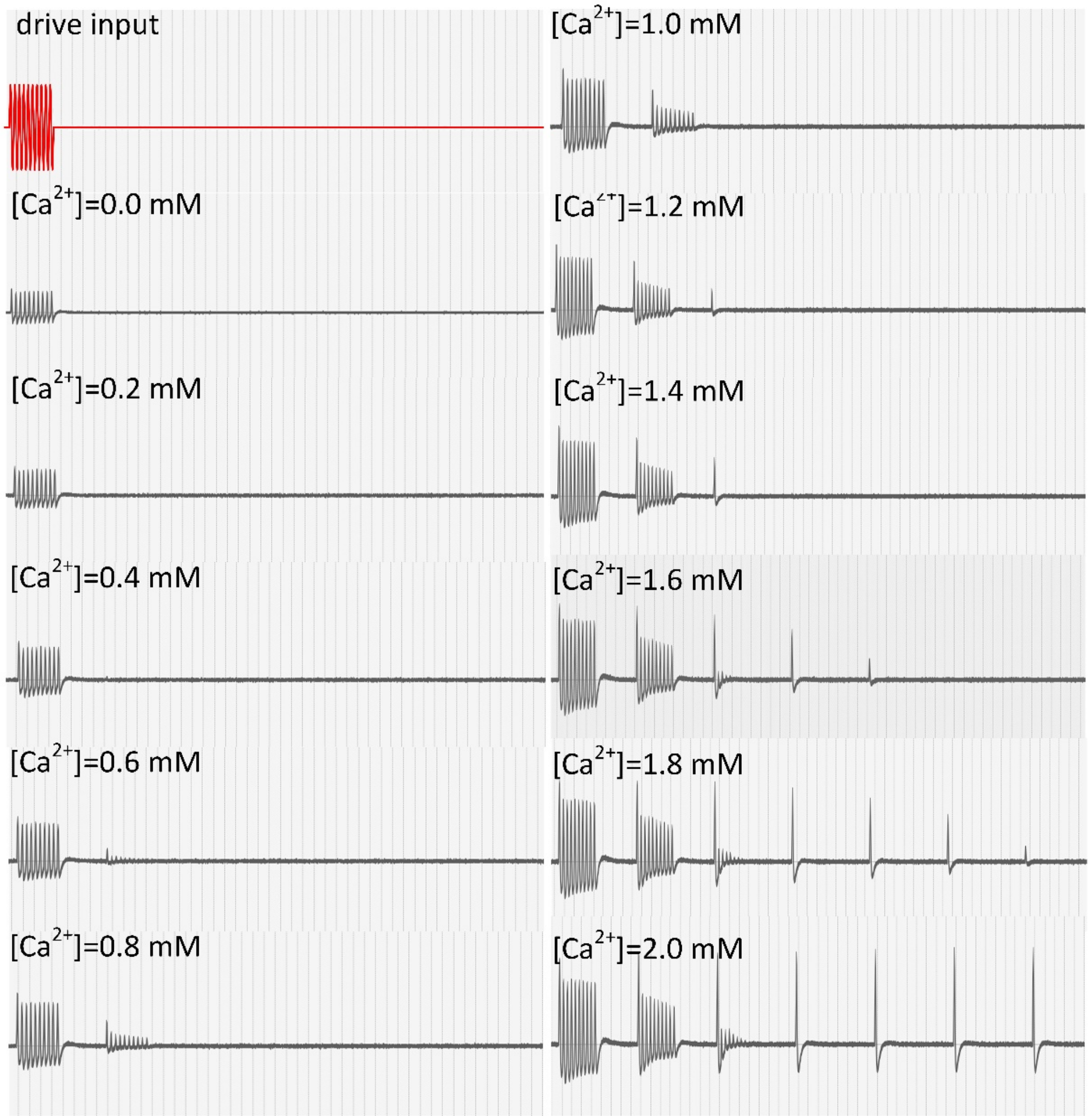

*Figure 31. The output signal (time evolution of the initial drive) recorded from the reservoir system depicted in Figure 30 for various calcium concentrations.[293]*

So far, a single report on the integration of the RC paradigm with a photoelectrochemical sensor.[169] Alizarine complexone-modified CdS, deposited by drop casting on ITO substrate, was used as a calcium ion sensor. Analysis was carried out twice: (i) determination of calcium based on the amplitude of the photocurrent and (ii) on the basis of the Fourier transform intensity of the reservoir output (Figure 31) [293]. It was indicated that application of the device RC approach increased the sensitivity by ca. an order of magnitude (Figure 32).

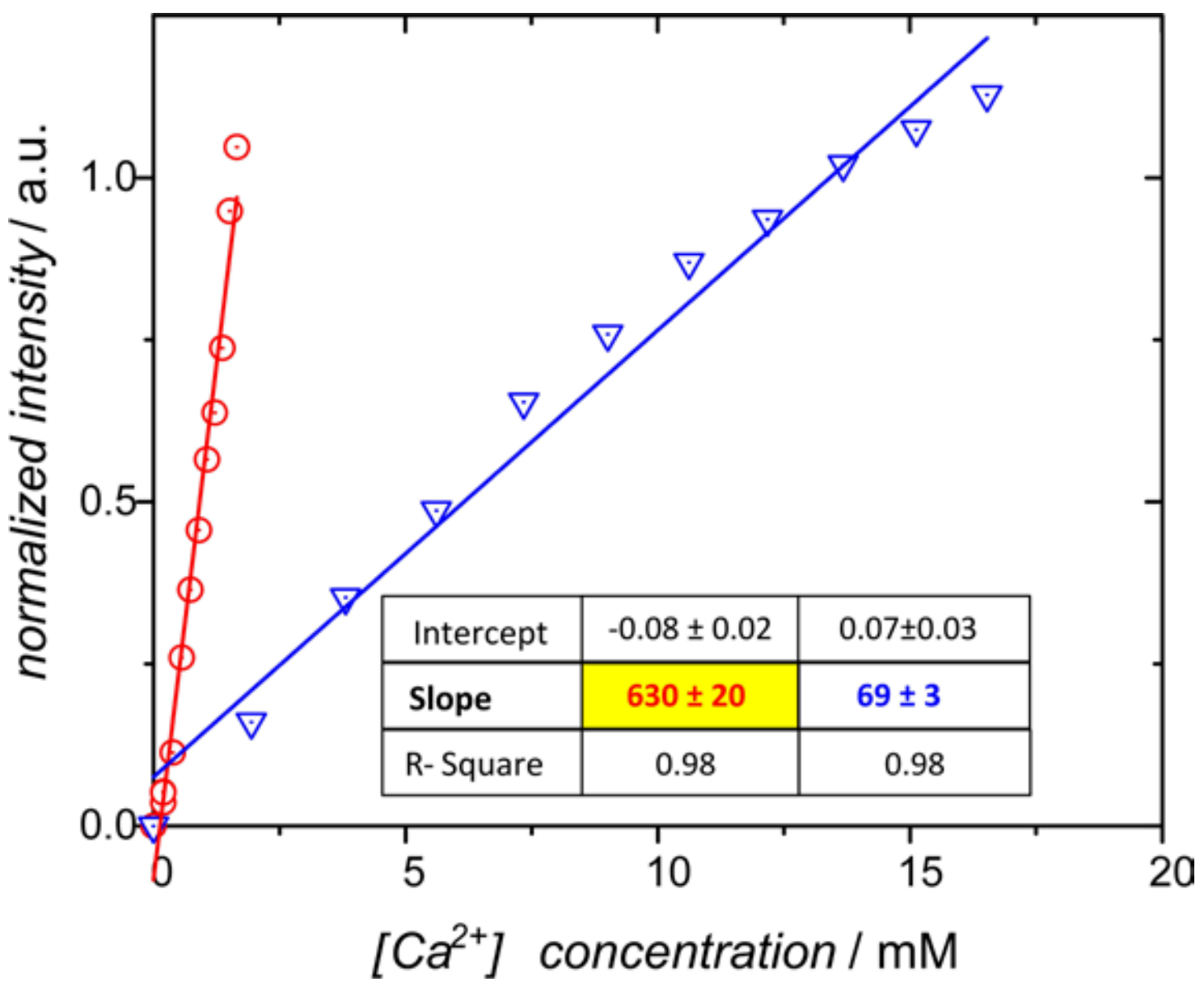

*Figure 32. Calibration curve for the same alizarine complexone-CdS electrode in the classical (blue) and RC (red) photoelectrochemical sensing system [293].*

Among all the photoelectrochemical sensing approaches presented in this review, most of them seem to be compatible with the RC approach. The only severe limitation can be the encountered in systems that consume the analyte in irreversible processes. Because of the much longer illumination time, a significant depletion of the analyte may occur, which will make analysis unreliable. On the other hand, if the active surface of the electrode is small and the volume is significantly large, the depletion of the sample in the reactive analyte will be insignificant.

Anyway, photoelectronic sensing systems in which the analyte plays the role of an electron transfer mediator, or modulated the efficiency of the photosensitized (e.g. via plasmon resonance) are the most suitable for integration with reservoir computing circuitry. They provide stable photocurrents, which can be easily used for modulation of the light source intensity. Other systems in which analyte, along with modulation of the photocurrent intensity, changes the recombination pathways (e.g. systems based on Z-scheme semiconductor assembles) may also benefit from integration with RC due to complex profiles of photocurrent intensity under pulsed illumination. In this case, the Fourier or Laplace transform should be used as post-processing routine in order to extract as much information as possible.

Along with the improvement of sensor performance, the integration of the RC principles into photoelectrochemical sensing systems may result in a new computational platform. In numerous cases, the known edge of the exact concentration of the analyte is not as important as the incorporation of the concentration range – this applies to quality control and preliminary medical diagnostics. RC can easily preprocess the analytical signal and yield clear GO/ NO GO messages valid in a predefined context, whereas the detailed data may arrive upon more precised signal analysis, even from the same measurement [294, 295]. It is important to mention here that in the best cases all this processing will be software-free.

## 8. Concluding remarks

Reservoir computing is an exotic computing paradigm, difficult to understand and implement, yet very powerful. The same concerns photoelectrochemical sensing. It is a niche among all sensing techniques that are rarely visited, mostly due to complex experimental requirements. It is also theoretically demanding that the development of a good photoelectrochemical sensing system, along with excellent knowledge of spectroscopy and electrochemistry, requires excellent fluency in photophysics and semiconductor nanophysics/nanochemistry. The combination of two exotic and difficult fields: reservoir computing and photoelectrochemical sensing is an overwhelming challenge, but the results, including excellent sensor performance and the pure beauty of the system computed using photophysical and electrochemical processes, compensate for all the efforts.

## Author information

**Corresponding Author**

* szacilow@agh.edu.pl, AGH University of Science and Technology, Academic Centre for Materials and Technology, al. Mickiewicza 30, 30-059 Kraków.

**Authors' Contributions**

The manuscript was written through contributions of all authors; they are listed in an alphabetical order. All authors have approved the final version of the manuscript. ‡These authors contributed equally to the manuscript.

**Acknowledgements**

Authors thank Dr. Dawid Przyczyna, Dr. Kacper Pilarczyk, Mr. Andrzej Blachecki and Mr. Vladimir Gorokh for numerous discussions and contribution in the experimental work on reservoir computing and photoelectrochemical sensors. Authors are indebted to Professor Zoran Konkoli for his guidance into the exotic land of reservoir computing and Professor Kapela Pilaka for numerous general comments and encouragement. Authors acknowledge the financial support from the Polish National Science Center within the OPUS programme (grant agreement No. UMO-2020/37/B/ST5/00663). This research was partly supported by program “Excellence initiative–research university” for the AGH University of Science and Technology and by JSPS Core-to-Core programme “Material intelligence: Exploiting intrinsic learning and optimization capabilities for intelligence system“.

**Abbreviations**

2D - two-dimensional

3D - three-dimensional

AA - ascorbic acid

Ab - antibody

ATO - anodic titanium oxide

bpy - 2 -2'-bipyridine

BSA - Bovine Serum Albumin

CB - conduction band

CEA - carci-noembryonic antigen

CL - chemiluminescence

DA - dopamine

dcbpy - 4 -4'-dicarboxy-2 -2'-bipyridine

DNA - deoxyribonucleic acid

dppz - dipyrido [3 -2-a:2′ -3′-c] phenazine

E2 - estradiol

FeTPPS - [meso-tetrakis(4-sulfonatophenyl)porphyrin] iron(III) monochloride

FRET - Förster resonance energy transfer

FTO - fluorine doped tin oxide

g-$C_3N_4$ - graphitic carbon nitrides

GCE - glassy carbon electrode

GSH - glutathione

GSSG - glutathione disulfide

HCR - hybridization chain reaction

HEPES - 4-(2-Hydroxyethyl)piperazine-1-ethanesulfonic acid

ITO - indium tin oxide

miRNA - micro ribonucleic acid

MOF - metal-organic framework

MXenes - a class of two-dimensional inorganic materials

NP - nanoparticle

NR - nanorod

NT - nanotube

OCP - open circuit potential

P3HT - poly(3-hexylthiophene)

PBS - phosphate-buffered saline

Pdots - polymeric dots

PEDOT:PSS - poly(3 -4-ethylenedioxythiophene) polystyrene sulfonate

PIRET - plasmon-induced resonance energy transfer

polyDA - polydopamine

ppy - 2-phenylpyridine

PSA - prostate-specific antigen

QD - quantum dot

RC - resevoir computing

RET - resonance energy transfer

RNA - ribonucleic acid

sCD146 - soluble cell adhesion molecule 146

SCE - standard calomel electrode

SPR - surface plasmon resonance

STA - streptavidin

SWCNT - single-walled carbon nanotube

TATA - DNA Hogness sequence - 5'-TATAAA-3'

TEA - triethanolamine

UV - ultraviolet

VB - valence band

VIS - visible

ZIF-8 - zeolite imidazolate framework 8

**Table 1. Photoactive materials using in charge transfer-based photoelectrochemical sensing**

| Sensing material | Substrate | Analyte | Environment | Linear detection range | Sensitivity / detection limit | Mechanism | E | Ref. |
|---|---|---|---|---|---|---|---|---|
| g-$C_3N_4$ | FTO | $Cu^{2+}$ | 0.1 M PBS (pH 6.9) | 1 to 100 nM | -/0.38 nM | Analyte as electron acceptor | -0.2 V (vs SCE) | [176] |
| CdS QDs | ITO | $Cu^{2+}$ | 0.1M TEA in PBS (pH 7.0) | 0.02 to 20.0 mM | -/$10^{-2}$ μM. | Analyte as electron acceptor | OCP | [178] |
| $Cu_2O$ | ITO | $Cu^{2+}$ | 0.1 M PBS (pH 7) | $10^{-8}$ to 1 mM | -/3.33 pM | Analyte as electron acceptor | -0.2 V (vs SCE) | [179] |
| $TiO_2$ NTs | Ti | $Pb^{2+}$ | 0.2 M $Na_2S$ | 0.1 to 1 μM | -/0.39 nM | Analyte as electron acceptor | 0 V (vs. Ag/AgCl) | [180] |
| ZnO NRs | ITO | $Cd^{2+}$ | 0.1 M PBS (pH 7.4) | 0.01 mM to 5 mM | -/ 3.3 μM | Analyte as electron acceptor | 0 V (vs SCE) | [182] |
| $TiO_2$ NTs | Ti | $Cd^{2+}$ | $H_2SO_4$ + $SeO_2$ | $10^{-6}$ to 1 mM | -/0.35 nM | Analyte as electron acceptor | 0 V (vs. Ag/AgCl) | [183] |
| CdS QDs | ITO | DA | 0.1M TEA + 0.1 M PBS (pH 9) | 0.02 to 50 μM (electrooxidation); 0.002 to 10 μM (alkaline solution oxidation) | -/8.0 × $10^{-3}$ μM and 2.0 × $10^{-4}$ μM | PolyDA as electron acceptor | - | [177] |
| $TiO_2$ ($TiO_2$NR array) | FTO | Glucose | 1M NaOH | 0.01 to 0.20 mM | 200 μA· mmol$^{-1}$ ·cm$^{-2}$ / 3.2 μM | Analyte as electron donor | -0.65 V (vs SCE) | [184] |
| α-$Fe_2O_3$ (Cubes) | ITO | Glucose | 0.1M PBS (pH = 7.4) | 0.2 to 2 mM | 32.302 μA· $mmol^{-1}$ $\cdot cm^{-2}$ / 0.015 mM | Analyte as electron donor | 0.3 V (vs. SCE) | [185] |
| $Co_3O_4$ | FTO | Glucose | 0.2M NaOH | 0.01 to 0.35 μM | 7753 μA· $mmol^{-1}$ $\cdot cm^{-2}$ / $3.6 \times 10^{-3}$ μM | Analyte as electron donor | 0.6 V (vs Ag/AgCl) | [186] |
| | | | | 0.1 to 0.35 μM | 2467 μA· $mmol^{-1}$ $\cdot cm^{-2}$ / - | | 0.6 V (vs Ag/AgCl) | |
| | | | | 0.2 to 0.35 μM | 873 μA· $mmol^{-1}$ $\cdot cm^{-2}$ / - | | OCP | |
| $BiVO_4$ (electrodeposited) | FTO | Glucose | 0.1M $NaNO_3$ | 0.5 μM to 5 mM | - / 0.13 μM | Analyte as electron donor | 0.15 V (vs Ag/AgCl) | [187] |
| $TiO_2$ATO | Ti | Glucose | 0.1M $KNO_3$ (pH 6.9) | 0 to 119 μM | 7.3 μA· $mmol^{-1}$ $\cdot cm^{-2}$ / 15.6 mM | Analyte as electron donor | 0.2 V (vs. SCE) | [189] |
| | | | | 380 to 665 μM | 7.5 μA· $mmol^{-1}$ $\cdot cm^{-2}$ / 7.8 mM | | 0.5 V (vs. SCE) | |
| | | | | 0 to 79 μM | 237 μA· $mmol^{-1}$ $\cdot cm^{-2}$ / 16.3 mM | | 1.0 V (vs. SCE) | |
| single-layer nano$MoS_2$ | Gold | DA | 0.1 M PBS (pH 7.0) | $1.0 \times 10^{-5}$ to 10 μM | - / $2.3 \times 10^{-6}$ μM | Analyte as electron donor | 0 V (vs. SCE) | [190] |
| $WO_3$ | FTO | DA | | 53 to 80 μM | - / 0.30 μM | | | [191] |
| | | | 0.1 M PBS (pH 7.0) | 85 to 155 µM | - / 0.30 µM | Analyte as electron donor | 0.71 V (vs. Ag/AgCl) | |
| ND-g-CN | ITO | Ciprofloxacin (CIP) | 0.1 M PBS (pH 7.0) | 60 to 19090 ng/L | - / 20 ng/L | Analyte as electron donor | 0 V (vs. Ag/AgCl) | [192] |
| ZnO/CdS | ITO | $Cu^{2+}$ | 0.1M PBS, (pH 7.0) with 0.1 M TEA | 0.02 to 40.0 µM | -/0.01 µM | Analyte as electron acceptor | OCP | [195] |
| $SnO_2$/CdS | FTO | $Cu^{2+}$ | 1 M $Na_2SO_3$ (pH 7.0) | 1.00–38.0 µM | -/0.55 µM | Analyte as electron acceptor | -0.2V | [196] |
| $WO_3$/CdS | FTO | $Cu^{2+}$ | 1 M $Na_2SO_3$ (pH 9.5) | 0.5 µM to 1 mM | -/0.06 µM | Analyte as electron acceptor | 0.8 V | [197] |
| $BiVO_4$/FeOOH | ITO | DA | 0.1M PBS, (pH 7.4) | 0.2–40 µM 40–1400 µM | -/0.09 µM | Analyte as electron acceptor | 0 V | [198] |
| $Cu_2O$/BiOI heterojunction | FTO | $H_2O_2$ | 0.1M PBS, (pH 7.4) | 1.99 µM – 17.54 mM | -/0.44 µM | Analyte as electron acceptor | -0.4 V | [199] |
| $Cu_2O$/ZnO | ITO | GSH | 0.1 M PBS (pH 7.0) | 1 to 10 µM | -/0.8 µM | Analyte as electron donor | 0.2 V (vs SCE) | [200] |
| $BiPO_4$/BiOCl | ITO | 4-chlorophenol | 0.1 M PBS (pH 7.0) | 20 to $3.38\times10^4$ ng/mL | -/6.78 ng/mL | Analyte as electron donor | OCP | [201] |
| P3HT/$TiO_2$ | glassy carbon | chlopyrifos (Organophosphorus Pesticide) | 0.1 M PBS (pH 7.0) | 0.2 to 16 µM | - | Analyte as electron donor | 0 V (vs SCE) | [203] |
| FeTPPS-$TiO_2$ | ITO | GSH | 0.1 M PBS (pH 7.0) | 0.05 to 2.4 mM | - | Analyte as electron donor | 0.2 V (vs SCE) | [204] |
| CdTe(QD) | FTO | $Cu^{2+}$ | $H_3BO_3$-$Na_2B_4O_7$ buffer (pH 6.0) | $8.0 \cdot 10^{-8}$ to $1.0 \cdot 10^{-4}$ M | $5.9 \cdot 10^{-9}$ M | Electron trapping | -0.2 V | [212] |
| porous g-$C_3N_4$/CuS | FTO | $S^{2-}$ | Tris-HCl (pH=7.4) | 50-700 µM | 0.06 µM | Electron trapping | | [213] |
| molecularly imprinted polymer/$TiO_2$@Au @CdS/ITO) | ITO | uric acid | 5 mM PBS (pH 7.0) | 1 nM to 9 µM | 170.0 µM $µA^{-1}$ $cm^{-2}$ /0.3 nM | Z-scheme/ uric acid molecules block light absorption and the capture of photogenerated holes | +0.4 V vs Ag/AgCl | [214] |
| ZnS/$Co_9S_8$ heterojunction | ITO | chlorpyrifos | 0.1 M PBS (pH = 7) | 0.05 ppb to 40 ppb | -/0.0166 ppb | Chrompyrifos block the Z-scheme Electron transfer in the Co9S8 CB of Co9S8 toward the ITO | 0.2 V vs Ag/AgCl | [215] |
| CdS QDs on the $AuNP@BiVO_4$ | FTO | PSA | 0.1 M $Na_2SO_4$ | 0.01-50 ng ml -1 | 9.4 pg ml-1 / 1.5 pg $mL^{-1}$ | PSA results in the formation of Z-scheme CdS QDs/ $AuNP@BiVO_4$ | 0 V vs SCE | [210] |
| $ZnIn_2S_4/WO_3$ | ITO | PSA | 0.10 M BPS (pH 7.4)/0.2M AA | 0.01–500 ng/ml | - /5 pg $ml^{-1}$ | PSA as insulated protein prevents the electron transfer/ Z-scheme | 0V vs SCE | [216] |
| BiOI-CdS | ITO | $Cu^{2+}$ | n 0.1 M $Na_2SO_4$ | 0.1-100 µM | -/0.02 µM | $Cu^{2+}$ act as acceptor in Z-scheme | -0.1 V vs SCE | [209] |
| $TiO_2$/Au/CdS QDs | ITO | $Pb^{2+}$ | 0.1 M phosphate buffer solution (PBS, pH 7.4) containing 0.1 M AA | 0.5 pM–10 nM | nA $pM^{-1}$/0.13 pM | Z-scheme/ $Pb^{2+}$ induced disassembly of the Z-scheme | -0.1 V vs SCE | [211] |
| I-BiOCl/N-GQD (i.e., nitrogen-doped graphene QD) heterojunction | | chlorpyrifos | 0.10 M $Na_2SO_4$ | 0.3−80 ng $mL^{-1}$ | -/0.01 ng $mL^{-1}$ | chlorpyrifos suppress the effective transfer of carriers in Z-scheme | -0.1 V vs SCE | [208] |
| Au NPs/BiOI nanosheets/B-$TiO_2$ NPs (photoactive matrix) + PbS/$Co_3O_4$ (signal label) | ITO | procalcitonin | PBS (pH 7.4) | 0.1 pg/mL to 50 ng/mL | 0.02 pg/mL | Signal off mechanism - antibody label signal adheres to analyte, leading to photocurrent depression | | [217] |
| $Ag_2S$/ITO electrode + $Ti_2C$ MXene | ITO | sCD146 – biomarker of lung cancer | | 0.1–1000 pg/mL | 18 fg/mL | sandwiched immune recognition structure resulting in signal switching for the whole photoelectrochemical - from cathodic to anodic photocurrent | | [218] |
| Au/$CeO_2$ core/shell NPs | Au | $H_2O_2$ | 100 mM PBS pH 7.4 | 4-2000 µM | 3 µM | | varying – from -500mV to +500mV vs. Ag/AgCl | [219] |

**Table2. Photoactive materials used in energy-transfer-based photoelectrochemical sensing.**

| Sensor | Substrate | Analyte | Environment | Linear detection range | Sensitivity/detection limit | Mechanism description | E | Ref. |
|---|---|---|---|---|---|---|---|---|
| Au@$SiO_2$ NPs | | alkaline phosphatase | | 0.04-400 $U\cdot L^{-1}$ | -/0.022 $U\cdot L^{-1}$ | | | [254] |
| Au/ZnO NRs | FTO | GSH | | 20 to 1000 μM | -/3.29 μM | | 0 V vs. Ag/AgCl | [234] |
| Au-NPs | ITO | Hydroquinone | 0.05 M phosphate buffer | 0.25-150 μM, | -/0.1 μM | hot electron injection from plasmonic excitation | 0.15 V vs. SCE | [232] |
| AuNPs/$Bi_4NbO_8Cl$ | ITO | cysteine | 0.1 mol $L^{-1}$ PBS (pH = 7.4) | $1x10^{-4}$ to $5\times10^{-3}$ mol $L^{-1}$ | -/$10^{-5}$ mol $L^{-1}$ | hot electron injection via SPR, diminishing charge trapping, fast charge separation and transfer | 0V (vs Ag/AgCl) | [221] |
| AuNP/g-$C_3N_4$ | glassy carbon | 4-chloro-1-naphthol (to | 0.1 M $Na_2SO_4$ solution | 2 to 100 mU mL–1 | -/1.0 mU $mL^{-1}$ | SPR-enhanced light harvesting | 0.2 V | [223] |
| | | monitor the activity of T4 polynucleotide kinase activity) | containing 0.1 M $Na_2S$ and 0.05 M $Na_2SO_3$ | | | and separation of photogenerated e–/h+ pairs | | |
| Ag/AgBr/BiOBr | ITO | PSA | PBS, 0.1 mM AA | 0.001–50 ngmL1 | -/0.25 $pgmL^{-1}$ | steric hindrance | | [229] |
| Au nanoclusters-Ag@$SiO_2$ Nanocomposites | FTO | Alkaline phosphatase activity assay | | 0.04 to 400 U·L−1 | -/0.022 $U{\cdot}L^{-1}$ | | | [222] |
| $TiO_2$ nanosheets | ITO | $Hg^{2+}$ | PBS with pH 8.0 with 2.0 mg $mL^{-1}$ of AA | 0.01–10 nM | -/2.5 pM | plasmon-induced charge separation local electrochemical field effect amplification | | [231] |
| Au/H-$TiO_2$ | FTO | glucose | 0.1 M $NaNO_3$ | - | - | Different response under UV or VIS light range (Au | +0.6V V (Ag/AgCl) | [227] |
| | | | | | | introduces VIS range absorption). hot hole transfer – oxidation of sugar molecules | | |
| $TiO_2$−Au@GM1 nanowires | FTO | cholera toxin subunit B | PBS (pH 7.4) | 0.167 to 16.7 nM | -/0.167 nM | Attenuation of electromagnetic field upon target binding<br>- | 0 V vs Ag/AgCl | [228] |
| Au nanoislands/$TiO_2$ | - | Steptavidin-modified gold NP (STA-AuNP) | $KClO_4$ (0.1 mol/L) solution | - | - | hot electron transfer | 0.3 V vs SCE | [230] |
| Au/g-$C_3N_4$ composite | ITO | 4-chlorophenol | PBS (0.1 M, pH 7.0) | 0.25 µM to 34.70 µM | -/0.08 µM | - Increased light absorption in Vis range<br>- enhanced charge separation due to SPR generated local electric field | −0.2 V | [233] |
| AuNP-pDNA/telomerase primer sequence/cDNA/ PFBT Pdots | ITO | telomerase activity | - | 100 to 800 and 2000 to 16 000 HeLa cells | -/30HeLa cells | quenching of the exciton states of Pdots due to SPR –generated local electric field from Au NPs | | [235] |
| hemin/G-quadruplex/CdS QDs | Au | single-stranded DNA | 10 mM HEPES buffer, pH = 9.0, which included 20 mM $KNO_3$, 200 mM, $NaNO_3$, 1 µM hemin, 0.1 µM 5, 0.57 mM luminol and 20 mM triethanolamine | 2−100 nM | -/2nM | chemiluminescence RET-generated photocurrents | - | [253] |
| hemin/G-quadruplex-glucose oxidase (GOx) conjugate/CdS QDs | Au | Glucose | 10 mM HEPES buffer, pH = 9.0, which included 20 mM $KNO_3$, 200 mM, $NaNO_3$, 1 | 5−40 mM | -/5 mM | chemiluminescence RET-generated photocurrents | - | [253] |
| | | | μM hemin, 0.1 μM 5, 0.57 mM luminol and 20 Mm triethanolamine | | | | | |
| C-DNA/CdS/ITO | ITO | DNA detection | 0.1 M Tris-HCl buffer of pH 9.0 containing 25 mM $H_2O_2$, 0.6 mM p-iodophenol and 0.6 mM luminol | 5 to 10000 fM | -/2.2 fM | catalytic hairpin assembly (CHA) amplification and porphyrin (FeTMPyP) Mediated Chemiluminescence | 0.2 V vs SCE | [252] |
| $CdS/TiO_2$ hybrid modified Au-PWE | Electrode of CdS / $TiO_2$ hybrid-modified porous | carci-noembryonic antigen (CEA) | - | 0.1 pg $mL^{-1}$– 5 μg $mL^{-1}$ | -/0.065 pg $mL^{-1}$ | CL | - | [249] |
| | Au-paper | | | | | | | |
| QDs alloyed with CdSeTe and $SiO_2$@Au nanocomposites | nitrogen-doped $TiO_2$ NTs ($TiO_2$:N-NT) | thrombin | Tris-HCl buffer (0.1 M, pH 7.4) containing 0.1 M AA | 10 fM to 50 pM | -/2.8 fM | exciton energy transfer | 0 V vs Ag/AgCl | [247] |
| Au NPs /DNA/CdS QDs | ITO | TATA-binding protein (TBP) | 0.10 M PBS solution containing 0.10 M AA with | 0.1 pg/mL to 20 ng/mL | -/0.05 pg/mL | Energy transfer | 0 V vs Ag/AgCl | [246] |
| $WO_3$/Au/ polyDA (PDA) | ITO | human epididymal protein 4 (HE4). | PBS solution (pH 7.4, 0.1 M) containing 0.2 M AA | 0.01 ng/mL to 200 ng/mL | -/1.56 pg/mL | plasmon resonance energy transfer (PRET) effect of | 0.75 V vs SCE | [255] |
| CdS QDs and AuNPs | ITO | DNA | 0.1 M PBS (pH 7.4, MNaCl = 0.1 | 0.005–5 pM | -/2 fM | RET | 0 V vs Ag/AgCl | [256] |
| | | | M) containing 0.08 M AA | | | | | |
| CdTe QDs and reduced graphene oxide-AuNPs nanocomposites | ITO | CEA | Tris-HCl buffered saline (0.1 M, pH 7.4) containing 0.01 M AA | 0.001 to 2.0 ng $mL^{-1}$ | -/0.47 pg $mL^{-1}$ | RET | -0.05 V vs Ag/AgCl | [257] |
| (poly(dimethyldiallylammonium chloride) $/CdS)_5$ multilayer film | ITO | GSH | PBS buffer (10 mM, pH 7.4 and pH 8.2) | 0.1–10 nM | 47 pM | CL emission of isoluminol-HO-$Co^{2+}$ | 0 V vs. Ag/AgCl | [258] |
| $CdS/MoS_2$ heterojunction | ITO | DNA | 0.1 M Tris–HCl buffer of pH 9.0 containing 20 mM H2O2, and 1.0 mM luminol | 0.001–100 pM | -/0.39 fM | enhanced CL excitation of luminol catalyzed by hemin-DNA complex | 0.2 V vs SCE | [259] |
| $TiO_2$ IOPCs–CdS:Mn | ITO | $H_2O_2$ | Triethanolamine (20 mM) and luminol (1 | 0.063 to 4 mM O | 0.35 µA $mM^{-1}cm^{-2}$ /19 µM | Enhance luminol CL RET | 0 V vs Ag/AgCl | [260] |
| | | | mM) in Tris-HCl buffer (0.1 M, pH 8.0) | | | | | |

**Table 3. Photoactive materials using in reactant transfer-based photoelectrochemical sensing.**

| Sensor | Substrate | Analyte | Environment | Linear detection range | Sensitivity/Detection limit | Mechanism | E | Ref. |
|---|---|---|---|---|---|---|---|---|
| $SnO_2$-NP, 3-mercaptopropyl-triethoxysilane, Au-NPs, DNA, $Ru(bpy)_2(dcbpy)^{2+}$) | ITO | DNA (oligonucleotides) | 0.1 M Tris–HCl, 40 mM triethanolamine | 0.1 nM to 8 nM | 0.94 fM | Altering electron-transfer tunnelling | 0.1 V, λ = 430 nm | [261] |
| CdS QDs, MEA/molecular beacon | ITO | Ag nanoclusters-DNA | - | 1 pM to 10 nM | 0.3 pM | Introduction of photoactive species | 0 V, 0.1 M AA, λ = 410 nm | [262] |
| SWCNTs, Ps-po-DNA, CdS QDs | ITO | mi-RNA-7f | 0.1 M PBS, 0.1 M AA | 50 fM to 100 pM | 34 fM | Release of photoactive species | -0.05 V, λ = 405 nm | [263] |
| Gate: $CdS/TiO_2$ NTs-DNA, Channel: PEDOT:PSS | Ti | mi-RNA | 0.1 M PBS, 0.1 M L-cysteine | 1 pM to 10 µM | 1 pM | Modification of transistor gate capitance | 0 V, λ = 425 nm | [264] |
| Au NPs, Cp-DNA, $[(ppy)_2Ir(dppz)]^+$ | ITO | DNA | 0.1 M PBS, 0.1 M triethanolamine | 25 fM to 0.1 nM | 9 fM | Amplification by intercalation indicator | 0.1 V, λ = 380 nm | [265] |
| Os polyvinylpyridine complex/CdS | graphite | anti-BSA antibody | $Na_3O_3PS$, 1-thioglycerol, $MgCl_2$, $Cd(NO_3)_2$ | up to 20 ng/mL | 2 ng/mL | Generation of photoactive species | 0.31 V, λ = 365 nm | [266] |
| graphene oxide/CEA/BSA/CdS | ITO | carcinoembryonic antigen | horseradish peroxide-Ab, 1.8 mM $Na_2S_2O_3$, 2.6 mM $H_2O_2$ | 2.5 ng/mL to 50 µg/mL | 0.72 ng/mL | Generation of photoactive species | -0.2 V, λ = 405 nm | [267] |
| $TiO_2$/BSA-$E_2$/$E_2$/TiP@$Cd^{2+}$-anti-$E_2$ | ITO | estradiol ($E_2$) | 0.1 M PBS, 0.1 M AA | 5 pg/mL to 4 ng/mL | 2 pg/mL | Generation of photoactive species | 0.1 V, λ = 430 nm | [268] |
| CdSe-ZnS QDs/trioctylphosphi | Au | glucose | 100 mM HEPES | 0.1 mM to 5 mM | µM | Consumption of electron acceptor | -0.35 V | [269] |
| ne oxide/1,4-benzenedithiol/glucose oxidase | | | | | | | | |
| ZnO IO/CS/Anti-AFP/AFP-CdS-glucose oxidase | FTO | alpha-fetoprotein | 0.1 M PBS | 0.1 ng/mL to 0.5 μg/mL | 10 pg/mL | Generation of electron donor | 0.6 V, Xe lamp, pH = 7.4 | [270] |
| $TiO_2$-TSPP/CS/Au NPs/anti-CEA/CEA | ITO | low-abundant proteins | 0.1 M PBS | 20 pg/mL to 40 ng/mL | 6 mg/mL | Generation of electron donor | 0 V, pH = 6.5 | [271] |
| $TiO_2$ NTs/CdS QDs immunogold labeled alkaline phosphatase | Ti | PSA | 0.1 M PBS, 0.1 M AA | - | 0.5 ng/mL | Generation of electron donor | 0 V, λ = 410 nm | [272] |
| g-$C_3N_4$-Au NPs/peptide-adenosine 5′-triphosphate/Phos-tag-biotin | ITO | protein kinase | 0.1 M PBS, 0.1 M AA | 0 unit/mL to 100 unit/mL | 0.015 unit/mL | Generation of electron donor | pH = 7.4 | [274] |
| $C_{60}$@$C_3N_4$, Au NPs, perylene tetracarboxylic acid | glassy carbon | thrombin | 0.1 M PBS, 2 mM AA | 10 fM to 10nM | 1.5 fM | Generation of electron donor | 0 V, λ = 365 nm, | [275] |
| CdTe QDs/C-DNA | ITO | DNA | 0.1 M PBS, 1.0 mM $H_2O_2$ | 1 pM to 1 nM | 0.8 pM | Generation of electron acceptor | -0.1 V, λ = 505 nm, pH = 7.4 | [276] |
| $TiO_2(CdSe_xTe_{1-x}/TiO_2)$ ($0 \leq x \leq 1$) | Ti | pentachlorophenol, trichlorophenol | 0.1 M NaOH ,0.1M PBS | 1 nM to 0.3 µM | 1 pM | Effect of steric hindrance | | [285] |
| CdSe/anatase $TiO_2$ | ITO | Ochratoxin A | PBS (pH 7) | 10.0 pg/mL–50.0 ng/mL | 2.0 pg/mL | Effect of steric hindrance | | [280] |
| CdS sensitized Fe-$TiO_2$ | ITO | Ochratoxin A | 0.25 M AA PBS (pH 7.4) | 0.001 ng $mL^{-1}$ to 75 ng $mL^{-1}$ | 0.22 pg $mL^{-1}$ | Effect of steric hindrance | | [281] |
| hybrid film consisting of oppositely charged polyelectrolytes and CdS QDs | ITO | protein | 0.01 M phosphate buffer, pH 7.4 | 0.5 pg/mL to 5.0 ng/mL | 0.5 pg/mL | Effect of steric hindrance | | [282] |
| g-$C_3N_4$ nanosheets (AuNP/g-$C_3N_4$) | glassy carbon | deoxyribozyme (DNAzyme) | 10 mM Tris-HCl buffer (pH 7.4) | 2 to 100 mU $mL^{-1}$ | 1.0 mU $mL^{-1}$ | Effect of steric hindrance | | [223] |
| carboxyl-functionalized graphene CdSe NPs. | ITO | $H_2O_2$ | 1M NaOH | 2 x $10^{-14}$ to 2 x $10^{-13}$ mM | 5.9 x $10^{-15}$ mM | Effect of steric hindrance | | [286] |